\documentclass[aps,prd,floatfix,twocolumn,superscriptaddress,preprintnumbers,nofootinbib,10pt]{revtex4-1}
\pdfoutput=1

\usepackage{graphicx}
\usepackage{amssymb}
\usepackage{amsmath,amssymb}
\usepackage{color}

\usepackage{slashed}

\graphicspath{{./}{figures/}}

\definecolor{darkblue}{rgb}{0,0,0.5}
\usepackage[colorlinks,linkcolor=darkblue,citecolor=darkblue,urlcolor=darkblue]{hyperref}

\allowdisplaybreaks

\begin{document}
\preprint{EFI-12-30}
\preprint{FERMILAB-PUB-12-590-T}
\preprint{MCTP-12-28}


\title{Indirect Probes of the MSSM after the Higgs Discovery}

\author{Wolfgang Altmannshofer}
\email{waltmann@fnal.gov}
\affiliation{Fermi National Accelerator Laboratory, 
P.O.~Box 500, Batavia, IL 60510, USA}

\author{Marcela Carena}
\email{carena@fnal.gov}
\affiliation{Fermi National Accelerator Laboratory, 
P.O.~Box 500, Batavia, IL 60510, USA}
\affiliation{Enrico Fermi Institute, 
University of Chicago, Chicago, IL 60637, USA}
\affiliation{Kavli Institute for Cosmological Physics, 
University of Chicago, Chicago, IL 60637, USA}

\author{Nausheen R. Shah}
\email{naushah@umich.edu}
\affiliation{Michigan Center for Theoretical Physics, Department of Physics, University of Michigan, 
Ann Arbor, MI 48109, USA}

\author{Felix Yu}
\email{felixyu@fnal.gov}
\affiliation{Fermi National Accelerator Laboratory, 
P.O.~Box 500, Batavia, IL 60510, USA}

\begin{abstract}
 
We study the minimal supersymmetric standard model (MSSM) with minimal
flavor violation (MFV), imposing constraints from flavor physics
observables and MSSM Higgs searches, in light of the recent discovery
of a 125 GeV Higgs boson by ATLAS and CMS.  We analyze the electroweak
vacuum stability conditions to further restrict the MSSM parameter
space. In addition, a connection to ultraviolet physics is shown via
an implementation of renormalization group running, which determines
the TeV-scale spectrum from a small set of minimal supergravity
parameters.  Finally, we investigate the impact from dark matter
direct detection searches.  Our work highlights the complementarity of
collider, flavor and dark matter probes in exploring the MSSM, and
shows that even in a MFV framework, flavor observables constrain the
MSSM parameter space well beyond the current reach of direct SUSY
particle searches.

\end{abstract}

\maketitle

\section{Introduction} \label{sec:intro}

Despite the null results for direct searches of supersymmetric
particles at the LHC~\cite{ATLAS:2012ds, ATLAS:2012dp, ATLAS:2012kr,
  ATLAS:2012ky, ATLAS:2012jp, ATLAS:2012ht, :2012fw, :2012uu, :2012yr,
  :2012ae, :2012ms, :2012tx, :2012ku, :2012gg, :2012ar, :2012si,
  :2012rz, :2012pq, Aad:2012hm, Chatrchyan:2012qka,
  Chatrchyan:2012mea, Chatrchyan:2012sa, :2012th, Chatrchyan:2012te,
  :2012mfa, :2012jx, :2012rg, :2012ew, CMS:2012un}, models of
supersymmetry (SUSY) remain among the most well-motivated and popular
extensions of the Standard Model (SM). Besides direct searches, there
exist numerous ways to indirectly probe SUSY models, {\it e.g.} with
low energy flavor observables, from dark matter direct detection
results, and through Higgs properties.  The discovery of a new
particle at the LHC with a mass of $\sim 125$~GeV compatible with a
SM-like Higgs boson~\cite{:2012gk,:2012gu} has far reaching
consequences for any model of New Physics (NP) with a non-standard
Higgs sector.  Indeed, the LHC Higgs results have motivated numerous
studies of their implications in the context of the Minimal
Supersymmetric Standard Model (MSSM) and its
variants~\cite{Hall:2011aa, Baer:2011ab, Feng:2011aa,
  Heinemeyer:2011aa, Arbey:2011ab, Arbey:2011aa, Draper:2011aa,
  Carena:2011aa, Ellwanger:2011aa, Buchmueller:2011ab, Akula:2011aa,
  Kadastik:2011aa, Cao:2011sn, Gunion:2012zd, King:2012is, Kang:2012sy,
  Aparicio:2012iw, Ellis:2012aa, Baer:2012uya, Cao:2012fz,
  Maiani:2012ij, Christensen:2012ei, Vasquez:2012hn, Ajaib:2012vc,
  Brummer:2012ns, Feng:2012jf, Carena:2012gp, Fowlie:2012im,
  Blum:2012ii, CahillRowley:2012rv, Benbrik:2012rm, Arbey:2012dq,
  Akula:2012kk, An:2012vp, Cao:2012yn, Giudice:2012pf,
  Buchmueller:2012hv, Espinosa:2012in, SchmidtHoberg:2012yy,
  Boudjema:2012in, Maiani:2012qf, Baer:2012mv, Drees:2012fb,
  Haisch:2012re}.

A SM-like Higgs with a mass of $M_h \simeq 125$~GeV can be
accommodated in the MSSM as long as stops are either very heavy or
strongly mixed.  Interestingly enough, large stop mixing unavoidably
leads to irreducible contributions to low energy flavor observables,
in particular in the Flavor Changing Neutral Current (FCNC) decays
$B_s \to \mu^+ \mu^-$ and $B \to X_s \gamma$, even if all soft masses
are flavor blind. Correspondingly, rare $B$ decays can be used to set
non-trivial constraints on MSSM parameters.

In this work, we discuss the status of the MSSM, in view of the recent
Higgs search results from the LHC~\cite{:2012gk,:2012gu} and the
Tevatron~\cite{:2012cn}, the recent strong limits on MSSM Higgs bosons
in $H/A \to b\bar b$
searches~\cite{CMS-PAS-HIG-12-026,CMS-PAS-HIG-12-027} and $H/A \to
\tau^+\tau^-$ searches~\cite{CMS-PAS-HIG-12-050} , the latest results
in $B$ physics, in particular the recent evidence for $B_s \to \mu^+
\mu^-$ from LHCb~\cite{1211.2674}, the updated results on $B \to \tau
\nu$ from Belle~\cite{Adachi:2012mm} and
BaBar~\cite{Collaboration:2012ju} as well as on $B \to X_s \gamma$
from BaBar~\cite{:2012iwb}, and also the updated Xenon100 bounds on
dark matter direct detection~\cite{:2012nq}.  We will assume that the
flavor structure of the SUSY particles is determined by the principle
of minimal flavor violation (MFV)~\cite{D'Ambrosio:2002ex,
  Chivukula:1987py, Hall:1990ac, Buras:2000dm}, {\it i.e.} we will
assume that the SM Yukawa couplings are the only sources of flavor
violation. This is motivated by the absence of any unambiguous
deviation from SM expectations in flavor observables.  We emphasize
that even in this restrictive framework, flavor observables play an
important role in constraining the viable parameter space of the
MSSM. Indeed, flavor bounds can be stronger than bounds from direct
searches for SUSY particles in various regions of parameter
space. This is particularly true for large values of $\tan \beta$,
where loop-induced flavor changing couplings of the heavy Higgs bosons
of the MSSM give enhanced contributions to FCNC processes.  In the
MSSM with large $\tan \beta$, direct searches of the heavy Higgs
bosons also become especially sensitive and, moreover, the exchange of
heavy Higgs bosons can also lead to large dark matter direct detection
cross sections, giving additional complementary means to probe this
region of parameter space.

In the large $\tan \beta$ regime, loop corrections to Higgs--fermion
couplings can be significant and it is mandatory to resum $\tan
\beta$--enhanced terms to obtain reliable predictions for any
observables that depend on these couplings in the MSSM.  We provide
comprehensive analytical expressions for all the relevant
loop-corrected Higgs couplings, loop corrections to the SM-like Higgs
mass, Higgs production and decay rates, contributions to flavor
observables, and dark matter direct detection cross sections,
consistently taking into account the most general structure of the
soft SUSY breaking terms compatible with the MFV ansatz. In
particular, we include effects from the bottom Yukawa coupling as well
as the tau Yukawa coupling, as they are relevant for large $\tan
\beta$. This goes beyond the analyses in~\cite{Carena:2006ai,
  Carena:2007aq, Carena:2008ue}, where bottom Yukawa effects in the
squark masses were neglected.

We also give a detailed treatment of gaugino loop contributions to FCNC
processes that can arise from a mass splitting between the left-handed
squarks of the first two and the third generations.  We highlight that
in order to discuss the gaugino contributions to FCNC processes in the
large $\tan\beta$ regime, both the squark mass splitting and the
alignment of this splitting with the quark Yukawas must be considered.

Putting together all the relevant experimental constraints coming from
current Higgs, flavor and dark matter sectors on the MSSM parameters,
we point out regions of the MFV MSSM parameter space where these
constraints are minimized. We also discuss the robustness of these
bounds and to which extent they can be avoided.  We take a
phenomenological approach and treat the MSSM parameters as free
parameters at the TeV scale: however, we augment this discussion with
a study of renormalization group equation (RGE) effects, assuming
minimal supergravity (mSUGRA)--like boundary conditions at a high
scale and monitoring the generic spectrum of SUSY particles and their
mass splittings induced by the running.

In Sec.~\ref{sec:MSSM}, we review the MSSM with minimal flavor
violation in the quark sector.  The impact of Higgs searches at the
LHC and Tevatron on the MSSM are analyzed in Sec.~\ref{sec:collider}.
We use the recent results indicating the presence of a SM-like Higgs
as well as dedicated searches for the additional Higgs bosons of the
MSSM.  In Sec.~\ref{sec:vacuum}, we study constraints on large $\mu
\tan \beta$ from vacuum stability considerations. Constraints from $B$
physics observables are analyzed in Sec.~\ref{sec:flavor}.  In
Sec.~\ref{sec:DM}, bounds on the MSSM from dark matter direct
detection are considered.  We conclude in Sec.~\ref{sec:concl}.

\section{The MSSM with Minimal Flavor Violation} \label{sec:MSSM}

In the following, we briefly review the MSSM with MFV.  Throughout
this work, in addition to MFV, we also assume minimal $CP$ violation,
{\it i.e.}  the phase of the CKM matrix is the only source of $CP$
violation, while all the MSSM parameters are $CP$ conserving.  We
discuss the MFV structure of the sfermion masses in
Sec.~\ref{sec:sfermion_spectrum}.  Aspects of the Higgs spectrum that
are relevant for our work are briefly reviewed in
Sec.~\ref{sec:Higgs_spectrum}.  In Sec.~\ref{sec:couplings}, we detail
the $\tan\beta$--enhanced loop corrections to the Higgs-fermion
couplings, allowing for the most general squark spectrum compatible
with our MFV and $CP$ conservation assumptions.

\subsection{Sfermion Spectrum} \label{sec:sfermion_spectrum}

The soft SUSY breaking terms that give mass to the squarks, {\it i.e.}
the soft masses, $m_Q^2$, $m_D^2$ and $m_U^2$, and trilinear
couplings, $A_d$ and $A_u$, are possible sources of flavor
violation. Generic flavor violating entries in these matrices are
strongly constrained by flavor physics data.  A simple approach to
address this ``SUSY flavor problem'' is the principle of minimal
flavor violation~\cite{D'Ambrosio:2002ex, Chivukula:1987py,
  Hall:1990ac, Buras:2000dm}, which states that the SM Yukawa
couplings are the only sources of flavor violation even in extensions
of the SM. In the context of the MSSM, this implies that the soft
terms can be expanded in powers of the Yukawa couplings. In the
super-Cabibbo-Kobayashi-Maskawa (super-CKM) basis, where squarks and
quarks are simultaneously rotated to obtain diagonal Yukawa couplings,
the soft masses are~\cite{D'Ambrosio:2002ex}
\begin{eqnarray}\label{eq:softMFV}
\hat m_Q^2 &=& \tilde m^2_Q \left( 1 + b_1 V^{\dagger} y_u^2 V + b_2 y_d^2 
\right. \nonumber \\
     & &\;\;\; \qquad \left. + b_3 (y_d^2 V^{\dagger} y_u^2 V + V^{\dagger} y_u^2 V y_d^2) 
\right)~, \nonumber \\
\hat m_U^2 &=& \tilde m^2_U \left( 1 + b_4 y_u^2\right)~, \nonumber \\
\hat m_D^2 &=& \tilde m^2_D \left( 1 + b_5 y_d^2\right)~,
\end{eqnarray}
where $y_u$ and $y_d$ are the diagonal up and down quark Yukawa
matrices and $V$ is the CKM matrix.  The soft mass $\hat m_Q^2$ enters
the left-left block of the down squark mass matrix, while $V \hat
m_Q^2 V^\dagger$ enters the up squark mass matrix.  The generic
structure in~(\ref{eq:softMFV}) is always generated by RGE running if
flavor blind boundary conditions are assumed at a high
scale~\cite{Paradisi:2008qh, Colangelo:2008qp}.  The parameters $b_i$
lead to splittings between the squark masses.  To be specific, the
parameters $b_4$ and $b_5$ generate a splitting between the masses of
the first two and the third generations of right-handed up and down
squarks, respectively, while the parameters $b_1$, $b_2$, and $b_3$
generate a splitting between the first two and the third generations
of left-handed squarks.  Note that the parameters $b_2$, $b_3$, and
$b_5$ only become important for large values of $\tan\beta$, where the
bottom Yukawa is $\mathcal{O}(1)$. As we are particularly interested
in the large $\tan\beta$ regime, in the following we will take all the
above masses as independent parameters and use $m_{Q_3}^2$,
$m_{D_3}^2$, and $m_{U_3}^2$ for the stop and sbottom masses and
$m_Q^2$, $m_D^2$, and $m_U^2$ for the masses of the first two
generations, which are degenerate to an excellent approximation in
this setup.  This is analogous to the pMSSM
framework~\cite{Berger:2008cq} frequently studied in the literature.

We stress, however, that the parameters $b_1$, $b_2$, and $b_3$ also
induce flavor violating entries in the left-handed squark mass
matrices.  These entries are proportional to small CKM angles and lead
to controlled but non-negligible contributions to FCNC processes.  In
fact, due to $SU(2)_L$ invariance, the left-left blocks of the up and
down squark mass matrices are related by a CKM rotation, and therefore
{\it any splitting} in the diagonal entries of the left-handed soft
masses $\hat m_Q^2$ unavoidably induces off-diagonal entries in the up
or down squark mass matrices.  Moreover, distinct flavor phenomenology
arises depending on which of the $b_1$, $b_2$, or $b_3$ parameters is
responsible for the splitting.  In particular, a splitting induced by
$b_1$ ($b_2$) is aligned in the up- (down-) sector and will only lead
to off-diagonal entries in the down (up) squark mass matrix. The
parameter $b_3$ induces flavor violation in both the up and down
squark masses matrices.  All flavor observables that we will discuss
in the following depend on the combination $b_1 + b_3 y_b^2$.  We thus
introduce one additional parameter
\begin{equation}
 \zeta = \frac{b_1 y_t^2 + b_3 y_b^2 y_t^2}{
b_1 y_t^2 + b_2 y_b^2 + 2 b_3 y_b^2 y_t^2} ~,
\end{equation}
which reflects the alignment of the splitting in the left-handed
squark masses and hence parametrizes the fraction of the splitting in
the masses leading to flavor violation in the down sector.  We assume
$\zeta$ is real in the following.\footnote{Note that while $b_1$ and
  $b_2$ have to be real due to hermiticity of the squark masses, $b_3$
  can in principle be complex. Indeed, as shown
  in~\cite{Colangelo:2008qp}, a tiny phase for $b_3$ is always
  generated during RGE running.}  We see that formally $\zeta = 1 +
\mathcal{O}(y_b^2)$. If we consider a splitting in the squark masses
that is radiatively induced through RGE running, then considering only
the top Yukawa in the running leads to $\zeta = 1$. Bottom Yukawa
effects become important for large $\tan\beta$ and can lead to $0 <
\zeta < 1$.  Typically we expect that $y_b$ is at most as large as
$y_t$, however, which implies $1/2<\zeta<1$.

We note that an expansion analogous to~(\ref{eq:softMFV}) also exists
for the trilinear couplings~\cite{D'Ambrosio:2002ex}. In particular,
higher order terms in the expansion can lead to flavor violating
trilinear terms.  Such terms only lead to corrections of the
holomorphic Higgs couplings, however. These corrections can induce
flavor changing neutral Higgs couplings, that are especially
interesting beyond MFV, where the corresponding effects can be
chirally enhanced~\cite{Crivellin:2010er, Crivellin:2011jt}.  In the
MFV framework considered here, these effects are less important
compared to contributions that are related to the loop-induced
non-holomorphic Higgs couplings.  The only relevant trilinear
couplings for our analysis are those for the third generation squarks,
$A_t$ and $A_b$, which we will take to be independent parameters.

For simplicity, we will also assume universal soft masses $m_L^2$ and
$m_E^2$, in the slepton sector.  The phenomenology of flavor
non-universalities in the lepton sector will be reserved for future
study.  The only relevant trilinear term in the slepton sector is the
tau trilinear coupling $A_\tau$, which, along with $A_t$ and $A_b$ and
all other parameters, we will take to be real.

\subsection{Higgs Spectrum} \label{sec:Higgs_spectrum}

The physical Higgs spectrum of the MSSM consists of two neutral scalar
bosons $h$ and $H$, one neutral pseudoscalar $A$, and a pair of
charged Higgs bosons $H^\pm$. At tree level, the full spectrum is
determined by only two real parameters: the mass of the pseudoscalar
Higgs, $M_A$, and the ratio of the two vacuum expectation values,
$\tan\beta = t_\beta = v_u/v_d$, with $v_u^2+v_d^2 = v^2 =
174^2$~GeV$^2$.  In the so-called decoupling limit, $M_A^2 \gg M_W^2$,
the masses of the Higgs bosons, $A$, $H$ and $H^\pm$, are
\begin{equation}
M_H^2 \simeq M_A^2 ~,~~ M_{H^\pm}^2 \simeq M_A^2 + M_W^2 ~.
\end{equation}
In this limit, the mass of the lightest Higgs $h$ is given at tree
level by
\begin{equation}
M_h^2 \simeq M_Z^2 \cos2\beta ~.
\end{equation}
As is well known, moderate or large values of $\tan \beta$ and large
1-loop corrections are required to lift $M_h$ up to phenomenologically
viable values.  Moreover, at large $\tan \beta$, the sbottom and stau
1-loop corrections can lower $M_h$ by a few GeV, which cannot be
neglected given the current Higgs mass precision data.  The dominant
stop, sbottom, and stau loop contributions for large $\tan \beta$ read
\begin{eqnarray} \label{eq:DeltaMhiggs}
\Delta M_h^2 \simeq&& \frac{3}{4\pi^2} \frac{m_t^4}{v^2} 
\left[ \log\left( \frac{m_{\tilde t}^2}{m_t^2} \right) + 
                  \frac{X_t^2}{m_{\tilde t}^2} - 
                  \frac{X_t^4}{12 m_{\tilde t}^4} \right] \nonumber  \\
&& - \frac{3}{48\pi^2} \frac{m_b^4}{v^2} ~\frac{t_\beta^4}{
(1+\epsilon_b t_\beta)^4}~\frac{\mu^4}{m_{\tilde b}^4} \nonumber \\
&& - \frac{1}{48\pi^2} \frac{m_\tau^4}{v^2} ~\frac{t_\beta^4}{
(1+\epsilon_\ell t_\beta)^4}~\frac{\mu^4}{m_{\tilde \tau}^4} \ ,
\end{eqnarray}
where $X_t = A_t - \mu/\tan \beta \approx A_t$ for large $\tan \beta$,
and $m_{\tilde{t}}$, $m_{\tilde{b}}$ and $m_{\tilde{\tau}}$ are the
average stop, sbottom, and stau masses, respectively.  The stop loop
corrections, reported in the first line of~(\ref{eq:DeltaMhiggs}), are
maximized for $A_t \simeq \sqrt{6} m_{\tilde t}$.  The contributions
from the sbottom and stau loops, in the second and third lines, always
reduce the light Higgs mass and can be particularly important for
large $\tan\beta$, large values of the Higgsino mass parameter, $\mu$,
and light sbottom or stau masses~\cite{Carena:1995wu}.  The
$\epsilon_i$ factors come from an all-order resummation of $\tan\beta$
enhanced corrections to the Higgs--fermion couplings and are discussed
in detail in Sec.~\ref{sec:couplings}.

The couplings of the lightest Higgs to SM fermions and gauge bosons
are mainly controlled by $\tan\beta$ and the angle $\alpha$ that
diagonalizes the mass matrix of the two scalar Higgs bosons.  If
\begin{equation} \label{eq:alpha}
\alpha = \beta - \pi/2 ~,
\end{equation}
the couplings of $h$ are exactly SM-like.  At the tree level,
Eq.~(\ref{eq:alpha}) holds up to corrections of order $M_Z^2/(t_\beta
M_A^2)$. Correspondingly, for large $\tan\beta$ and moderately heavy
$M_A$, the couplings of $h$ are already SM-like to a good
approximation. At 1-loop, Eq.~(\ref{eq:alpha}) gets corrected by an
additional term $\sim \lambda_7 v^2/M_A^2$, where $\lambda_7$ is a
loop-induced quartic Higgs coupling that reads
\begin{eqnarray}
\lambda_7 &\simeq& \frac{3}{96\pi^2} \frac{m_t^4}{v^4} 
\frac{\mu A_t}{m_{\tilde t}^2} 
\left( \frac{A_t^2}{m_{\tilde t}^2}- 6\right)\nonumber \\
&& +\frac{3}{96\pi^2} \frac{m_b^4}{v^4} 
\frac{t_\beta^4}{(1+\epsilon_b t_\beta)^4} 
\frac{\mu^3 A_b}{m_{\tilde b}^4} \nonumber \\
&& +\frac{1}{96\pi^2} \frac{m_\tau^4}{v^4} 
\frac{t_\beta^4}{(1+\epsilon_\ell t_\beta)^4} 
\frac{\mu^3 A_\tau}{m_{\tilde \tau}^4}~.
\end{eqnarray}
If $\lambda_7$ is sizable, corrections to the light Higgs couplings
can become relevant, as discussed in the next section, and are
constrained by the SM Higgs searches at the LHC.

\subsection{Higgs Couplings to Fermions} \label{sec:couplings}

At tree level, the MSSM Higgs sector is a 2 Higgs doublet model of
type II, where only $H_u$ couples to right-handed up quarks and only
$H_d$ couples to right-handed down quarks and leptons. The Yukawa
interactions thus have the following form
\begin{eqnarray}
\mathcal{L}_{\rm Yuk} &=& (y_u)_{ij} ~H_u \bar Q_i U_j + 
(y_d)_{ij} ~H_d \bar Q_i D_j \nonumber \\
&& + (y_\ell)_{ij} ~H_d \bar L_i E_j  ~+ {\rm h.c.}~.
\end{eqnarray}
As a consequence, the couplings of the neutral Higgs bosons to
fermions are flavor diagonal in the mass eigenstate basis. At the loop
level, however, ``wrong'' Higgs couplings are generated and lead to
potentially large threshold corrections to the masses of down type
quarks and leptons~\cite{Hempfling:1993kv, Hall:1993gn, Carena:1994bv,
  Dobrescu:2010mk} as well as CKM matrix
elements~\cite{Blazek:1995nv}. They also significantly modify charged
Higgs couplings to quarks~\cite{Carena:1999py,Carena:2000uj} and
generate flavor changing neutral Higgs
couplings~\cite{Hamzaoui:1998nu, Babu:1999hn, Isidori:2001fv,
  Dedes:2002er, Buras:2002vd, Hofer:2009xb, Crivellin:2010er, Crivellin:2011jt}.  All
these effects become particularly relevant in the large $\tan\beta$
regime, where the inherent 1-loop suppression can be partly
compensated.  In the following, we analyze the form of the neutral and
charged Higgs couplings with fermions in the phenomenologically
motivated limit, $v^2 \ll M_{\rm SUSY}^2$ (see~\cite{Hofer:2009xb} for
a discussion of the regime $v^2 \sim M_{\rm SUSY}^2$).  We
consistently take into account the most generic MFV structure of the
squark masses as discussed in Sec.~\ref{sec:sfermion_spectrum}. In
particular, we consider splittings between the first two and the third
generation squarks in the left-handed as well as right-handed sectors.

Once the 1-loop corrections are taken into account and we have
diagonalized the quark mass matrices, the couplings of the neutral
Higgs mass eigenstates to fermions have the following generic form
\begin{eqnarray} \label{eq:interactions}
\mathcal{L}_{\rm int} &\supset& \sum_{q,q^\prime}\frac{m_q}{\sqrt{2}
  v} (\bar q^\prime_L q_R) \left( \xi_{q^\prime q}^h h + \xi_{q^\prime
  q}^H H + i \xi_{q^\prime q}^A A \right) \\ &+&
\sum_{\ell}\frac{m_\ell}{\sqrt{2} v} (\bar \ell_L \ell_R) \left(
\xi_\ell^h h + \xi_\ell^H H + i \xi_\ell^A A \right) + \text{ h.c. ,}
\nonumber
\end{eqnarray}
where we have neglected flavor changing couplings to leptons, which
are not relevant for our analysis.  Using the notation $\xi_{qq}^i =
\xi_q^i$, and again for large $\tan \beta$, the flavor conserving
couplings of the heavy scalar and pseudoscalar Higgses, normalized to
their respective SM Yukawas, are
\begin{eqnarray}
-\xi_q^H &\simeq& \xi_q^A \simeq \frac{1}{t_\beta} - \epsilon_q \ , \quad 
\text{ for } q = u,c,t ~, \\
\xi_q^H &\simeq& \xi_q^A \simeq \frac{t_\beta}{1+\epsilon_q t_\beta} \ , \quad
\text{ for } q = d,s,b ~, \\
\xi_\ell^H &\simeq& \xi_\ell^A \simeq \frac{t_\beta}{1+\epsilon_\ell t_\beta} ~.
\end{eqnarray}
In the above expressions, $\tan \beta$-enhanced corrections to the
couplings are resummed to all orders and the $\epsilon_i$ factors
parametrize the loop induced ``wrong'' Higgs couplings.  The exact
form of each $\epsilon_i$ in terms of MSSM parameters is discussed at
the end of this section. Since we assume the MSSM parameters to be
$CP$ conserving, all $\epsilon_i$ parameters are real.

Among the flavor changing Higgs couplings only the coupling of a right
handed bottom with a left-handed strange quark will be relevant in the
following discussion. Normalized to the bottom Yukawa of the SM, we
have
\begin{equation} \label{eq:xi_sb}
\xi_{sb}^H \simeq \xi_{sb}^A \simeq \frac{\epsilon_{\rm FC} ~ t^2_\beta}{
(1+\epsilon_b t_\beta)(1+\epsilon_0 t_\beta)}~  V_{tb} V_{ts}^*~, \\
\end{equation}
where $\epsilon_0$ is defined as $\epsilon_0 = \epsilon_b -
\epsilon_{\rm FC}$, and $\epsilon_{\rm FC}$ is discussed in detail
below.

The couplings of the light Higgs boson, $h$, are exactly SM-like in
the decoupling limit: $\xi_{q}^h = \xi_\ell^h = 1$ and $\xi_{q^\prime
  q}^h = 0$ for $q^\prime \neq q$. While non-standard effects in the
couplings to up-type quarks are generically tiny even away from the
decoupling limit, corrections to the couplings with down-type quarks
and leptons decouple very slowly and can be relevant.  We have
\begin{equation}
\xi_f^h = -\frac{s_\alpha}{c_\beta} ~ \frac{1 - \epsilon_f / t_\alpha}{
1+\epsilon_f t_\beta}~, ~~~ \text{for}~ f = d,s,b,\ell~.
\end{equation}

The couplings of the physical charged Higgs bosons to fermions can be
written as
\begin{eqnarray}
\mathcal{L}_{\rm int} &\supset& 
\sum_{q,q^\prime} \frac{m_q}{v} (\bar q^\prime_L q_R) 
\xi_{q^\prime q}^\pm H^\pm  \\ 
&+& \sum_{\ell} \frac{m_\ell}{v} (\bar \nu^\ell_L \ell_R) 
\xi_{\nu\ell}^\pm H^\pm  + \text{ h.c. .} \nonumber
\end{eqnarray}
For the couplings relevant to our analysis, we have 
\begin{eqnarray}
\frac{\xi_{tb}^\pm}{V_{tb}} &=& \frac{t_\beta}{1+\epsilon_b t_\beta} ~,~~ \qquad \qquad
\frac{\xi_{us}^\pm}{V_{us}} = \frac{t_\beta}{1+\epsilon_s t_\beta} ~, \\
\frac{\xi_{ub}^\pm}{V_{ub}} &=& \frac{\xi_{cb}^\pm}{V_{cb}} = 
\frac{t_\beta}{1+\epsilon_0 t_\beta} ~,~~ \quad \;
\xi_{\nu \ell}^\pm = \frac{t_\beta}{1+\epsilon_\ell t_\beta} ~, \\
\frac{\xi_{st}^\pm}{V_{ts}^*}& =& \frac{1}{t_\beta} - \epsilon_0^\prime + 
\epsilon_{\rm FC}^\prime 
\frac{\epsilon_{\rm FC} t_\beta}{1 + \epsilon_0 t_\beta}~,
\end{eqnarray}
where $V_{ij}$ are CKM matrix elements and $\epsilon_0^\prime =
\epsilon_t - \epsilon_{\rm FC}^\prime$. The parameter $\epsilon_{\rm
  FC}^\prime$ is the up-sector analogue of $\epsilon_{\rm FC}$.

As already mentioned, the various $\epsilon$ factors in the above
expressions parametrize loop-induced non-holomorphic Higgs
couplings. They arise from Higgsino-squark loops, gluino-squark loops
and wino-sfermion loops.  We do not explicitly state the typically
negligible contributions coming from bino-sfermion loops; however,
they are included in our numerical analysis.

For the bottom quark, we can decompose $\epsilon_b =
\epsilon_b^{\tilde g} + \epsilon_b^{\tilde W} + \epsilon_b^{\tilde
  H}$, where these contributions are
\begin{eqnarray}
\epsilon_b^{\tilde g} &=& \frac{\alpha_s}{4 \pi} \frac{8}{3} 
~\mu M_3~ g(M_3^2,m_{Q_3}^2,m_{D_3}^2) ~, \\
\epsilon_b^{\tilde W} &=& -\frac{\alpha_2}{4 \pi} \frac{3}{2} 
~\mu M_2~ g(\mu^2,M_2^2,m_{Q_3}^2) ~, \\
\label{eq:epsilon_bH}
\epsilon_b^{\tilde H} &=& \frac{\alpha_2}{4\pi} \frac{m_t^2}{2M_W^2} 
~\mu A_t~ g(\mu^2,m_{Q_3}^2,m_{U_3}^2) ~. 
\end{eqnarray}
The loop function $g$ is listed in the appendix, and has units of
(GeV)$^{-2}$.  Hence, the $\epsilon$ factors generally exhibit
non-decoupling as the SUSY mass scale increases.  In particular,
rescaling all the SUSY mass parameters, {\it i.e.} the squark masses,
gaugino masses, the Higgsino mass parameter and the trilinear coupling
by a common, arbitrarily large factor, leaves the $\epsilon$
parameters invariant. For a degenerate SUSY spectrum with mass $\tilde
m$, we obtain $g(\tilde m^2,\tilde m^2,\tilde m^2) = 1/2 \tilde m^2$.
Our sign convention is that the left-right mixing entries in the top
and bottom squark mass matrices are given by $m_t (A_t - \mu
\cot\beta)$ and $m_b(A_b - \mu \tan\beta)$, respectively.
Furthermore, the gluino mass $M_3$ is always positive in our
convention.

For the strange and down quarks, the Higgsino contribution is highly
suppressed by small Yukawa couplings or CKM angles and only the gluino
and wino loops are relevant: $\epsilon_{s,d} = \epsilon_{s,d}^{\tilde
  g} + \epsilon_{s,d}^{\tilde W}$, where the $\epsilon_{s,d}^i$ can be
easily obtained from the corresponding $\epsilon_b^i$ expressions,
replacing third generation squark masses with second or first
generation squark masses.

For leptons, only the wino (and the bino) loops give contributions,
and $\epsilon_\ell^{\tilde W}$ is given by $\epsilon_b^{\tilde W}$
with the sbottom masses replaced by the slepton masses.

In case of the top quark, analogous to the bottom quark, we consider
the gluino, wino, and Higgsino contributions: $\epsilon_t =
\epsilon_t^{\tilde g} + \epsilon_t^{\tilde W}+ \epsilon_t^{\tilde
  H}$. The expressions for $\epsilon_t^{\tilde g}$ and
$\epsilon_t^{\tilde W}$ are trivially obtained from the corresponding
$\epsilon_b^i$ by replacing the relevant squark masses. The Higgsino
contribution is explicitly given by
\begin{equation}
\epsilon_t^{\tilde H} = -\frac{\alpha_2}{4\pi} \frac{m_b^2}{2M_W^2} 
~\frac{t^2_\beta}{(1+\epsilon_b t_\beta)^2}~ 
~\mu A_b~ g(\mu^2,m_{Q_3}^2,m_{D_3}^2) ~. 
\end{equation}
Here, $\epsilon_t^{\tilde H}$ is suppressed by the bottom quark mass
and only becomes relevant for large values of $\tan\beta$.

The flavor changing couplings, $\epsilon_{\rm FC}$ and $\epsilon_{\rm
  FC}^\prime$, can be decomposed as
\begin{eqnarray} \label{eq:epsilon_FC}
\epsilon_{\rm FC} &=& \epsilon_b^{\tilde H} + \zeta
\epsilon_{\rm FC}^{\tilde g} + \zeta \epsilon_{\rm FC}^{\tilde W} ~, \\
\epsilon_{\rm FC}^\prime &=& \epsilon_t^{\tilde H} + (1-\zeta)
\epsilon_{\rm FC}^{\prime \, \tilde g} + (1-\zeta) \epsilon_{\rm FC}^{\tilde W} ~,
\end{eqnarray}
with
\begin{eqnarray}
\label{eq:epsilon_FCg}
\epsilon_{\rm FC}^{\tilde g} &=& \frac{\alpha_s}{4\pi} \frac{8}{3} 
~\mu M_{\tilde g} \nonumber \\
&& \times \big(  g(M_3^2,m_{Q_3}^2,m_{D_3}^2) -  g(M_3^2,m_Q^2,m_{D_3}^2) \big)  ~, \\
\label{eq:epsilon_FCw}
\epsilon_{\rm FC}^{\tilde W} &=& -\frac{\alpha_2}{4 \pi} \frac{3}{2} 
~\mu M_2 \nonumber \\
&& \times \big( g(\mu^2,M_2^2,m_{Q_3}^2) - g(\mu^2,M_2^2,m_Q^2) \big) ~,
\end{eqnarray}
and $\epsilon_{\rm FC}^{\prime \, \tilde g}$ is obtained from
$\epsilon_{\rm FC}^{\tilde g}$ by replacing the right-handed sbottom
mass, $m_{D_3}$, with the right-handed stop mass, $m_{U_3}$.  The
$\epsilon_b^{\tilde H}$ and $\epsilon_t^{\tilde H}$ expressions were
already given above.  Note that $m_{D_3}$ enters both loop functions
in (\ref{eq:epsilon_FCg}) and hence, in general, $\epsilon_{\rm
  FC}^{\tilde g} \neq \epsilon_b^{\tilde g} - \epsilon_s^{\tilde g}$,
in contrast to the case where all right-handed down squarks have the
same mass, $m_{D_3} = m_D$.  Clearly, a splitting between the third
and the first two generations of left-handed squarks induces non-zero
$\epsilon_{\rm FC}^{\tilde g}$, $\epsilon_{\rm FC}^{\tilde W}$ and/or
$\epsilon_{\rm FC}^{\prime \, \tilde g}$.  

As already described in Sec.~\ref{sec:sfermion_spectrum}, $\zeta$
parametrizes the alignment of the left-handed squark mass with the
quark masses.  The case $\zeta = 1$ corresponds to a $m_Q^2$ that is
aligned in the up sector such that the mass splitting between the
first two and the third generations leads to off-diagonal entries only
in the down squark mass matrix. This in turn implies maximal flavor
changing gaugino loop corrections to the Higgs--down quark
couplings. The case $\zeta = 0$ corresponds to alignment in the down
sector, with no off-diagonal entries appearing in the down squark mass
matrix. Generically, from RGE running, we expect $1/2 < \zeta < 1$.

\section{Higgs Collider Searches} \label{sec:collider}

\subsection{SM-like 125 GeV  Higgs } \label{sec:h}

The LHC experiments, ATLAS and CMS, have recently discovered a new
particle with a mass of about 125~GeV~\cite{:2012gk, :2012gu}.  This
discovery is based on results from SM Higgs searches in the $\gamma
\gamma$, $ZZ$ and $ WW$ channels.  The observed signals indicate that
the new particle is a boson with spin 0 or 2, and overall, they are in
reasonable agreement with expectations for a SM Higgs. Other searches
in the $\tau^+ \tau^-$ and $bb$ decay channels are also being pursued,
but more statistics are needed in order to make conclusive statements.

The most visible feature of the extracted signal strength in all the
different channels under study is an enhancement in the $\gamma
\gamma$ decay rate in comparison to the SM rate.  The decay
rates into $WW$ and $ZZ$ gauge bosons are consistent with the SM
values at the $1\sigma$ level.  The present experimental uncertainties
in the signal strength in the various production and decay channels
allow for many new physics alternatives.  In particular, within
supersymmetric extensions, it is possible to enhance or suppress the
gluon fusion production with light stops, depending on the amount of
mixing in the stop sector. It is also possible to suppress gluon
fusion with light sbottoms that have sizable mixing driven by large
values of $\mu \tan \beta$. In all cases, enhancement of gluon fusion
implies a suppression of the $h \to \gamma \gamma$ decay rate, and
vice-versa.  The overall effective $gg \to h \to \gamma\gamma$ rate,
however, is governed by the enhancement or suppression of the gluon
fusion production cross section.

To achieve a net enhancement in the $h \to \gamma \gamma$ rate,
uncorrelated with a simultaneous enhancement in the $h \to WW/ZZ$
rates coming from an enhanced gluon fusion production or reduced $h
\to b\bar b$ partial width, the existence of new, light, charged
colorless particles running in the loop is required. In the MSSM, the
only two options are charginos, which only contribute for $\tan \beta
\sim 1$ (disfavored by a 125 GeV Higgs mass), and light staus with
large mixing, {\it{i.e.}} large $ \mu \tan \beta$. A detailed
discussion of the possible deviations from SM values of the production
and decay rates for a SM-like Higgs in the MSSM can be found
in~\cite{Carena:2011aa, Carena:2012gp}. Possible correlations with
flavor observables have very recently been studied
in~\cite{Haisch:2012re}.

While it is very interesting to investigate deviations from SM
expectations in Higgs data that would point towards new SUSY particles
within the reach of the LHC, we take a different approach in this work
by assuming a Higgs boson with approximately SM-like properties. We
concentrate on possible signatures of new physics that may appear in
$B$ physics observables, direct non-SM Higgs searches and dark matter
direct detection searches within the MSSM with MFV, while fulfilling
the requirement of a 125 GeV SM-like Higgs.  In this way, we show
indirect effects from SUSY particles in flavor and Higgs physics in
regions of parameter space beyond the present reach of the LHC.

\subsection{Searches for Heavy Scalars and Pseudoscalars} \label{sec:HA}

Searches for the heavy neutral Higgs bosons of the MSSM have been
performed in the $H/A \to bb$ and $H/A \to \tau^+ \tau^-$ channels
both at the Tevatron~\cite{Aaltonen:2009vf, Abazov:2011jh,
  Abazov:2011qz, Aaltonen:2012zh} and the LHC~\cite{Chatrchyan:2012vp,
  ATLAS-CONF-2012-094, CMS-PAS-HIG-12-026, CMS-PAS-HIG-12-027,
  CMS-PAS-HIG-12-050}.

Searches also exist for light charged Higgs bosons in top decays at
both the Tevatron~\cite{Aaltonen:2009ke, Abazov:2009aa} and the
LHC~\cite{Aad:2012tj, ATLAS-CONF-2011-094, :2012cwb}.  For the MSSM
scenarios considered in this work, however, the corresponding bounds
are not competitive with the bounds from searches of the neutral Higgs
bosons.

In the large $\tan\beta$ regime, the cross sections for the heavy
scalar and pseudoscalar Higgses rescale according to
\begin{eqnarray}
&&\sigma_{bb \to H} \simeq \sigma_{bb \to A} \simeq \sigma_{bb \to h}^{\rm SM} \times 
\frac{t_\beta^2}{(1 + \epsilon_b t_\beta)^2} ~, \\
&&\sigma_{gg \to H} \simeq \sigma_{gg \to A} \simeq \sigma^{tt,~\rm SM}_{gg \to h} \times 
\left(\frac{1}{t_\beta} - \epsilon_t \right)^2 \nonumber \\ 
&& ~~ + \sigma^{tb,~\rm SM}_{gg \to h} \times 
\frac{1 - \epsilon_t t_\beta}{1 + \epsilon_b t_\beta} + \sigma^{bb,~\rm SM}_{gg \to h} \times 
\frac{t_\beta^2}{(1 + \epsilon_b t_\beta)^2} \ ,
\end{eqnarray}
evaluated at a common mass for all Higgs bosons.  For large
$\tan\beta$, the $\sigma_{bb \to H/A}$ production cross sections can
dominate over gluon fusion.  We use \verb|HIGLU|~\cite{Spira:1995mt}
and \verb|bbh@nnlo|~\cite{Harlander:2003ai} to compute the respective
SM cross sections $\sigma^{i,~\rm SM}_{gg \to h}$ and $\sigma_{bb \to
  h}^{\rm SM}$ at the LHC.

The most important decay modes of the heavy Higgs bosons are $H,A \to
bb$ and $H,A \to \tau^+ \tau^-$.  The corresponding partial widths can
be written as
\begin{eqnarray}
 \Gamma_{Hbb} \simeq \Gamma_{Abb} &\simeq& \Gamma_{h bb}^{\rm SM} \times 
\frac{t_\beta^2}{(1 + \epsilon_b t_\beta)^2} ~,\\
 \Gamma_{H\tau\tau} \simeq \Gamma_{A\tau\tau} &\simeq& \Gamma_{h \tau\tau}^{\rm SM} 
\times \frac{t_\beta^2}{(1 + \epsilon_\tau t_\beta)^2} ~,
\end{eqnarray}
where $\Gamma_{h ff}^{\rm SM}$ are the corresponding decay widths of a
Higgs boson with the same mass as $H$ and $A$ and with SM-like
couplings to $bb$ and $\tau^+ \tau^-$. In our numerical analysis, we
compute $\Gamma_{h ff}^{\rm SM}$ using
\verb|HDECAY|~\cite{Djouadi:1997yw}.

Note that the main dependence of the production cross sections and
branching ratios is on $\tan\beta$ and the heavy Higgs
masses. Dependence on other MSSM parameters enters only at the loop
level through the $\tan\beta$ resummation factors $\epsilon_i$.

\begin{figure}[tb]
\centering
\includegraphics[width=0.45\textwidth]{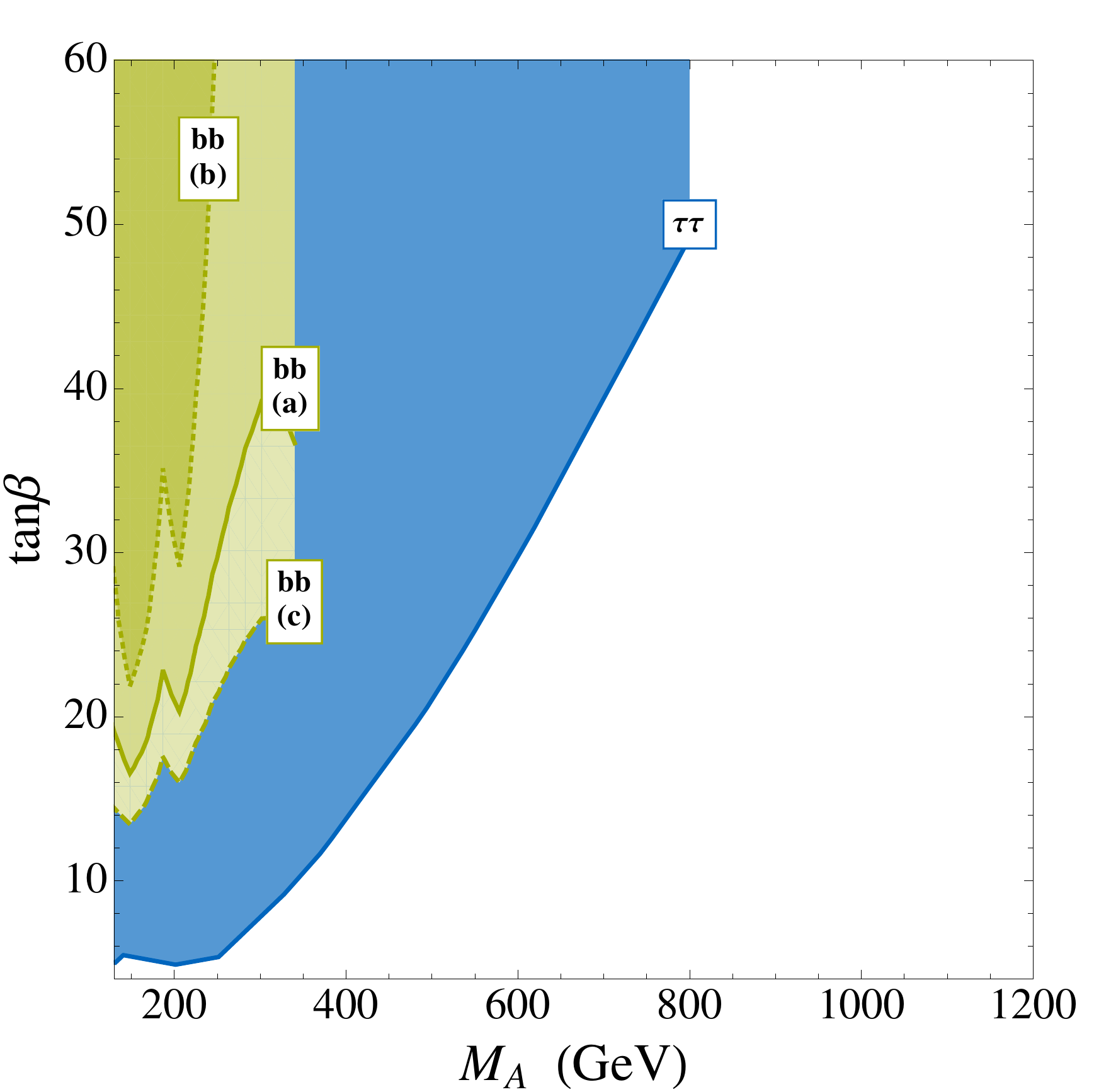}
\caption{Constraints in the $M_A$--$\tan\beta$ plane from direct
  searches of the neutral MSSM Higgs bosons at CMS and ATLAS. The
  solid, dotted and dashed lines correspond to scenarios (a), (b), and
  (c) as defined in Tab.~\ref{tab:scenarios}. The blue (green) regions
  are excluded by searches in the $\tau^+ \tau^-$ ($bb$) channel.}
\label{fig:collider}
\end{figure}

\begin{table}[tb]
\renewcommand{\arraystretch}{1.7}
\renewcommand{\tabcolsep}{6pt}
\small
\begin{center}
\begin{tabular}{|c|ccccc|}
\hline
Scenario & (a) & (b) & (c) & (d) & (e)\\
\hline\hline
$\mu$ [TeV]& 1 & 4 & -1.5 & 1 & -1.5\\
sign($A_t$) & + & + & + & - & - \\
\hline
\end{tabular}
\end{center}
\caption{Illustrative MSSM scenarios discussed in the text. All
  sfermion masses are set to a common value 2 TeV, the gaugino masses
  to $6 M_1 = 3 M_2 = M_3 = 1.5$~TeV. The trilinear couplings $A_t =
  A_b = A_\tau$ are set such that the lightest Higgs mass is $M_h =
  125$~GeV.}
\label{tab:scenarios} 
\end{table}

In our framework, the most important constraints come from the CMS
bounds in the $\tau^+ \tau^-$ channel~\cite{CMS-PAS-HIG-12-050}, which
are available up to masses of $M_A = 800$~GeV and the $b \bar b$
channel~\cite{CMS-PAS-HIG-12-026,CMS-PAS-HIG-12-027} which cover heavy
Higgs masses up to $M_A < 350$~GeV.  Our estimates for the excluded
regions from the $H/A \to b\bar b$ searches are shown in
Fig.~\ref{fig:collider} in yellow-green and labeled with $bb$.  We set
all sfermion masses to 2 TeV and the gaugino masses to $6 M_1 = 3 M_2
= M_3 = 1.5$ TeV. The solid, dotted and dashed contours correspond to
a Higgsino mass parameter $\mu = 1$ TeV (scenario a), $4$ TeV
(scenario b) and $-1.5$ TeV (scenario c), respectively. For every
point in the $M_A$--$\tan\beta$ plane, the trilinear couplings $A_t =
A_b = A_\tau$ are positive and chosen such that the lightest Higgs
mass, computed using \verb|FeynHiggs|~\cite{Heinemeyer:1998yj}, is
$M_h = 125$~GeV.\footnote{The Higgs mass, $M_h$, is not a monotonic
  function in $A_t$ and for a given sign of $A_t$ there are typically
  two choices of $A_t$ that lead to $M_h = 125$~GeV. We always take
  the $A_t$ that is smaller in magnitude.} The respective choices for
$\mu$ lead to representative values for the $\tan\beta$ resummation
factors of $\epsilon_b^{\tilde g} + \epsilon_b^{\tilde W} \simeq
3.3\times 10^{-3}$, $1.5\times 10^{-2}$ and $-5.1 \times 10^{-3}$.

As is well known, the bounds in the $M_A$--$\tan\beta$ plane coming
from the $\tau^+ \tau^-$ channel are robust against variations of the
MSSM parameters. Indeed, the dependence of the production cross
section on $\epsilon_b$ is largely cancelled by the corresponding
dependence of the BR$(A,H \to \tau^+ \tau^-)$~\cite{Carena:2002qg,
  Carena:2005ek}.  In Fig.~\ref{fig:collider}, we therefore simply
report in blue the $\tan\beta$ bounds obtained
in~\cite{CMS-PAS-HIG-12-050} in the so-called $M_h^{\rm max}$
scenario.  We checked explicitly that the constraints are largely
independent of the scenarios in Tab.~\ref{tab:scenarios}. We find that
the constraints can only be weakened mildly for large $\tan \beta$ and
$M_A$ if the MSSM parameters are such that $\epsilon_b$ is sizable and
positive, as in scenario (b)~\footnote{The CMS results in the $\tau^+ \tau^-$ channel are only
  available as constraints in the $M_A$--$\tan\beta$ plane for the
  $M_h^{\rm max}$ scenario. We translate these constraints into bounds
  on the corresponding $\sigma \times $BR and then reinterpret the
  cross section bounds as constraints in the $M_A$--$\tan\beta$ plane
  for various choices of the other MSSM parameters summarized in
  Tab.~\ref{tab:scenarios}.  We assume constant efficiencies
  throughout this procedure.}.  We note however, that in the region
with low $\tan\beta$, the bounds do depend to some extent on the SUSY
spectrum, in particular the neutralino and chargino spectrum. Indeed,
for low $\tan\beta$, the heavy scalar and pseudoscalar Higgs bosons
can have sizable branching ratios in neutralinos or charginos if these
decays are kinematically allowed.  The $M_h^{\rm max}$ scenario
considered in~\cite{CMS-PAS-HIG-12-050} contains light neutralinos
with $M_{\chi_1} \simeq 95$ GeV. For small $\tan\beta$, the obtained
bounds from the searches in the $\tau^+ \tau^-$ channel are therefore
slightly weaker compared to scenarios with heavier neutralinos.

The CMS searches in the $bb$ channel~\cite{CMS-PAS-HIG-12-026,
  CMS-PAS-HIG-12-027} are not yet competitive with the $\tau^+ \tau^-$
searches, but might become important for large $M_A$ in the
future~\cite{Carena:2012rw}.  Compared to the $\tau^+ \tau^-$
searches, the bounds coming from the $bb$ searches show a stronger
dependence on the remaining MSSM parameters~\cite{Carena:2002qg,
  Carena:2005ek}. In particular, for large negative $\mu$, the bounds
become significantly stronger, while for large positive $\mu$, the
bounds can be weakened considerably.  Note that for large negative
$\mu$ and large $\tan\beta$, however, constraints from vacuum
stability and perturbativity of the bottom Yukawa have to be taken
into account.

Since the theoretical precision of the light Higgs mass prediction in
the MSSM allows shifts of a few GeV, we checked the extent to which
the $H/A \to \tau^+ \tau^-$ and $H/A \to b\bar b$ bounds depend on the
exact value of the Higgs mass assumed in our analysis, $M_h =
125$~GeV.  We find that varying the light Higgs mass in the range
$122$~GeV $< M_h < 128$~GeV does not change the constraints from $H$
and $A$ searches significantly.

\section{Vacuum Stability} \label{sec:vacuum}

\begin{figure*}[tb]
\centering
\includegraphics[width=0.45\textwidth]{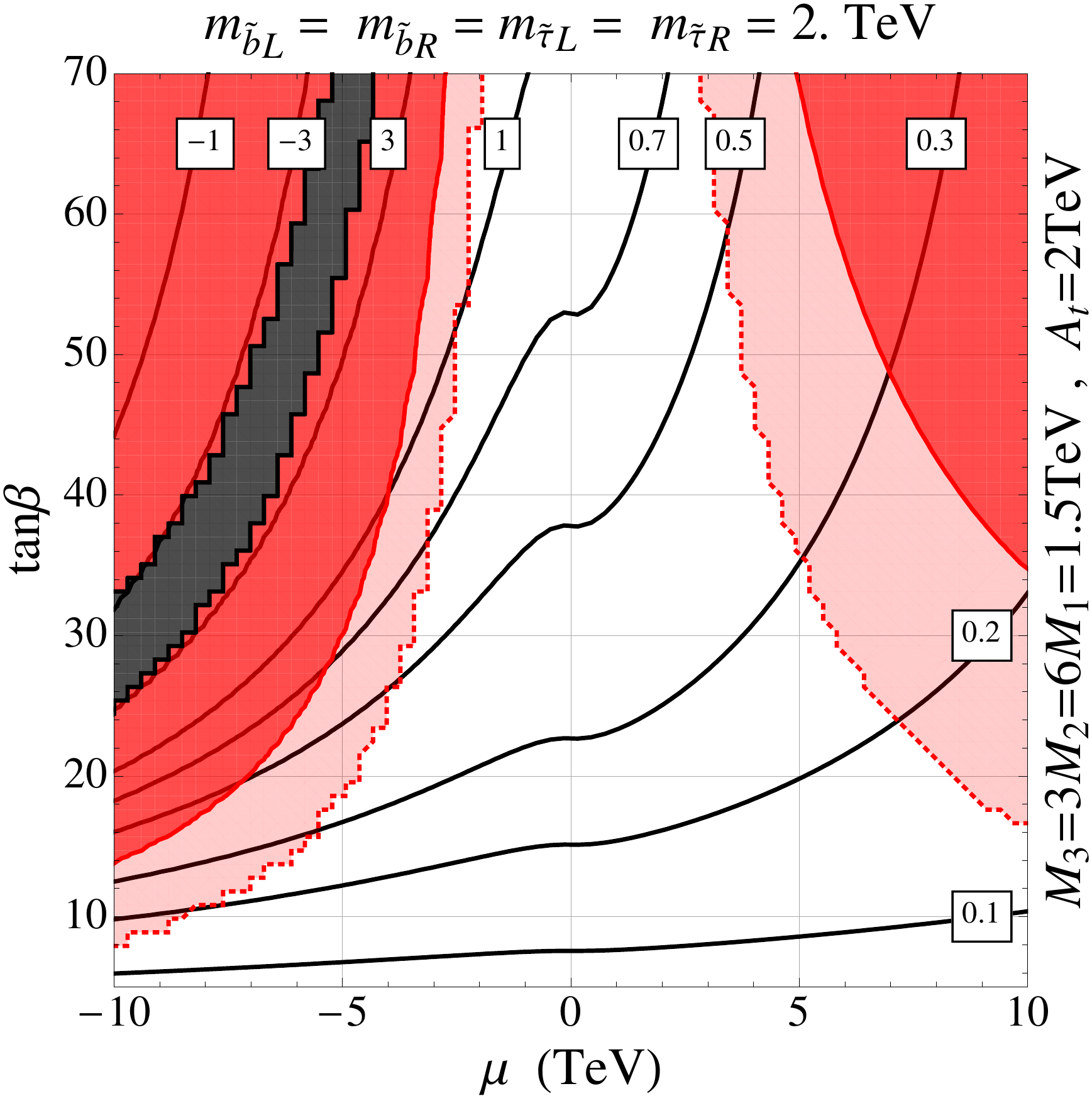} ~~~~
\includegraphics[width=0.45\textwidth]{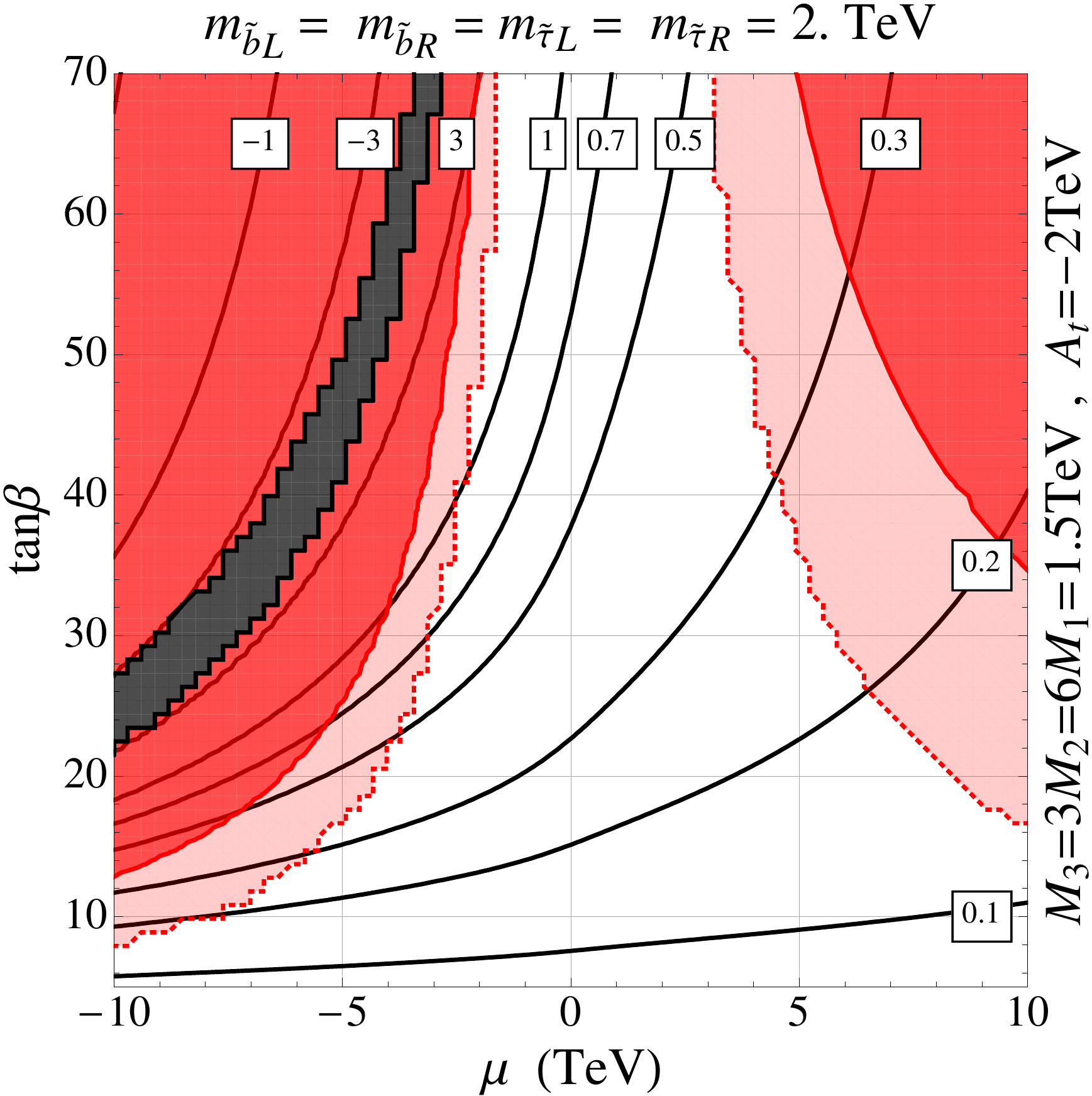}
\caption{Constraints from vacuum stability in the $\mu$--$\tan\beta$
  plane.  We set the sbottom and stau soft masses to 2~TeV and the
  gaugino masses to $6 M_1 = 3 M_2 = M_3 = 1.5$~TeV. In the left
  (right) plot, the trilinear coupling of the stops is $A_t = 2$~TeV
  ($A_t = -2$~TeV). The labeled contours show the values of the bottom
  Yukawa coupling.  In the light red (light gray) regions, a charge
  and color breaking vacuum exists that is deeper than the electroweak
  breaking vacuum, but the electroweak vacuum has a lifetime that is
  longer than the age of the universe.  In the dark red (gray)
  regions, the electroweak vacuum is not stable on cosmological time
  scales.  Finally, in the black regions, one of the sbottoms becomes
  tachyonic.}
\label{fig:vacuum_stability}
\end{figure*}

Independent of experimental searches, large values of $\mu$ can 
be constrained based on vacuum stability considerations, particularly
if $\tan\beta$ is also large. Indeed, large values of $\mu \tan\beta$
can lead to charge and color breaking minima in the scalar potential
of the MSSM~\cite{Hisano:2010re}. 

The trilinear couplings of the up-type Higgs, $H_u$, with sbottoms,
$\tilde b_L$ and $\tilde b_R$, and staus, $\tilde \tau_L$ and $\tilde
\tau_R$, are controlled by
\begin{eqnarray} \label{eq:potential}
\mathcal{L} &\supset&\; \frac{m_\tau}{v} 
\frac{\mu \tan\beta}{1 + \epsilon_\tau \tan\beta} 
(H_u^0 \tilde \tau_L^* \tilde \tau_R) + \nonumber \\
&+& \frac{m_b}{v} 
\frac{\mu \tan\beta}{1 + \epsilon_b \tan\beta} 
(H_u^0 \tilde b_L^* \tilde b_R) + \text{ h.c.} ~.
\end{eqnarray}
For trilinear couplings that are large compared to the sbottom or stau
masses, minima with non-zero vevs for the sbottom and/or stau fields
can arise in addition to the standard electroweak minimum. If these
minima are deeper than the electroweak minimum, the electroweak
minimum becomes unstable and can decay. The corresponding regions of
parameter space are only viable as long as the lifetime of the
electroweak minimum is longer than the age of the universe. This
corresponds to requiring that the bounce action, $B$, of the tunneling
process is $B \gtrsim 400$~\cite{Claudson:1983et, Kusenko:1996jn}.

Our phenomenological flavor analysis is largely independent of the
values of the stau masses, and non-zero stau vevs can always be
avoided if $m_{\tilde \tau_L}$ and $m_{\tilde \tau_R}$ are large
enough. Nonetheless, we will consider a scenario where the squark and
slepton masses are the same order and thus include both sbottoms and
staus in the following analysis.

Starting with the MSSM scalar potential, we restrict ourselves to
terms that contain only the up-type Higgs, sbottoms, and staus, which
are the degrees of freedom most relevant for large $\mu \tan\beta$.
We consider three cases: (i) only terms with the up-type Higgs and
staus, (ii) only up-type Higgs and sbottoms, and (iii) up-type Higgs,
staus and sbottoms simultaneously.  In each case, we search for
additional minima in field space and estimate the bounce action for
tunneling from the electroweak minimum into the deepest minimum of the
potential.  In the end, we apply the strongest of the three bounds.

For each case, the second vacuum generally has separately nearly
degenerate stau vevs and nearly equal sbottom vevs. In the case of the
sbottoms, this is expected from the $SU(3)$ $D$-terms in the scalar
potential:
\begin{equation}
\mathcal{L} \supset 
\frac{g_s^2}{6}\left( |\tilde b_L|^2 - |\tilde b_R|^2 \right)^2 \ .
\end{equation}
We can clearly see that at least for the 3-dimensional parameter space
in case (ii), deviations from equality of the squark/slepton fields
along the path chosen to compute the action would come at the expense
of large contributions from the $D$ terms.  Therefore, to obtain an
analytical estimate for the bounce action, we consider a straight path
in field space connecting the electroweak minimum and the charge
and/or color breaking minimum.

We then approximate the potential along the straight line by a
triangle and use the analytical expressions in~\cite{Duncan:1992ai} to
calculate the bounce action.  We construct the triangle such that for
a few chosen parameter points, the obtained bounce action agrees
approximately with the bounce action from the analytic expression of
the potential solved numerically by a standard overshoot/undershoot
method.  We further crosschecked our results with
\verb|CosmoTransitions|~\cite{Wainwright:2011kj} taking into account
the up-type Higgs, the down-type Higgs, the sbottoms, and the staus.
Overall, we find good agreement with our approximate analytical
approach.

The constraints thus derived in the $\mu$--$\tan\beta$ plane are shown
in Fig.~\ref{fig:vacuum_stability}. We fix the SUSY masses as in the
scenarios considered above, namely we assume degenerate sfermion
masses with $\tilde m = 2$~TeV and gaugino masses with $6 M_1 = 3 M_2 = M_3
= 1.5$~TeV. The trilinear couplings we set to $A_t = 2$~TeV in the
left and to $A_t = -2$~TeV in the right plot. In the white region, the
electroweak minimum is the deepest minimum in the potential and
therefore absolutely stable. In the light red (light gray) region, a
charge (and possibly color) breaking minimum exists that is deeper
than the electroweak minimum, but the electroweak minimum has a
lifetime longer than the age of the universe. In the dark red (gray)
region the lifetime of the electroweak minimum is shorter than the age
of the universe. Finally, in the black region, one of the sbottoms is
tachyonic.

The solid lines labeled in the plots show contours of constant bottom
Yukawa couplings in the $\mu$--$\tan\beta$ plane.  For large and
negative $\mu \tan\beta$, close to the region where one of the
sbottoms becomes tachyonic, the bottom Yukawa coupling becomes non
perturbatively large.

We observe that large negative values for $\mu$ are strongly
constrained by the requirement of vacuum stability.  This is because
the $\tan\beta$ resummation factor, $\epsilon_b$,
in~(\ref{eq:potential}) is linearly proportional to $\mu$. It
increases the trilinear coupling of the up-type Higgs with sbottoms
for negative values of $\mu$ and can lead to a deep second minimum
mainly in the field direction of the sbottoms. In particular, we find
that values of $\mu \tan\beta$ negative and large enough that the
bottom Yukawa changes its sign (the parameter space in the
upper left corner of the plot beyond the region excluded by tachyonic
sbottoms) are excluded by the requirement of vacuum stability.  For
positive values of $\mu$, the coupling of the up-type Higgs with
sbottoms is reduced while its coupling with staus is slightly enhanced
by the $\epsilon_\tau$ term. In this region of parameter space,
constraints come typically from a second minimum in the stau
direction.  Positive values for $\mu$ are less constrained than
negative ones, and the allowed region for $\mu$ can be extended above
$\mu > 10$~TeV for sufficiently heavy staus.

The viable regions of parameter space can be enlarged slightly when we
allow for a splitting between the masses of the left- and right-handed
sbottoms and staus. Nevertheless, we do not find any regions of
parameter space where both vacuum stability and $\epsilon_b \tan \beta
< -1$~(which flips the sign of the tree level bottom Yukawa and hence
changes the typical sign of the SUSY contribution to $B$ observables)
can be achieved simultaneously.  In the end, we see that the scenarios
discussed in the previous section are all compatible with bounds from
vacuum stability.

\section{\texorpdfstring{\boldmath $B$}{B} 
Physics Observables} \label{sec:flavor}

Flavor observables play a crucial role in determining viable regions
of parameter space of SUSY models. This is true both under the MFV
assumption~\cite{Bertolini:1990if,Goto:1998qv,Carena:2006ai,
  Isidori:2006pk, Altmannshofer:2007cs, Carena:2007aq, Domingo:2007dx,
  Wick:2008sz, Carena:2008ue, Altmannshofer:2010zt} and if new sources
of flavor violation are allowed~\cite{Gabbiani:1996hi, Baek:2001kh,
  Foster:2005wb, Giudice:2008uk, Altmannshofer:2009ne, Crivellin:2010ys,
  Barbieri:2011ci, Crivellin:2011sj, Calibbi:2011dn, Elor:2012ig}.

Of particular importance in the MFV setup are rare $B$ decays that are
helicity suppressed in the SM, because SUSY contributions to these
decays can be enhanced by $\tan\beta$ factors. Interesting processes
include the tree level decay $B \to \tau \nu$, the purely leptonic
decay $B_s \to \mu^+ \mu^-$, and the radiative decay $B \to X_s
\gamma$.  Additional constraints on the SUSY parameter space can be
also derived from the $(g-2)$ of the muon.  The $(g-2)_\mu$ bound
becomes particularly important if sleptons are only moderately heavy,
which is a scenario that we do not consider here.

\subsection{\texorpdfstring{\boldmath $B \to 
\tau \nu$}{B --> tau nu}, \texorpdfstring{\boldmath $B \to D^{(*)}
    \tau \nu$}{B --> D(*) tau nu} and \texorpdfstring{\boldmath $K \to
    \mu \nu$}{K --> mu nu}}
\label{sec:Btaunu}

The decay $B \to \tau \nu$ is a sensitive probe of extended Higgs
sectors as it can be modified by charged Higgs exchanges at tree
level~\cite{Hou:1992sy}.  The most important inputs for the SM
prediction are the CKM element $|V_{ub}|$ and the $B$ meson decay
constant.  Using the PDG value $|V_{ub}| = (3.89 \pm 0.44) \times
10^{-3}$~\cite{Nakamura:2010zzi}, a conservative average over direct
determinations from inclusive and exclusive semi-leptonic $B$ decays,
and an average of recent precise lattice determinations of the decay
constant $f_B = (190 \pm 4)$~MeV~\cite{McNeile:2011ng, Bazavov:2011aa,
  Na:2012kp, Davies:2012qf}, we find
\begin{equation} \label{eq:Btaunu_SM}
{\rm BR}(B \to \tau \nu)_{\rm SM} = (0.97 \pm 0.22) \times 10^{-4}~.
\end{equation}
While previous experimental data gave values for the branching ratio
more than $2\sigma$ above the SM prediction, a recent result from
Belle~\cite{Adachi:2012mm} has a much lower central value.  An average
of all the available data from BaBar~\cite{Aubert:2008ac,
  Collaboration:2012ju} and Belle~\cite{Hara:2010dk, Adachi:2012mm}
gives
\begin{equation} \label{eq:Btaunu_WA}
{\rm BR}(B \to \tau \nu)_{\rm exp} = (1.16 \pm 0.22) \times 10^{-4}~.
\end{equation}
This value is in very good agreement with the SM but still leaves room
for NP contributions.

Closely related decay modes that are also sensitive to charged Higgs
effects are the $B \to D \tau \nu$ and $B \to D^* \tau \nu$
decays~\cite{Tanaka:1994ay, Nierste:2008qe, Kamenik:2008tj,
  Fajfer:2012vx, Becirevic:2012jf}. While predictions of the
corresponding branching ratios suffer from large hadronic
uncertainties coming from the $B \to D$ and $B \to D^*$ form factors,
the ratios BR($B \to D \tau \nu$)/BR($B \to D \ell \nu$) and BR($B \to
D^* \tau \nu$)/BR($B \to D^* \ell \nu$), where $\ell = e$ or $\mu$,
can be predicted with reasonable accuracy in the
SM~\cite{Fajfer:2012vx, Bailey:2012jg}. Interestingly, recent results
from BaBar~\cite{Lees:2012xj} on these ratios are around $2\sigma$
above the SM predictions in both decay modes. Older results from
Belle~\cite{Bozek:2010xy} give similar central values but with much
larger uncertainties.

Another interesting observable in this context is
$R_{\mu23}$~\cite{Antonelli:2010yf} that probes the tree level charged
Higgs exchange in the $K \to \mu \nu$ decay. The much smaller
sensitivity of $K \to \mu \nu$ to charged Higgs effects compared to
the $B$ decays is compensated by its extremely high experimental
precision and the excellent control on theoretical uncertainties
giving~\cite{Antonelli:2010yf}
\begin{equation}\label{eq:Rmu23_Exp}
R_{\mu23} = 0.999 \pm 0.007 ~.
\end{equation}

All the mentioned tree level decays depend in similar ways on possible
new physics contributions in the MSSM with MFV.  Defining
\begin{equation}
X_{B(K)}^2 = \frac{1}{M_{H^\pm}^2} 
\frac{t_\beta^2}{(1+\epsilon_{0(s)} t_\beta)(1+\epsilon_\ell t_\beta)} ~,
\end{equation}
we can write
\begin{eqnarray} \label{eq:RBtaunu}
R_{B\tau\nu} &=& \frac{{\rm BR}(B \to \tau \nu)}{
{\rm BR}(B \to \tau \nu)_{\rm SM}} \nonumber \\
&=& \Big( 1 - m^2_{B^+} X_B^2 \Big)^2 ~,\\
\label{eq:RBDtaunu}
R_{D\tau\nu} &=& \frac{{\rm BR}(B \to D\tau \nu)}{
{\rm BR}(B \to D\tau \nu)_{\rm SM}} \nonumber \\ 
&=& \Big( 1 - 1.5\, m_\tau m_b X_B^2 + 1.0 \, m_\tau^2 m_b^2 X_B^4 \Big) ~,\\ 
\label{eq:RBDstaunu}
R_{D^*\tau\nu} &=& \frac{{\rm BR}(B \to D^*\tau \nu)}{
{\rm BR}(B \to D^*\tau \nu)_{\rm SM}} \nonumber \\ 
&=& \Big( 1 - 0.12\, m_\tau m_b X_B^2 + 0.05\, m_\tau^2 m_b^2 X_B^4 \Big) ~,\\
\label{eq:Rmu23}
R_{\mu23} &=& \frac{{\rm BR}(K \to \mu \nu)}{
{\rm BR}(K \to \mu \nu)_{\rm SM}} \nonumber\\
&=& \Big( 1 - m^2_{K^+} X_K^2 \Big) ~.
\end{eqnarray}

\begin{figure}[tb]
\centering \includegraphics[width=0.5\textwidth]{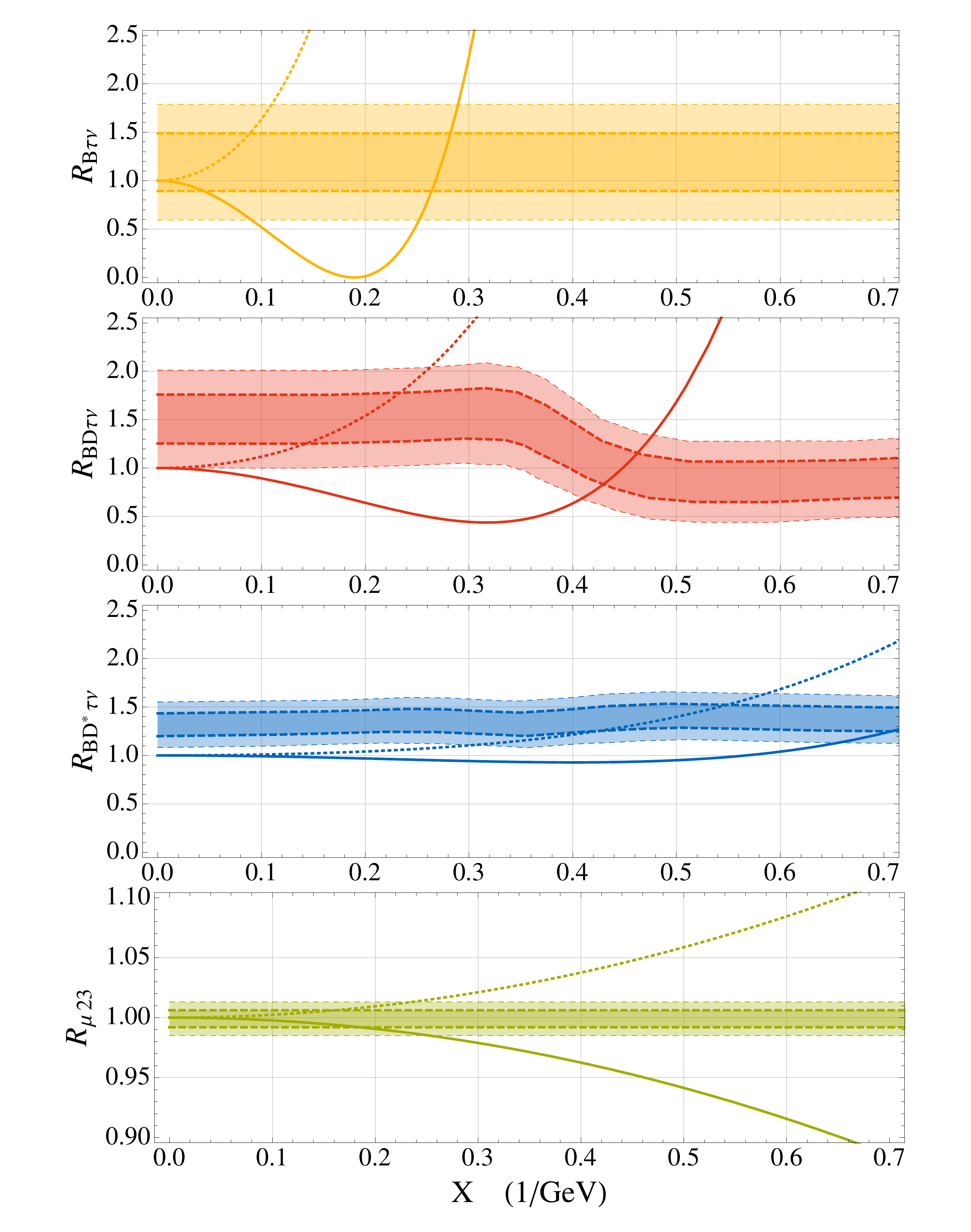}
\caption{Branching ratios of the decays $B \to \tau \nu$, $B \to D
  \tau \nu$, $B \to D^* \tau \nu$, and $K \to \mu \nu$ as functions of
  $X_B$, $X_K = \sqrt{|X_{B, K}^2|}$ as appropriate and which are
  defined in the main text.  The dashed bands show the $1\sigma$ and
  $2\sigma$ experimental ranges. The solid (dotted) lines are the
  theory predictions for positive (negative) $X^2$ giving destructive
  (constructive) interference with the SM amplitude.}
\label{fig:R}
\end{figure}

\begin{figure*}[tb]
\centering
\includegraphics[width=0.45\textwidth]{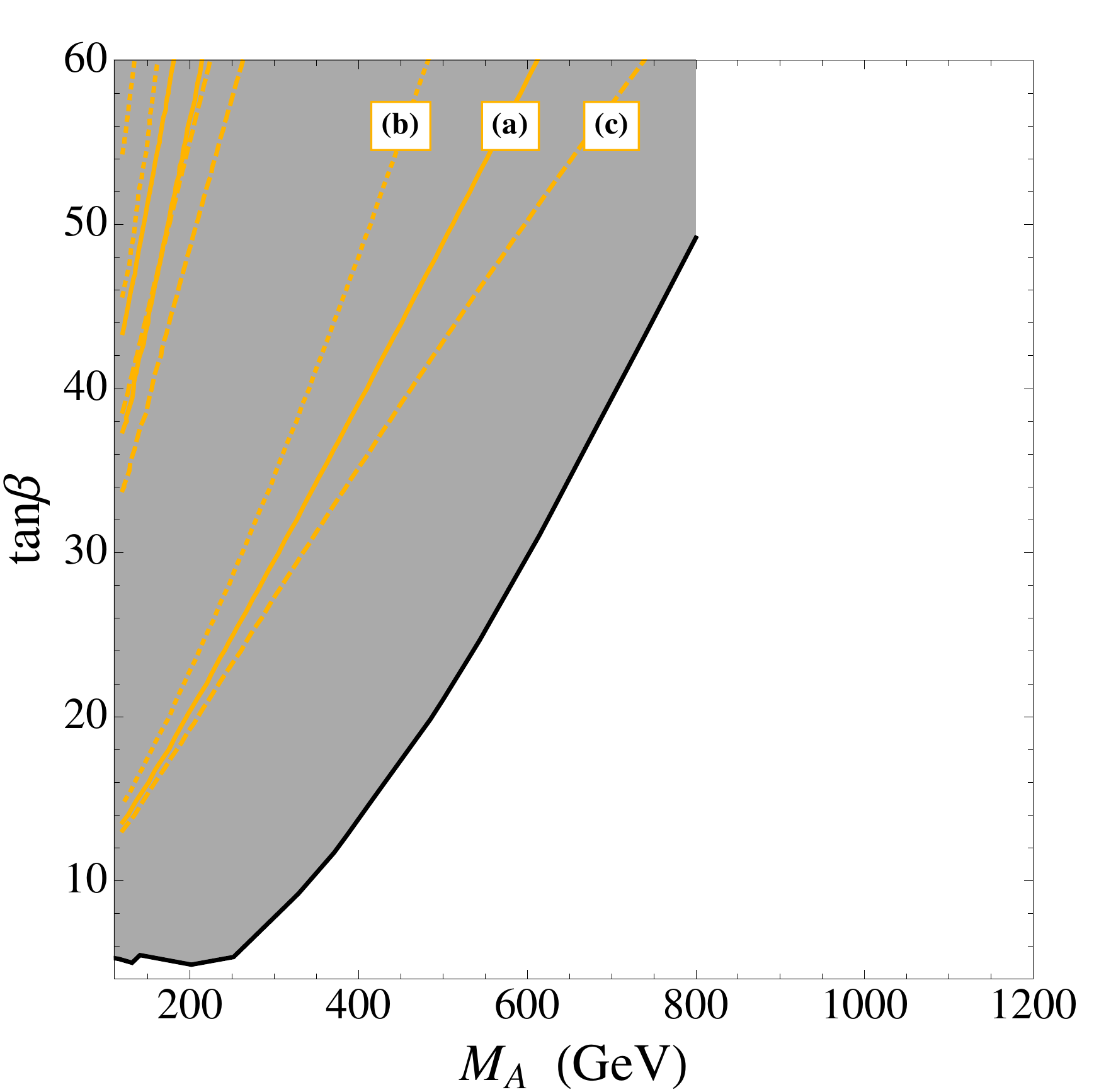} ~~~~
\includegraphics[width=0.45\textwidth]{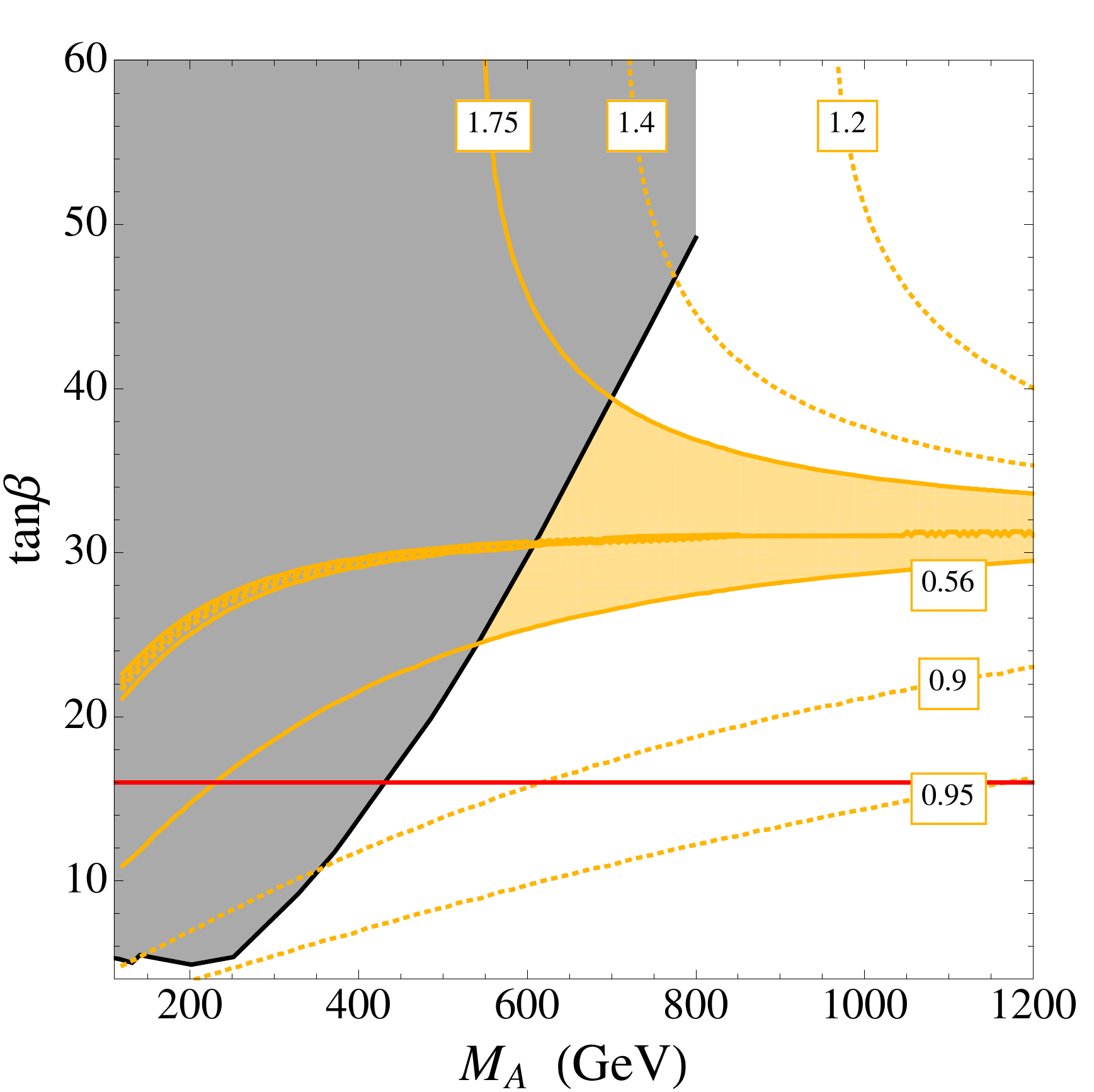}
\caption{Constraints in the $M_A$--$\tan\beta$ plane from the tree
  level $B \to \tau \nu$ decay. The constraint from direct heavy Higgs
  searches is also shown in gray.  The yellow solid, dotted and dashed
  contours in the left plot correspond to scenarios (a), (b), and (c)
  defined in Tab.~\ref{tab:scenarios}.  The right plot shows a
  scenario with $\mu = -8$ TeV, leading to a large negative
  $\epsilon_0$ such that the charged Higgs contribution interferes
  constructively with the SM in the region with $\tan\beta \gtrsim
  30$. The labeled contours indicate values for $R_{B\tau\nu}$. Above
  the red horizontal line, the electroweak vacuum has a lifetime
  shorter than the age of the universe.}
\label{fig:Btaunu}
\end{figure*}

In Fig.~\ref{fig:R} we show these ratios as function of $X_{B, K} =
\sqrt{|X_{B, K}^2|}$ both for positive $X_i^2$ (solid lines) and
negative $X_i^2$ (dotted lines) in comparison with the experimental
1$\sigma$ and 2$\sigma$ ranges (dashed bands)
from~(\ref{eq:Btaunu_SM})-(\ref{eq:Rmu23_Exp})
and~\cite{Fajfer:2012vx}-\cite{Lees:2012xj}.  Here, positive $X_i^2$
illustrates destructive interference with the SM, while negative
$X_i^2$ illustrates constructive interference with the SM.

We observe that agreement of theory and experiment in all three $B$
observables is impossible to achieve. In particular the tensions in $B
\to D \tau \nu$ and $B \to D^* \tau \nu$ cannot be addressed in the
context of the MSSM with MFV, but require more radical
approaches~\cite{Fajfer:2012jt, Crivellin:2012ye, Datta:2012qk,
  Deshpande:2012rr,Celis:2012dk}.

Considering MSSM contributions to $K \to \mu \nu$ and $B \to \tau
\nu$, we observe that generally, $X_B$ and $X_K$ are equal to a good
approximation.  The only way to induce a difference is through a
splitting between the right-handed strange squark mass and the
right-handed bottom squark mass which enter the corresponding
$\epsilon$ factors in the definitions of $X_B$ and $X_K$. As discussed
in Sec.~\ref{sec:sfermion_spectrum}, such a splitting is compatible
with the MFV ansatz for the squark spectrum as long as $\tan\beta$ is
large enough that $y_b$ effects cannot be neglected.  However, we find
that even for a large mass splitting $X_B \simeq X_K$ holds, except
for regions of parameter space with large and negative $\mu$, such
that $\epsilon_{0(s)} \tan\beta \sim \mathcal{O}(-1)$.  Such regions
of parameter space are strongly constrained by perturbativity of the
bottom Yukawa and vacuum stability considerations, as discussed in
Sec.~\ref{sec:vacuum}.  If $X_B \simeq X_K$, then the $B \to \tau\nu$
decay gives stronger constraints than $K \to \mu \nu$.\footnote{For
  the special range $0.25 \lesssim X_B, X_K \lesssim 0.30$ and
  $\epsilon_0 \tan \beta > -1$, the $K \to \mu \nu$ constraint is
  stronger than $B \to \tau \nu$, but this parameter region is
  excluded by direct searches as discussed in the main text.}

In the following, we therefore concentrate on the constraint from $B
\to \tau \nu$ on the MSSM parameter space.  Apart from corners of
parameter space with very large and negative $\epsilon_0 \tan \beta <
-1$, the charged Higgs contribution interferes destructively with the
SM ($X_B^2 > 0$), and leads to constraints in the $M_{H^\pm}$--$\tan
\beta$ plane.  These constraints depend on other SUSY parameters only
through the loop-induced $\tan \beta$ resummation factors $\epsilon_i$
and are therefore robust in large parts of parameter space.

The yellow lines in the left plot of
Fig.~\ref{fig:Btaunu} show the $B \to \tau \nu$ constraints in the
$M_A$--$\tan\beta$ plane corresponding to the 3 choices of MSSM
parameters (a), (b), and (c) given in Tab.~\ref{tab:scenarios} and
already discussed in Sec.~\ref{sec:HA}. For comparison, the constraint
from direct searches in the $\tau^+ \tau^-$ channel is also shown in
gray.  There are also a narrow strips of small Higgs masses and large
$\tan \beta$ where the NP contribution to the $B \to \tau \nu$
amplitude is twice as large as the SM contribution.  This in turn
implies that this region of parameters is in principle allowed by the
experimental data on $B \to \tau \nu$.  It is in strong tension,
however, with the results from $B \to D \tau \nu$, $B \to D^* \tau
\nu$, and $K \to \mu \nu$ and furthermore is excluded by direct
searches.

The dependence of the $B \to \tau \nu$ constraints on the $\tan\beta$
resummation factors is stronger than the one of the direct searches in
the $\tau^+ \tau^-$ channel, especially for large values of $\tan
\beta$.  For large values of $\tan\beta$ and a positive (negative)
value of $\epsilon_0$ the constraint can be weakened (strengthened)
considerably.  As $\epsilon_0$ does not depend on $A_t$, the
constraint from $B \to \tau \nu$ is to a large extent insensitive to
the exact value of the light Higgs mass.  Constraints from direct MSSM
Higgs searches are generically stronger for $M_A < 800$~GeV.  While
the latest results from direct MSSM Higgs searches in the $\tau^+
\tau^-$ channel at the LHC end at $M_A = 800$~GeV, obviously no such
restriction exists for the $B \to \tau \nu$ constraints.  Only very
large values of $\tan\beta \gtrsim 60$, however, are typically probed
by $B \to \tau\nu$ for such large heavy Higgs masses.

In corners of parameter space with very large negative $\mu$, we can
have $\epsilon_0 \tan\beta < -1$ for values of $\tan\beta$ that are
not extremely large and when the bottom Yukawa is
perturbative~\cite{Altmannshofer:2010zt}. Such a situation is shown in
the right plot of Fig.~\ref{fig:Btaunu}, where $\mu = -8$~TeV and the
resulting $\epsilon_0 \simeq -0.03$.  For $\tan \beta \lesssim 30$,
the charged Higgs still interferes destructively with the SM. For
larger values of $\tan\beta \gtrsim 30$, however, the sign of $X_B^2$
flips, the interference becomes constructive, and the branching ratio
is always enhanced. This behavior can be seen from the values of
$R_{B\tau\nu}$ indicated with the dotted contours in the right plot of
Fig.~\ref{fig:Btaunu}.

Note that vacuum stability requirements, however, strongly constrain
very large and negative values of $\mu$.  As discussed in
Sec.~\ref{sec:vacuum}, we do not find viable regions of parameter
space where the bottom Yukawa has a negative sign with respect to the
SM one, {\it i.e.} with $\epsilon_b \tan\beta < -1$. For $B \to \tau
\nu$, the relevant parameter combination is $\epsilon_0 \tan\beta$.
The horizontal red line in the right plot of Fig.~\ref{fig:Btaunu}
marks the upper bound on $\tan\beta$ in the scenario with $\mu =
-8$~TeV, such that the electroweak vacuum remains stable on timescales
of the age of the universe. Therefore, we see that  $\epsilon_0
\tan\beta < -1$ is also excluded by vacuum stability considerations.
This conclusion holds beyond the discussed $\mu = -8$ TeV example.

\subsection{\texorpdfstring{\boldmath $B_s \to \mu^+ \mu^-$}{Bs --> mu+ mu-}} 
\label{sec:Bsmumu}

The $B_s \to \mu^+ \mu^-$ decay is a flavor changing neutral current
process and correspondingly only induced at the loop level, both in
the SM and the MSSM.  In the SM, $B_s \to \mu^+ \mu^-$ is also
helicity suppressed by the muon mass, resulting in a tiny SM
prediction, at the level of $10^{-9}$. Using the recently given
precise value for the $B_s$ meson decay constant $f_{B_s} = (227 \pm
4)$~MeV~\cite{Davies:2012qf} which is an average of several lattice
determinations~\cite{McNeile:2011ng, Bazavov:2011aa, Na:2012kp}, and
taking into account the effect of the large width difference in the
$B_s$ meson system~\cite{DeBruyn:2012wj, DeBruyn:2012wk}, we have the
branching ratio extracted from an untagged rate
as~\cite{Altmannshofer:2012az} (see also \cite{Buras:2012ru})
\begin{equation}
{\rm BR}(B_s \to \mu^+ \mu^-)_{\rm SM} = (3.32 \pm 0.17) \times 10^{-9}~.
\end{equation} 
Experimental searches for that decay have been carried out at
D0~\cite{Abazov:2010fs} and CDF~\cite{Aaltonen:2011fi}, and are
ongoing at ATLAS~\cite{Aad:2012pn}, CMS~\cite{Chatrchyan:2012rg}, and
LHCb~\cite{Aaij:2012ac,1211.2674}.  Very recently, the LHCb
collaboration reported first evidence for the $B_s \to \mu^+\mu^-$
decay~\cite{1211.2674}.  LHCb finds for the branching ratio the
following value
\begin{equation}
{\rm BR}(B_s \to \mu^+ \mu^-)_\text{exp} = (3.2^{~+1.4~+0.5}_{~-1.2~-0.3}) \times 10^{-9} ~,
\end{equation}
and gives the following two sided $95\%$ C.L. bound
\begin{equation} \label{eq:Bsmm_exp}
1.1 \times 10^{-9} < {\rm BR}(B_s \to \mu^+ \mu^-)_\text{exp} < 6.4 \times 10^{-9} ~.
\end{equation}
We use this bound in our analysis.  Note that the upper bound
in~(\ref{eq:Bsmm_exp}) is considerably {\it weaker} than the official
combination of the previous LHCb result~\cite{Aaij:2012ac} with the
ATLAS and CMS bounds~\cite{Bsmm_LHC}.

For large values of $\tan\beta$, order of magnitude enhancements of
the BR($B_s \to \mu^+ \mu^-$) are possible in the
MSSM~\cite{Choudhury:1998ze, Babu:1999hn}.  In the large $\tan\beta$
limit, the CP averaged branching ratio in the MFV MSSM can be written to a good
approximation as
\begin{equation} \label{eq:Rbsmumu}
R_{B_s\mu\mu}=\frac{{\rm BR}(B_s \to \mu^+ \mu^-)}{
{\rm BR}(B_s \to \mu^+ \mu^-)_\text{SM}} \simeq 
\left|\mathcal{A}\right|^2 + \left|1- \mathcal{A}\right|^2 ~.
\end{equation}
The MSSM contribution $\mathcal{A}$ is dominated by so-called Higgs
penguins, {\it i.e.} the exchange of the heavy scalar $H$ and
pseudoscalar $A$ with their 1-loop induced flavor changing $b \to s$
couplings, that are parametrized by $\epsilon_{\rm FC}$ given in
(\ref{eq:xi_sb}).  We find
\begin{equation} \label{eq:A}
\mathcal{A} = \frac{4\pi}{\alpha_2} \frac{m_{B_s}^2}{4 M_A^2} 
\frac{\epsilon_{\rm FC} ~ t^3_\beta}{
(1+\epsilon_b t_\beta)(1+\epsilon_0 t_\beta)(1+\epsilon_\ell t_\beta)} ~ 
\frac{1}{Y_0}~.
\end{equation}
The SM loop function $Y_0$ depends on the top mass and is
approximately $Y_0 \simeq 0.96$. Note that the MSSM contributions to
$B_s \to \mu^+ \mu^-$ do not decouple with the scale of the SUSY
particles, but with the masses of the heavy scalar and pseudoscalar
Higgs bosons $M_H^2 \simeq M_A^2$. Due to the strong enhancement by
$\tan^3\beta$, the large $\tan\beta$ regime of the MSSM is highly
constrained by the current experimental results on BR$(B_s \to \mu^+
\mu^-)$. We remark, however, that $\epsilon_{\rm FC}$ in the numerator
of~(\ref{eq:A}) is a sum of several terms (see~(\ref{eq:epsilon_FC}))
each of which depend strongly on several MSSM parameters. In addition,
cancellations among the different terms can occur in certain regions
of parameter space, rendering the $B_s \to \mu^+ \mu^-$ constraint
very model dependent, even in the restrictive framework of MFV.
Additional contributions to $B_s \to \mu^+\mu^-$ can arise from
charged Higgs loops~\cite{Logan:2000iv}.  They interfere destructively
with the SM contribution and scale as $(\tan\beta)^2/M_{H^\pm}^2$.
Typically, their effect is considerably smaller compared to the SUSY
contribution in~(\ref{eq:A}).

We stress that there is a simple mathematical lower bound of $R_{B_s
  \mu \mu} = 1/2$ in~(\ref{eq:Rbsmumu}) that is saturated for
$\mathcal{A} = 1/2$.  In this case, the SUSY contribution partially
cancels the SM amplitude, but simultaneously generates a
non-interfering piece that cannot be canceled.  This lower limit
provides a significant threshold for experiments searching for BR$(B_s
\to \mu^+ \mu^-)$: not only is the SM branching fraction a meaningful
value to test experimentally, but the potential observation of the
branching fraction below one half of the SM value would strongly
indicate NP and imply departure from the MSSM with MFV.
Note that the current $2\sigma$ {\it lower bound} from LHCb on the 
branching ratio is below 1/2 of the SM value and therefore does 
not lead to constraints in our framework, yet.

\begin{figure}[tb]
\centering
\includegraphics[width=0.45\textwidth]{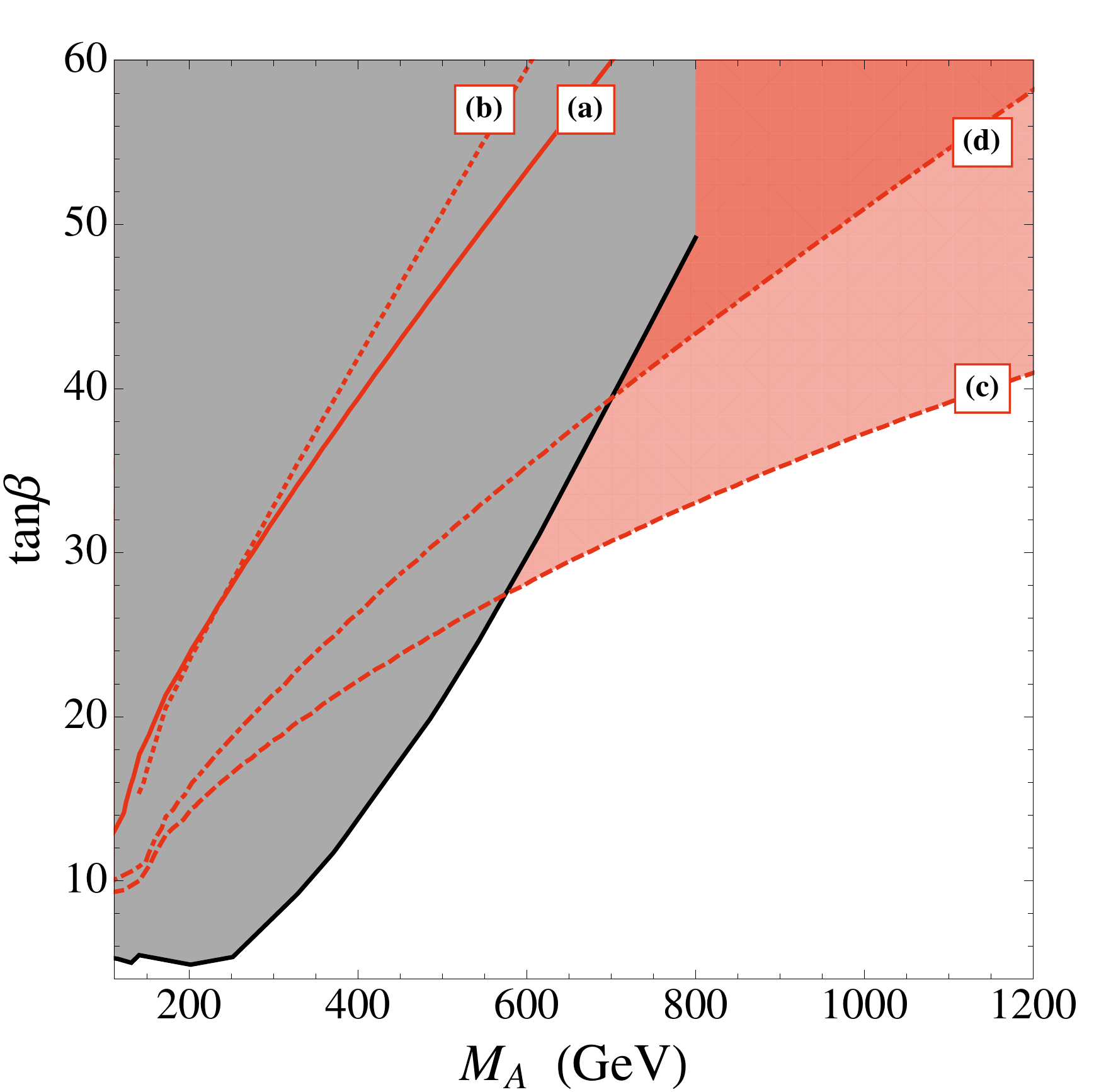}
\caption{Constraints in the $M_A$--$\tan\beta$ plane from the $B_s \to
  \mu^+ \mu^-$ decay.  The red solid, dotted, dashed and dash-dotted
  contours correspond to scenarios (a), (b), (c) and (d), as described
  in the text.  The gray region is excluded by direct searches of MSSM
  Higgs bosons in the $H/A \to \tau^+ \tau^-$ channel.}
\label{fig:Bsmumu1}
\end{figure}

\begin{figure*}[tb]
\centering
\includegraphics[width=0.45\textwidth]{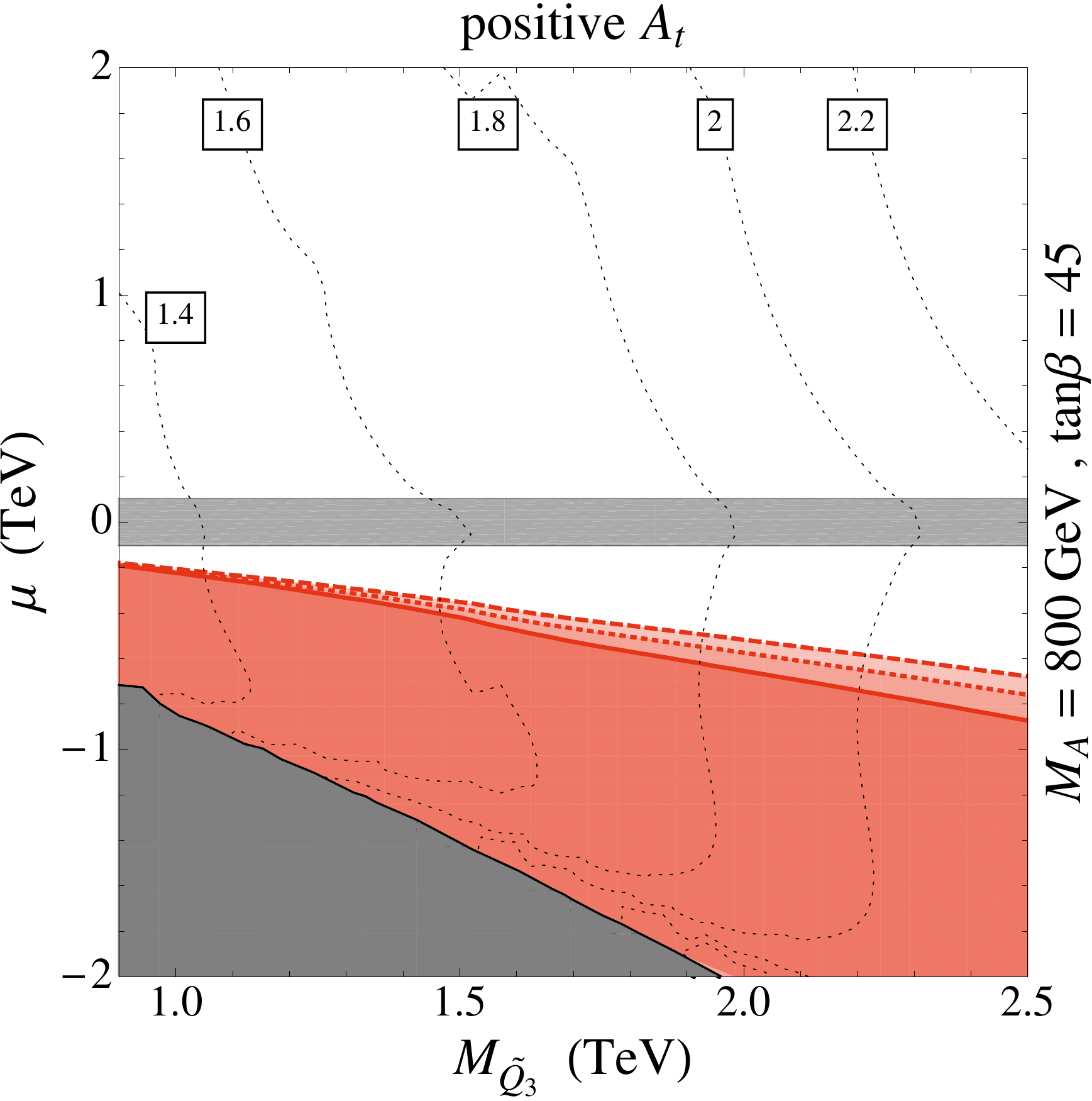} ~~~~
\includegraphics[width=0.45\textwidth]{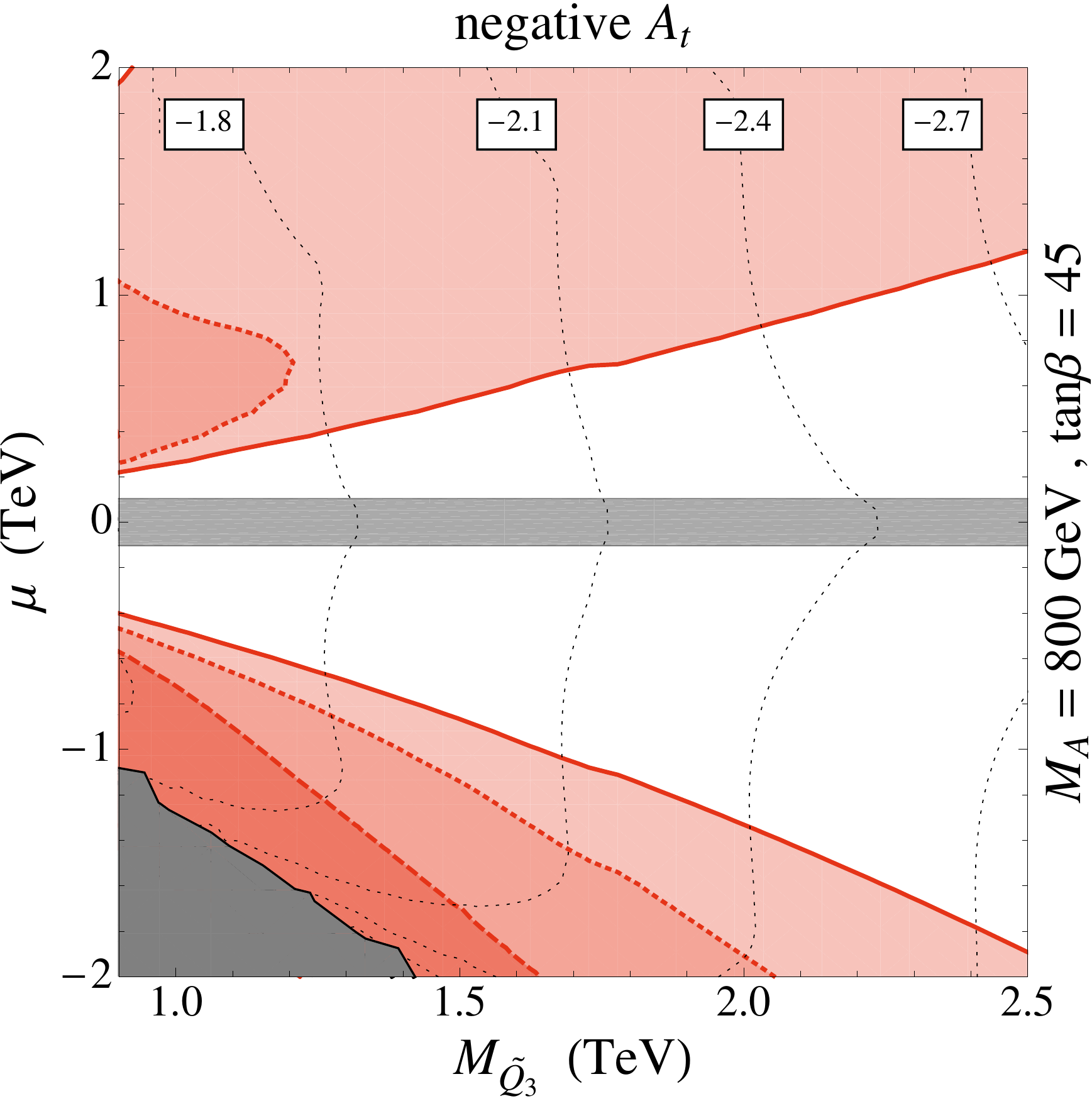}
\caption{Constraints in the $m_{Q_3}$--$\mu$ plane from the $B_s \to
  \mu^+ \mu^-$ decay, with fixed $M_3 = 3M_2 = 6M_1 = 1.5$ TeV,
  $M_A=800$ GeV and $\tan \beta = 45$. The solid bounded regions
  correspond to a degenerate squark spectrum.  The dashed and dotted
  bounded regions correspond to choosing the first two squark
  generations 50\% heavier than the third generation squark masses,
  with an alignment of $\zeta = 1$ and $\zeta = 0.5$, respectively.
  The gray horizontal band corresponds to the constraint from direct
  searches of charginos at LEP.  The vertical dotted lines show
  contours of constant $A_t$ such that $M_h = 125$~GeV. In the gray
  regions in the lower left corners, the lightest Higgs mass is always
  below $M_h < 125$~GeV, taking into account a 3 GeV theory
  uncertainty.}
\label{fig:Bsmumu2}
\end{figure*}

\begin{figure}[tb]
\centering
\includegraphics[width=0.45\textwidth]{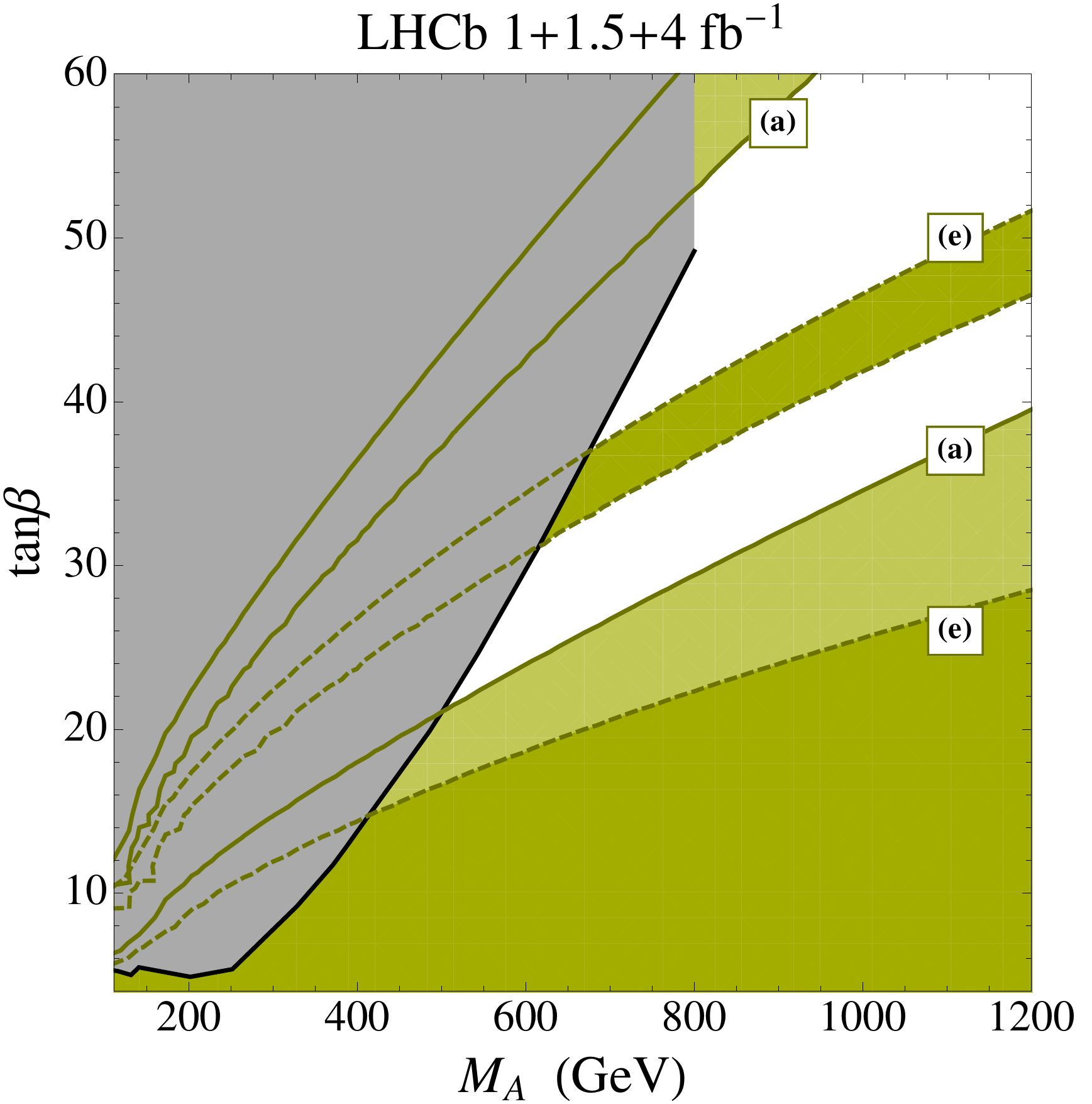}
\caption{The $M_A$--$\tan\beta$ plane in view of projected constraints
  from the BR$(B_s \to\mu^+ \mu^-)$, assuming a future $\pm 0.5 \times
  10^{-9}$ uncertainty in the measurement with the SM prediction as the central value.  The
  green shaded regions between and below the solid and dashed contours
  correspond to values for $\tan\beta$ and $M_A$ {\it allowed} in
  scenarios (a) and (e), as defined in Tab.~\ref{tab:scenarios}.  The
  gray region is excluded by current direct searches of MSSM Higgs
  bosons in the $H/A \to \tau^+ \tau^-$ channel.}
\label{fig:Bsmumu1_projection}
\end{figure}

\begin{figure*}[tb]
\centering
\includegraphics[width=0.45\textwidth]{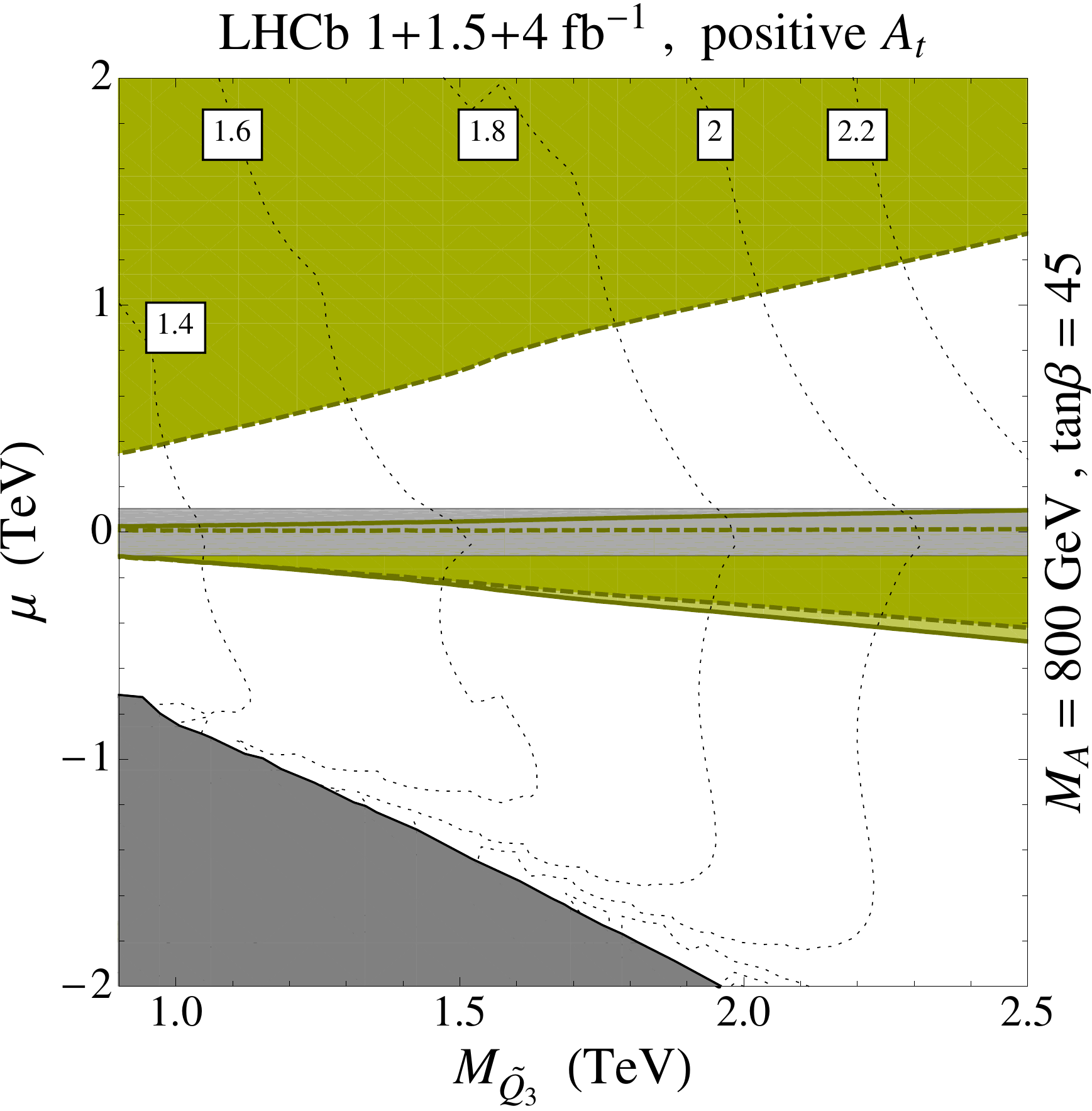} ~~~~
\includegraphics[width=0.45\textwidth]{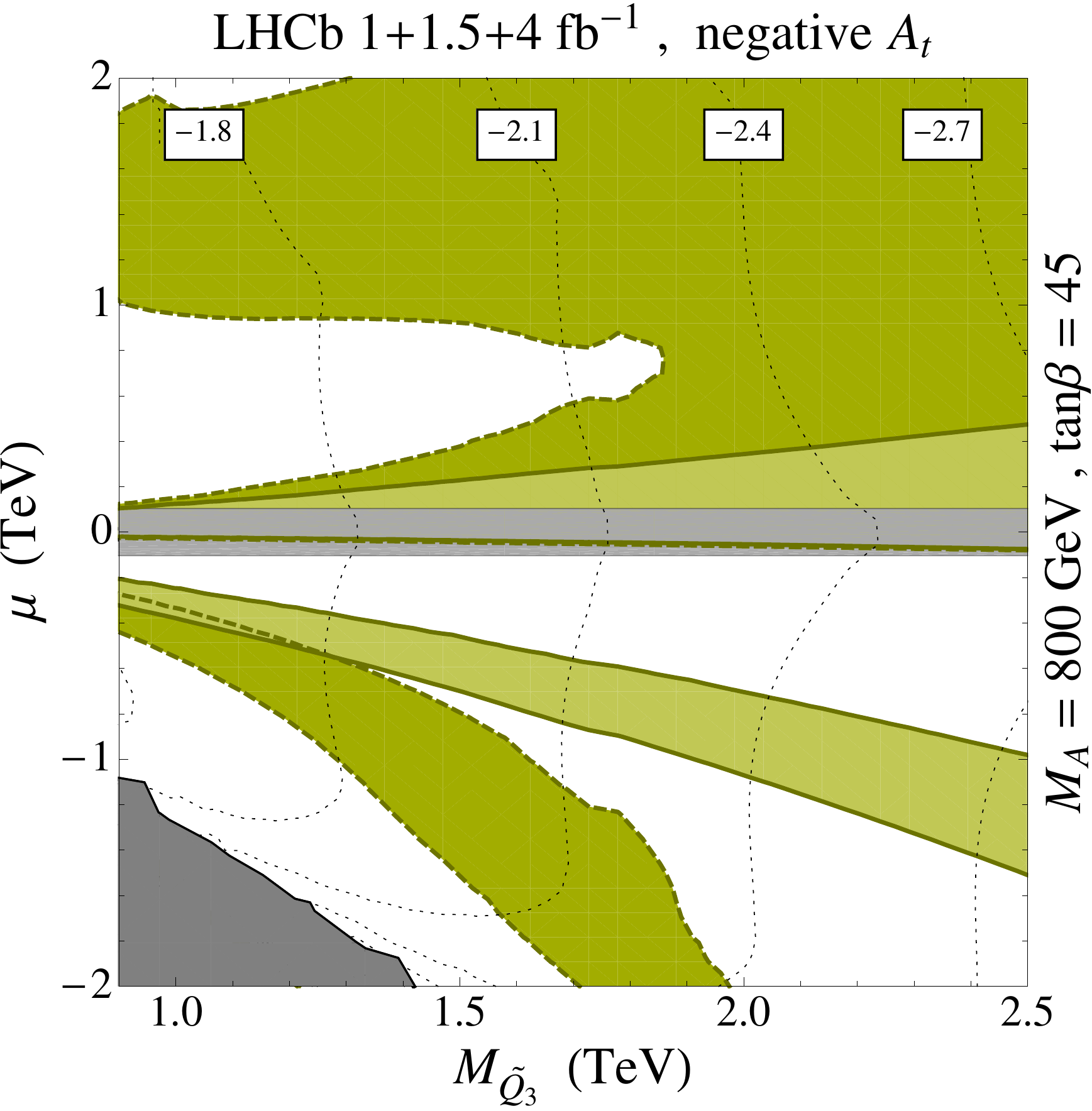}
\caption{The $m_{Q_3}$--$\mu$ plane in view of projected constraints
  from the BR$(B_s \to\mu^+ \mu^-)$, assuming a future $\pm 0.5 \times
  10^{-9}$ uncertainty in the measurement with the SM prediction as the central value. We
  fixed $M_3 = 3M_2 = 6M_1 = 1.5$ TeV, $M_A=800$ GeV and $\tan \beta =
  45$. The green shaded regions between the solid contours correspond
  to values for $m_{Q_3}$ and $\mu$ {\it allowed} for a degenerate
  squark spectrum.  The green shaded regions between and above the
  dashed contours are allowed if the first two squark generations are
  50\% heavier than the third generation squark masses, with an
  alignment of $\zeta = 1$.  The gray horizontal band corresponds to
  the constraint from direct searches of charginos at LEP.  The
  vertical dotted lines show contours of constant $A_t$ such that $M_h
  = 125$~GeV. In the gray regions in the lower left corners, the
  lightest Higgs mass is always below $M_h < 125$~GeV, taking into
  account a 3 GeV theory uncertainty.}
\label{fig:Bsmumu2_projection}
\end{figure*}

In Fig.~\ref{fig:Bsmumu1}, we show the constraints from $B_s \to \mu^+
\mu^-$ in the $M_A$--$\tan\beta$ plane. The red solid, dotted and
dashed contours correspond to scenarios (a), (b), and (c) of
Tab.~\ref{tab:scenarios}. The dash-dotted contour corresponds to
scenario (d), with all MSSM parameters as for the solid contour, but
with a negative sign for the trilinear coupling.  For comparison, the
constraints from direct searches are again shown in gray.  As
expected, we observe a very strong dependence of the $B_s \to \mu^+
\mu^-$ bounds on the choices of the remaining MSSM parameters,
particularly the sign of $\mu A_t$.  Note that in the considered
scenarios, we assume degenerate squarks such that the only term
entering $\epsilon_{\rm FC}$ is from the irreducible Higgsino loop
contribution, $\epsilon_b^{\tilde H}$, whose sign is dictated by $\mu
A_t$. For positive (negative) $\mu A_t$ the NP contribution interferes
destructively (constructively) with the SM amplitude. Since the lower
bound on BR$(B_s \to \mu^+ \mu^-)$ from LHCb is still below half of
the SM value, destructively interfering NP is much less constrained
than constructively interfering NP.

The plots of Fig.~\ref{fig:Bsmumu2} show in red the constraints from
$B_s \to \mu^+ \mu^-$ in the plane of the third generation squark
masses and the Higgsino mass parameter $\mu$. The gray horizontal band
corresponds to the constraint from direct searches of charginos at LEP
that exclude $|\mu| \lesssim 100$~GeV~\cite{Abdallah:2003xe,
  Abbiendi:2003sc}.  In these plots, we fix $M_A = 800$~GeV,
$\tan\beta = 45$ (fully compatible with the $B \to \tau \nu$
constraint and not yet constrained by direct searches), and gaugino masses
with $6 M_1 = 3 M_2 = M_3 = 1.5$~TeV.  As in all the other plots, we
vary the trilinear couplings $A_t = A_b = A_\tau$ throughout the plot
such that the lightest Higgs mass is $M_h = 125$~GeV. The values for
$A_t$ are indicated in the plots by the vertical dotted contours. The
two plots correspond to positive and negative values of the $A$-terms.
In the gray region in the lower left corners of the plots, the sbottom
loop corrections to the lightest Higgs mass become so large that the
lightest Higgs mass is always below $M_h < 125$~GeV for any value of
$A_t$, taking into account a 3 GeV theory uncertainty.  We checked
that varying the light Higgs mass between $122$~GeV $< M_h < 128$~GeV
can change the values of $A_t$ by around $25\%$ in each direction and
therefore can affect the constraints derived from $B_s \to \mu^+
\mu^-$ at a quantitative level. However, the qualitative picture of
the constraints and the interplay of the SUSY contributions to $B_s
\to \mu^+ \mu^-$, as discussed below, are unaffected by this
variation.

The solid contours are obtained under the assumption that the masses
of the first two generation squarks are equal to the third generation,
while for the dashed and dotted contours we assume the first two
generations to be heavier by 50\%.  For the dashed contours, we assume
the splitting for the left-handed squarks to be fully aligned in the
up-sector, such that gaugino-squark loops also contribute to
$\epsilon_{\rm FC}$ with $\zeta = 1$ (see~(\ref{eq:epsilon_FC})
and~(\ref{eq:epsilon_FCg})).  We set $\zeta = 0.5$ for the dotted
contours, such that only half of the squark mass splitting induces
flavor violation in the down-sector.  For negative $A_t$, the obtained
bounds show a strong dependence on the value of $\zeta$.  The BR$(B_s
\to \mu^+ \mu^-)$ bounds in Fig.~\ref{fig:Bsmumu2} clearly display the
non-decoupling behavior mentioned above.  Due to this non-decoupling,
the BR$(B_s \to \mu^+ \mu^-)$ results can constrain SUSY parameter
space in regions that are beyond the current and expected future reach
of direct searches.

A crucial element of our analysis is the viability of the cancellation
of the SUSY contribution to the $B_s \to \mu^+ \mu^-$ branching ratio.
This cancellation is driven by the presence of $\epsilon_{\rm FC}$
in~(\ref{eq:A}), which is schematically given in~(\ref{eq:epsilon_FC})
and its various contributions are detailed in
(\ref{eq:epsilon_bH}),~(\ref{eq:epsilon_FCg})
and~(\ref{eq:epsilon_FCw}).  First, in the following discussion, we
neglect the wino contribution given by~(\ref{eq:epsilon_FCw}), which
is generally smaller than the gluino contribution. This is due to the
smallness of $M_2$ and $\alpha$ in~(\ref{eq:epsilon_FCw}) compared to
$M_3$ and $\alpha_s$ in~(\ref{eq:epsilon_FCg})~(of course, our
numerical analysis always includes the wino contribution).  Since each
SUSY contribution is proportional to $\mu$, we see that switching the
sign of $\mu$ changes the relative sign between the SUSY and SM
amplitudes.  Furthermore, by switching the sign of $A_t$, between the
left and right panels of Fig.~(\ref{fig:Bsmumu2}), we change the
relative sign between the gluino contribution and the Higgsino
contribution.  Thus, for a particular choice of sign$(A_t)$ and
sign$(\mu)$, we can exploit a cancellation between the gluino
vs. Higgsino loop, diminishing the magnitude of the SUSY contribution,
and a second cancellation between the overall SUSY contribution and
the SM amplitude.  In particular, even if the magnitude of the SUSY
contribution is by itself larger than the SM contribution, we can
exercise the second cancellation where the SUSY amplitude overshoots
the SM one.

These cancellations are clearly in effect in the left and right panels
of Fig.~\ref{fig:Bsmumu2}.  We first focus on the regions bounded by
solid lines, which correspond to degenerate squark masses. This
implies that the SUSY contribution dominantly arises from
$\epsilon_b^{\tilde{H}}$ in~(\ref{eq:epsilon_bH}).  In the upper half
of the left panel corresponding to positive $A_t$ and positive $\mu$,
the SUSY contribution cancels with the SM contribution and always
leads to a BR$(B_s \to \mu^+ \mu^-)$ below the current bound.  In the
lower half of the left panel, with positive $A_t$ and negative $\mu$,
the Higgsino contribution adds constructively with the SM
contribution, leading to significant constraints. In the upper half of
the right panel, the Higgsino contribution also adds constructively
with the SM, leading again to a bound. This bound is less stringent
compared to the positive $A_t$ and negative $\mu$ case, because for
positive $\mu$, the $\epsilon_b$ and $\epsilon_0$ terms
in~(\ref{eq:A}) lead to a suppression of the SUSY amplitude. Finally,
in the lower half of the right panel, with negative $A_t$ and negative
$\mu$, the Higgsino contribution interferes destructively with the SM.
The constraint is non-vanishing, however, because for negative $\mu$,
the $\tan\beta$ resummation factors, given in (\ref{eq:A}), enhance
the SUSY amplitude such that it can be more than twice as large as the
SM amplitude.

When we include squark splitting, we further strengthen the SUSY
contribution for positive $A_t$, because the gluino and Higgsino
contributions add constructively. Hence the overall SUSY+SM
interference is more restricted. The bounds due to this splitting in
the masses are shown by the regions enclosed by the dashed and dotted
lines in Fig.~(\ref{fig:Bsmumu2}). For negative $A_t$, shown in the
right panel, the gluino contribution partially cancels the Higgsino
contribution, leading to a weaker constraint.  The effect of the
gluino contributions decreases for larger gluino mass, $M_3$.

In tandem, the complementary views provided by the different panels of
Figs.~\ref{fig:Bsmumu1} and~\ref{fig:Bsmumu2} clearly demonstrate that
certain choices of SUSY parameters relax the constraints considerably.
For example, with $M_A = 800$ GeV and $\tan \beta = 45$, the region of
parameter space with positive $\mu$ and positive $A_t$ is robustly
unconstrained from the $B_s \to \mu^+ \mu^-$ limit.  Moving from top
to bottom along a constant $A_t$ contour in the left plot of
Fig.~\ref{fig:Bsmumu2} corresponds to a rapid coverage of the $\tan
\beta$ vs. $M_A$ plane from the (b) to (a) to (c) exclusion regions.

Regions of parameter space with destructive interference between SM
and SUSY amplitudes ({\it i.e.} the regions with positive $\mu A_t$)
will be constrained significantly if a lower bound of BR$(B_s \to
\mu^+\mu^-)$ above half of the SM prediction is established in the
future.  We illustrate this in the plots of
Figs.~\ref{fig:Bsmumu1_projection} and~\ref{fig:Bsmumu2_projection},
which assume a measurement of BR$(B_s \to \mu^+\mu^-)$ at the SM
expectation as a central value with an experimental uncertainty of
$\pm 0.5 \times 10^{-9}$.  Such a precision is expected to be achieved
by LHCb at the end of the 13 TeV run with a combined analysis of 1
fb$^{-1}$ of 7 TeV data, 1.5 fb$^{-1}$ of 8 TeV data, and 4 fb$^{-1}$
of 13 TeV data~\cite{Bediaga:1443882}.  The plots in
Figs.~\ref{fig:Bsmumu1_projection} and~\ref{fig:Bsmumu2_projection}
show in green the regions in the $M_A$--$\tan\beta$ and
$m_{Q_3}$--$\mu$ planes that are {\it allowed} by the expected results
on the $B_s \to \mu^+\mu^-$ decay.  As shown in
Fig.~\ref{fig:Bsmumu1_projection}, apart from the allowed regions with
large $M_A$ and small $\tan\beta$, there are also strips with large
$M_A$ and large $\tan\beta$ where the expected bounds from $B_s \to
\mu^+\mu^-$ can be avoided.  In these regions, the SUSY amplitude has
approximately the same size as the SM amplitude but is opposite in
sign.  According to~(\ref{eq:Rbsmumu}), this leads to a branching
ratio close to the SM prediction.

For the example parameter point with $M_A = 800$ GeV and $\tan\beta =
45$, the projected lower bound on BR$(B_s \to \mu^+\mu^-)$ leads to
very strong constraints in the $m_{Q_3}$--$\mu$ plane for positive
$\mu A_t$.  Indeed, for $M_A = 800$ GeV and $\tan\beta = 45$, and
given the assumed experimental precision, charged Higgs loop
contributions to $B_s \to \mu^+\mu^-$ already lead to a non-negligible
suppression~\cite{Logan:2000iv}, leaving hardly any room for
destructively interfering SUSY contributions.  Only if the SUSY
contribution is so large that $\mathcal{A} \simeq -1$ does the parameter
space open up again. The corresponding regions that are excluded by
the assumed lower bound are clearly visible in the white region of the
upper half of the left plot and the upper white region in the lower
half of the right plot in Fig.~\ref{fig:Bsmumu2_projection}.

\subsection{\texorpdfstring{\boldmath $B \to X_s \gamma$}
{B --> Xs gamma}} 
\label{sec:bsgamma}

The loop induced $B \to X_s \gamma$ decay is also highly sensitive to
NP effects coming from SUSY particles. The NNLO SM prediction for the
branching ratio reads~\cite{Misiak:2006zs} (see
also~\cite{Becher:2006pu,Benzke:2010js})
\begin{equation}
{\rm BR}(B \to X_s \gamma)_{\rm SM} = (3.15 \pm 0.23) \times 10^{-4}~.
\end{equation}
On the experimental side, BaBar recently presented updated results for
the branching ratio~\cite{:2012iwb}.  Including this, the new
world average reads~\cite{Amhis:2012bh}
\begin{equation}
{\rm BR}(B \to X_s \gamma)_{\rm exp} = (3.43 \pm 0.22) \times 10^{-4}~,
\end{equation}
which is slightly lower than the previous world average and is in very
good agreement with the SM prediction. In the MSSM with minimal flavor
violation and no new sources of $CP$ violation, the branching ratio
can be written as~\cite{Freitas:2008vh}
\begin{eqnarray}
R_{bs\gamma} &=& \frac{{\rm BR}(B \to X_s \gamma)}{
{\rm BR}(B \to X_s \gamma)_{\rm SM}} \;,\nonumber \\
&&\nonumber\\
&\simeq& 1 - 2.55~C_7^{\rm NP} - 0.61~C_8^{\rm NP} + 0.74 C_7^{\rm NP}C_8^{\rm NP} 
\nonumber \\
&& \;\;\;+ 1.57~(C_7^{\rm NP})^2 + 0.11~(C_8^{\rm NP})^2 ~,
\end{eqnarray}
where $C_{7,8}^{\rm NP}$ are the NP contributions to the magnetic and
chromo-magnetic $b \to s \gamma$ operators evaluated at the scale
160~GeV.

Apart from the $B \to X_s \gamma$ decay, the modifications of the
Wilson coefficients $C_7$ and $C_8$ also enter predictions of
observables in the $B \to K^* \ell^+\ell^-$ decay.  In our MSSM setup
with minimal flavor and $CP$ violation, we only have real NP
contributions to $C_7$ and $C_8$.  In this framework, the experimental
data on $B \to K^* \ell^+\ell^-$ does not put additional restrictions,
once the bounds from BR$(B \to X_s \gamma)$ are taken into
account~\cite{Altmannshofer:2011gn, Altmannshofer:2012az}.  Therefore,
we focus only on the $B \to X_s \gamma$ decay.

The SUSY contributions to $C_{7,8}^{\rm NP}$ come from charged
Higgs--top loops, neutral Higgs--bottom loops, Higgsino--stop loops,
and gaugino--squark loops. As with the Higgs--fermion couplings, we
take into account the most generic MFV structure of the squark masses
and consistently consider splittings between the first two and the
third generation squarks in the left-handed as well as the right
handed sector.  The resulting dominant MSSM contributions to $C_{7,8}$
read
\begin{widetext}
\begin{eqnarray} \label{eq:C78Higgs}
C_{7,8}^{H} &=& 
\left(\frac{1-\epsilon_0^\prime t_\beta}{1 + \epsilon_b t_\beta} + \frac{\epsilon_{\rm FC}^\prime \epsilon_{\rm FC} t_\beta^2}{(1+\epsilon_b t_\beta)(1+\epsilon_0 t_\beta)} \right) \frac{m_t^2}{2 M_{H^\pm}^2} h_{7,8}(r_t) + \frac{\epsilon_{\rm FC} t_\beta^3}{(1+\epsilon_b t_\beta)^2(1+\epsilon_0 t_\beta)} \frac{m_b^2}{2 M_A^2} z_{7,8} ~, \\
\label{eq:C78Higgsino}
C_{7,8}^{\tilde H} &=& -\frac{t_\beta}{1 + \epsilon_b t_\beta} 
~\frac{m_t^2}{2} A_t \mu~ f^{\tilde H}_{7,8}(m_{Q_3}^2,m_{U_3}^2,\mu^2) ~, \\
 \label{eq:C78gluino}
\frac{g_2^2}{g_3^2}~C_{7,8}^{\tilde g} &=& 
\frac{t_\beta}{1 + \epsilon_0 t_\beta} 
~M_W^2 \mu M_3~\zeta~ \Big( f^{\tilde g}_{7,8}(m_{Q}^2,m_{D_3}^2,M_3^2) 
- f^{\tilde g}_{7,8}(m_{Q_3}^2,m_{D_3}^2,M_3^2) \Big)  \nonumber \\
&& - \frac{\epsilon_{\rm FC} t_\beta^2}{(1+\epsilon_b t_\beta)(1+\epsilon_0 t_\beta)} ~M_W^2 \mu M_3~f^{\tilde g}_{7,8}(m_{Q_3}^2,m_{D_3}^2,M_3^2) ~, \\
\label{eq:C78Wino}
C_{7,8}^{\tilde W} &=& \frac{t_\beta}{1 + \epsilon_0 t_\beta} 
~M_W^2 \mu M_2~\zeta~ \Big( f^{\tilde W}_{7,8}(M_2^2,\mu^2,m_{Q}^2) 
- f^{\tilde W}_{7,8}(M_2^2,\mu^2,m_{Q_3}^2) \Big)  \nonumber \\
&& - \frac{\epsilon_{\rm FC} t_\beta^2}{(1+\epsilon_b t_\beta)(1+\epsilon_0 t_\beta)} ~M_W^2 \mu M_2~f^{\tilde W}_{7,8}(M_2^2,\mu^2,m_{Q_3}^2) ~.
\end{eqnarray}
\end{widetext}
The first term in~(\ref{eq:C78Higgs}) corresponds to contributions
from a charged Higgs loop.  The loop functions, $h_{7,8}$ depend on
the ratio of the top mass and the charged Higgs mass, $r_t =
m_t^2/M_{H^\pm}^2$, and for $r_t=1$ are given by $h_7(1) = -7/18$ and
$h_8(1) = -1/3$. Their full analytical expressions can be found in the
appendix. The second term in~(\ref{eq:C78Higgs}) arises from neutral
heavy Higgs loops. It is strongly suppressed by the bottom quark mass
and is only important for very large $\tan\beta$. The loop functions, 
$z_{7,8}$, depend on the ratio of the bottom mass and the charged 
Higgs mass and since $m_b^2/M_{H^\pm}^2 \ll 1$, they are very well 
approximated by  $z_7 =-\frac{1}{18}$ and $z_8 = \frac{1}{6}$.

\begin{figure}[tb]
\centering
\raisebox{8pt}{\includegraphics[width=0.45\textwidth]{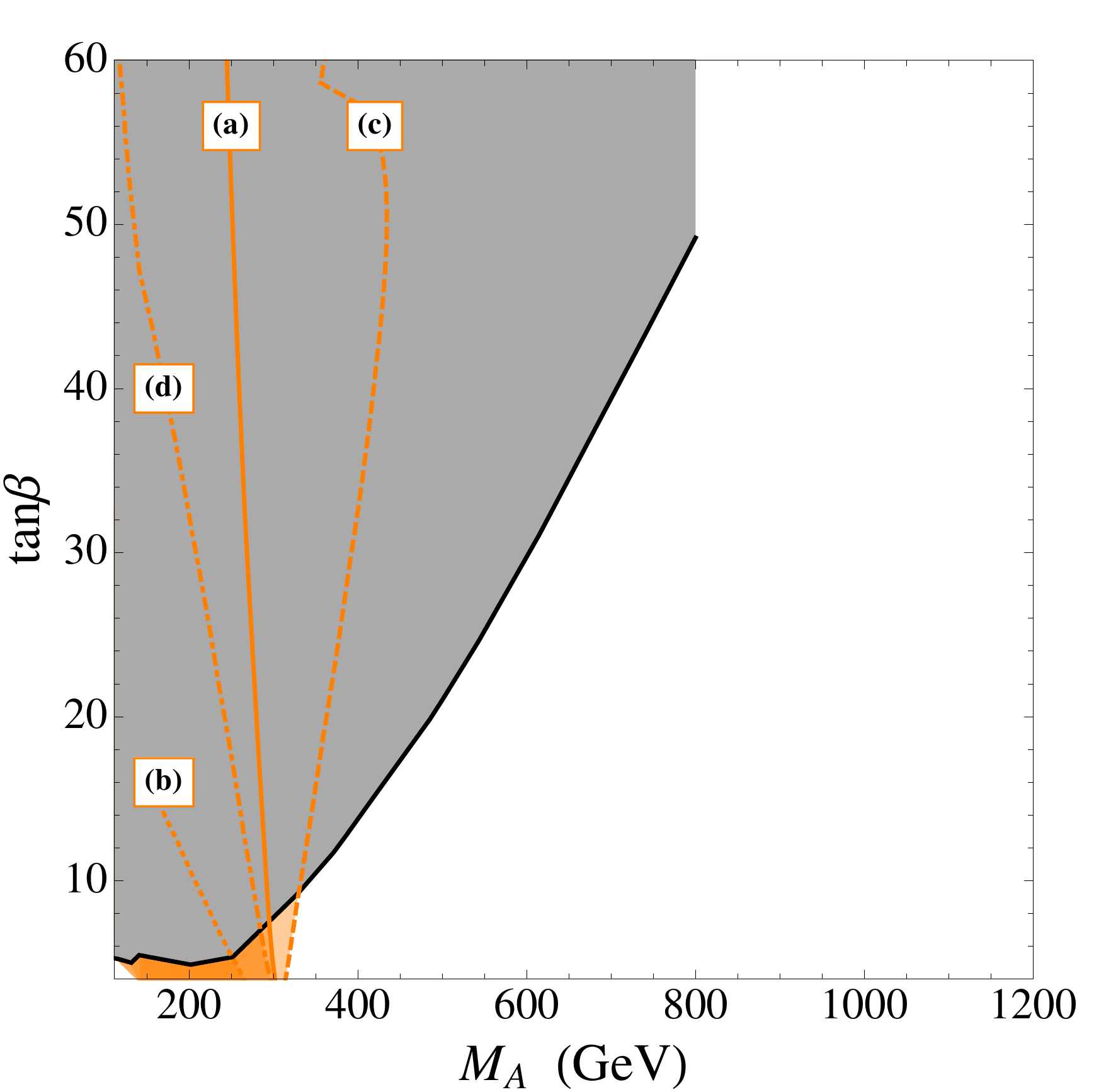}}
\caption{Constraints in the $M_A$--$\tan\beta$ plane from the $B \to
  X_s \gamma$ decay. The orange solid, dotted, dashed, and dash-dotted
  contours correspond to scenarios (a), (b), (c), and (d) as described
  in the text.  The gray region is excluded by direct searches of MSSM
  Higgs bosons in the $H/A \to \tau^+ \tau^-$ channel.}
\label{fig:bsgamma1}
\end{figure}

\begin{figure*}[tb]
\centering
\includegraphics[width=0.45\textwidth]{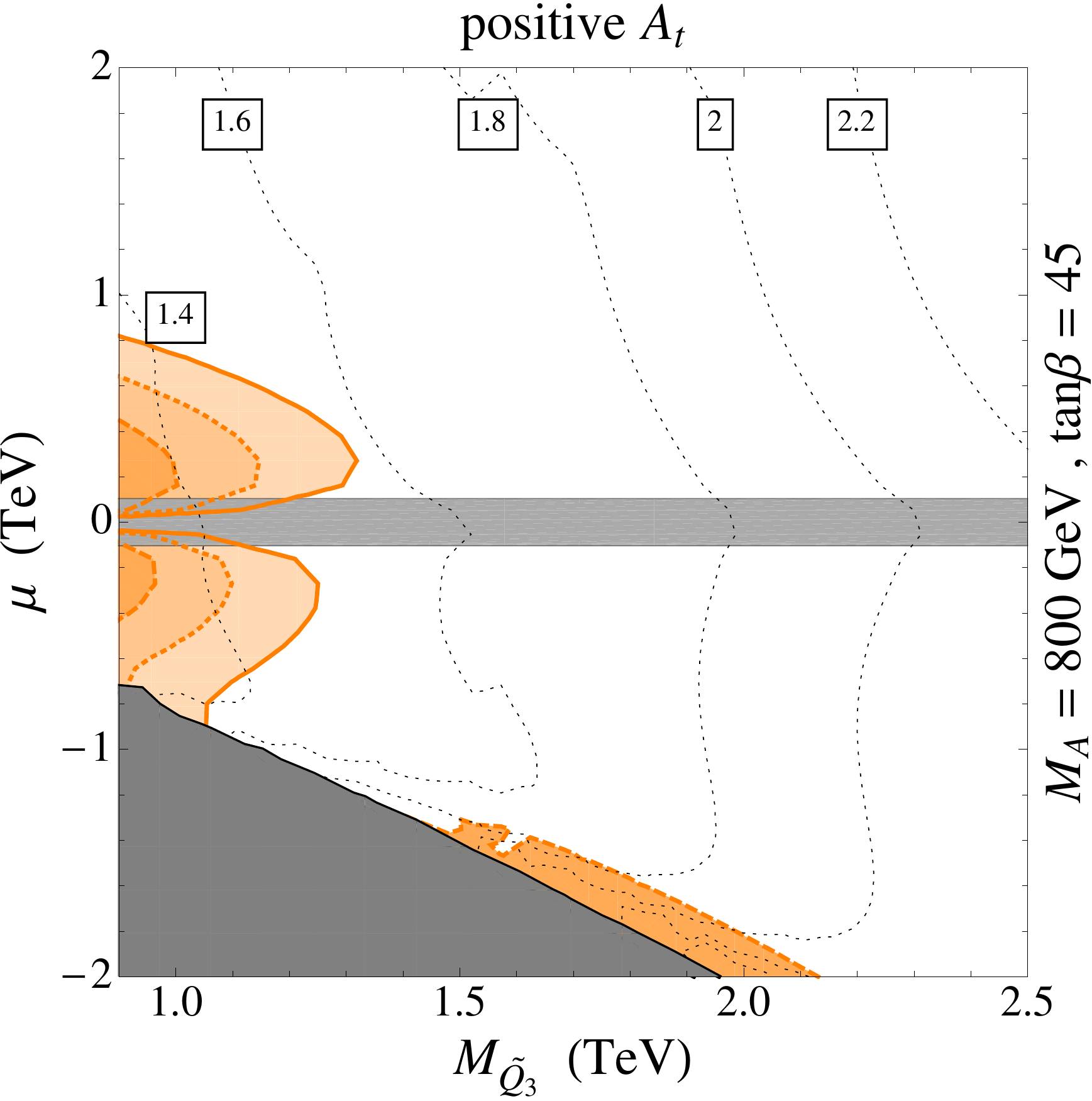} ~~~~
\includegraphics[width=0.45\textwidth]{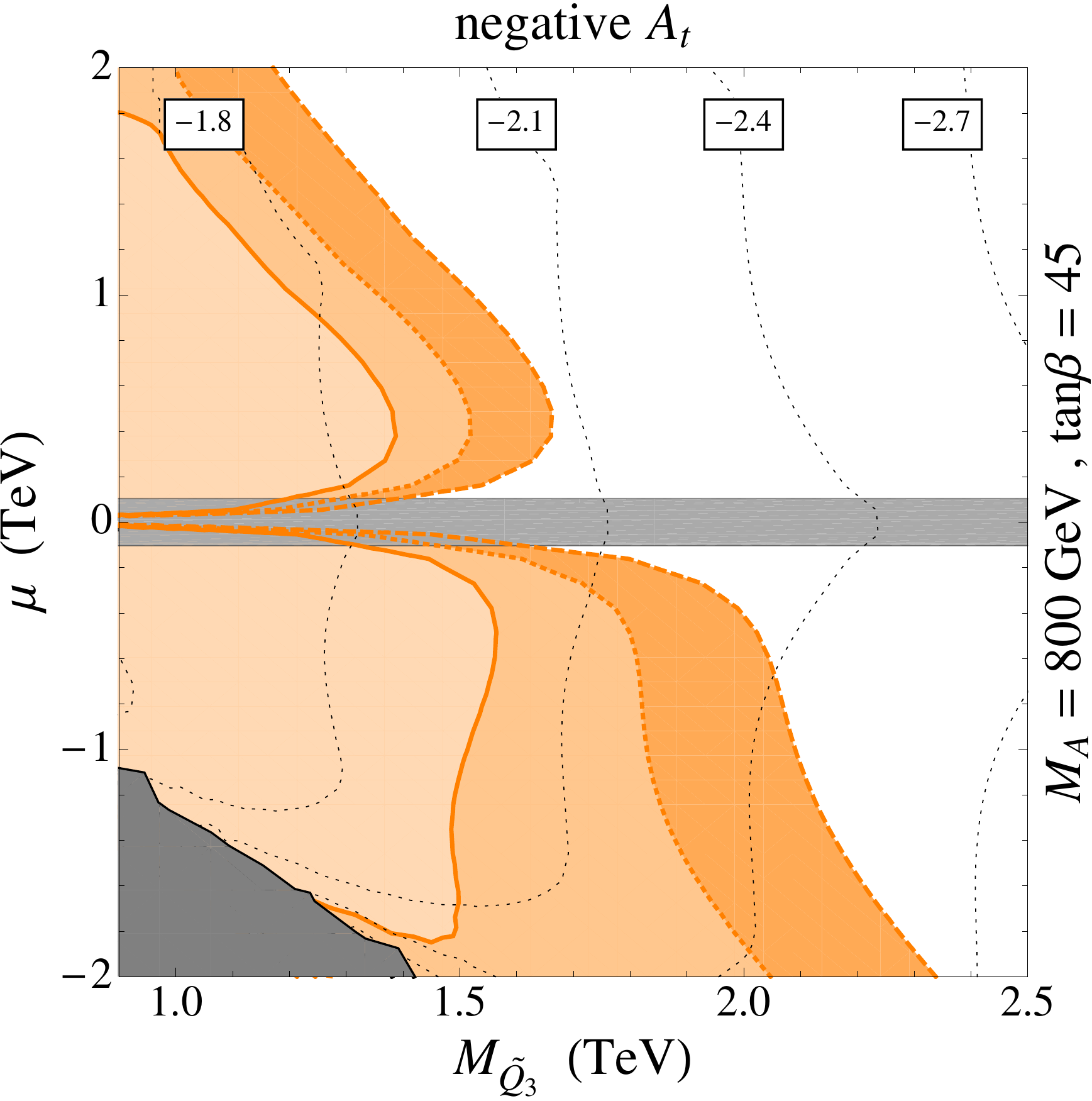}
\caption{Constraints in the $m_{Q_3}$--$\mu$ plane from the $B \to X_s
  \gamma$ decay, for fixed $M_3 = 3M_2 = 6M_1 = 1.5$ TeV. The solid
  bounded regions correspond to a degenerate squark spectrum.  The
  dashed and dotted bounded regions correspond to choosing the first
  two squark generations 50\% heavier than the third generation squark
  masses, with an alignment of $\zeta = 1$ and $\zeta = 0.5$,
  respectively.  The gray horizontal band corresponds to the
  constraint from direct searches of charginos at LEP. The vertical
  dotted lines show contours of constant $A_t$ such that $M_h =
  125$~GeV. In the gray regions in the lower left corners the lightest
  Higgs mass is always below $M_h < 125$~GeV, taking into
  account a 3 GeV theory uncertainty.}
\label{fig:bsgamma2}
\end{figure*}

Contributions from Higgsino--stop, gluino--down squark, and Wino--down
squark loops are shown
in~(\ref{eq:C78Higgsino}),~(\ref{eq:C78gluino}),
and~(\ref{eq:C78Wino}), respectively.  We do not write the typically
negligible bino contributions.  

For a degenerate SUSY spectrum with mass $\tilde m$, the loop
functions entering the Higgsino and gaugino contributions reduce to
\begin{eqnarray}
f_7^{\tilde H} \to \frac{5}{36} \frac{1}{\tilde m^4} &,& \quad 
f_7^{\tilde g} \to -\frac{2}{27} \frac{1}{\tilde m^4} \ , \quad
f_7^{\tilde W} \to -\frac{7}{24} \frac{1}{\tilde m^4} \ , \nonumber \\
f_8^{\tilde H} \to \frac{1}{12} \frac{1}{\tilde m^4} &,& \quad 
f_8^{\tilde g} \to -\frac{5}{18} \frac{1}{\tilde m^4} \ , \quad
f_8^{\tilde W} \to -\frac{1}{8}  \frac{1}{\tilde m^4} \ . \nonumber
\end{eqnarray}
Their full analytical expressions are collected in the appendix. In
contrast to the Higgs penguin contributions to $B_s \to \mu^+ \mu-$,
the SUSY loop contributions to $b\to s\gamma$ do decouple with the
SUSY scale. 

The first terms in~(\ref{eq:C78gluino}) and~(\ref{eq:C78Wino})
correspond to 1-loop flavor changing gaugino contributions. They
vanish for $m_{Q_3} = m_Q$, {\it i.e.} if there is no splitting
between the first two and the third generations of left-handed squark
masses. In the presence of a splitting, the parameter $\zeta$ again
parametrizes the alignment of the left-handed squark mass matrix. As
mentioned before, if the splitting is generated by RGE running we
expect $1/2 < \zeta < 1$. The second terms in~(\ref{eq:C78gluino})
and~(\ref{eq:C78Wino}) are formally 2-loop contributions but they can
be relevant for large $\tan\beta$. They {\it do not} vanish for
degenerate masses~\cite{Hofer:2009xb, Foster:2005wb}.

Similarly to $B_s \to \mu^+ \mu^-$, the MSSM contribution to $B \to
X_s \gamma$ is a sum of several terms that depend sensitively on many
parameters, particularly the signs of $\mu$ and $A_t$.

In Fig.~\ref{fig:bsgamma1}, we show in orange the constraints from $B
\to X_s \gamma$ in the $M_A$--$\tan\beta$ plane obtained analogous to
the $B_s \to \mu^+ \mu^-$ constraints discussed previously.  The plots
of Fig.~\ref{fig:bsgamma2} show the $B \to X_s \gamma$ constraints in
the plane of the third generation squark masses and the Higgsino mass
parameter $\mu$, again in complete analogy to the $B_s \to \mu^+
\mu^-$ constraints.

We can again see the connection between the constraints in the $\tan
\beta$ vs. $M_A$ plane, given in Fig.~\ref{fig:bsgamma1}, and the
$\mu$ vs. $m_{Q_3}$ plane, given in Fig.~\ref{fig:bsgamma2}.  The
squark masses are fixed to 2~TeV in Fig.~\ref{fig:bsgamma1}.  This
causes the stop-chargino contribution to be essentially negligible,
and hence we are only constrained by the Higgs contribution in the low
$M_A$ and large $\tan \beta$ regions.  For heavy squarks and low
$\tan\beta$, the bound on the charged Higgs mass is approximately
independent of the other SUSY parameters and is given by $M_{H^\pm}
\gtrsim 300$~GeV.  For large $\tan\beta$, the resummation factors
in~(\ref{eq:C78Higgs}) become relevant.  The most important effect
arises from the factors $\epsilon_0^\prime$ and $\epsilon_b$ in the
first term in~(\ref{eq:C78Higgs}).  For negative $\mu$,
$\epsilon_0^\prime$ and $\epsilon_b$ are negative and therefore the
bounds become stronger for larger $\tan\beta$ in scenario (c). For
positive $\mu$ (scenarios a, b, and d) instead, the bounds are relaxed
for large $\tan\beta$. As the dominant gluino contribution to
$\epsilon_0^\prime$ and $\epsilon_b$ grows with $\mu$ the $B \to X_s
\gamma$ constraint is weakest in scenario (b) that has the largest
$\mu=4$ TeV.  For the heavy squark masses chosen in
Fig.~\ref{fig:bsgamma1}, the direct searches for MSSM Higgs bosons
give stronger constraints compared to $B \to X_s \gamma$ except for
small values of $\tan\beta$.

In the plots of Fig.~\ref{fig:bsgamma2}, the variation of the squark
masses allows the stop-chargino contribution to become important for
small $m_{Q_3}$, demonstrating that the $\tan \beta$ vs. $M_A$
projection insufficiently illustrates the $B \to X_s \gamma$
constraint.  Partial cancellations are again in effect, and we
describe the relative signs of the various contributions in the
following.  Apart from extreme regions of parameter space, the charged
Higgs contribution interferes constructively with the SM and enhances
BR$(B \to X_s \gamma)$. However, for the case shown, $M_A = 800$ GeV,
this contribution is small.  For positive (negative) $(\mu A_t)$, the
Higgsino loop contribution come with same (opposite) sign with respect
to the SM. Among the gaugino contributions, the dominant one is
typically the 1-loop gluino contribution. If a splitting in the left-handed squark masses is induced radiatively, its sign depends, for
positive $M_3$, only on the sign of $\mu$. For positive (negative)
$\mu$, gluinos interfere destructively (constructively) with the SM.

The plots of Fig.~\ref{fig:bsgamma2} clearly show the decoupling
behavior of the MSSM contributions to the $b \to s \gamma$
transition. For a degenerate squark spectrum ($m_{Q_3} = m_Q = m_{U_3}
= m_U = m_{D_3} = m_D = \tilde m$) and a heavy charged Higgs, the
bound from BR$(B \to X_s \gamma)$ hardly constrains the MSSM parameter
space beyond squark masses that are already excluded by direct SUSY
searches, namely $\tilde m \gtrsim \mathcal{O}(1~{\rm TeV})$.  In the
presence of a mass splitting between the first two and the third
generations of squarks, the $B \to X_s \gamma$ constraint can become
relevant for negative $A_t$, since the gluino and Higgsino
contributions add constructively. Squark masses significantly above
1 TeV can be probed in that case. For positive values of $A_t$, on
the other hand, the gluino and Higgsino loops partially cancel and the
bound from $B \to X_s \gamma$ is barely relevant.

\subsection{Discussion of RGE Effects} \label{sec:RGEs}

\begin{figure*}[tb]
\centering
\includegraphics[width=0.30\textwidth]{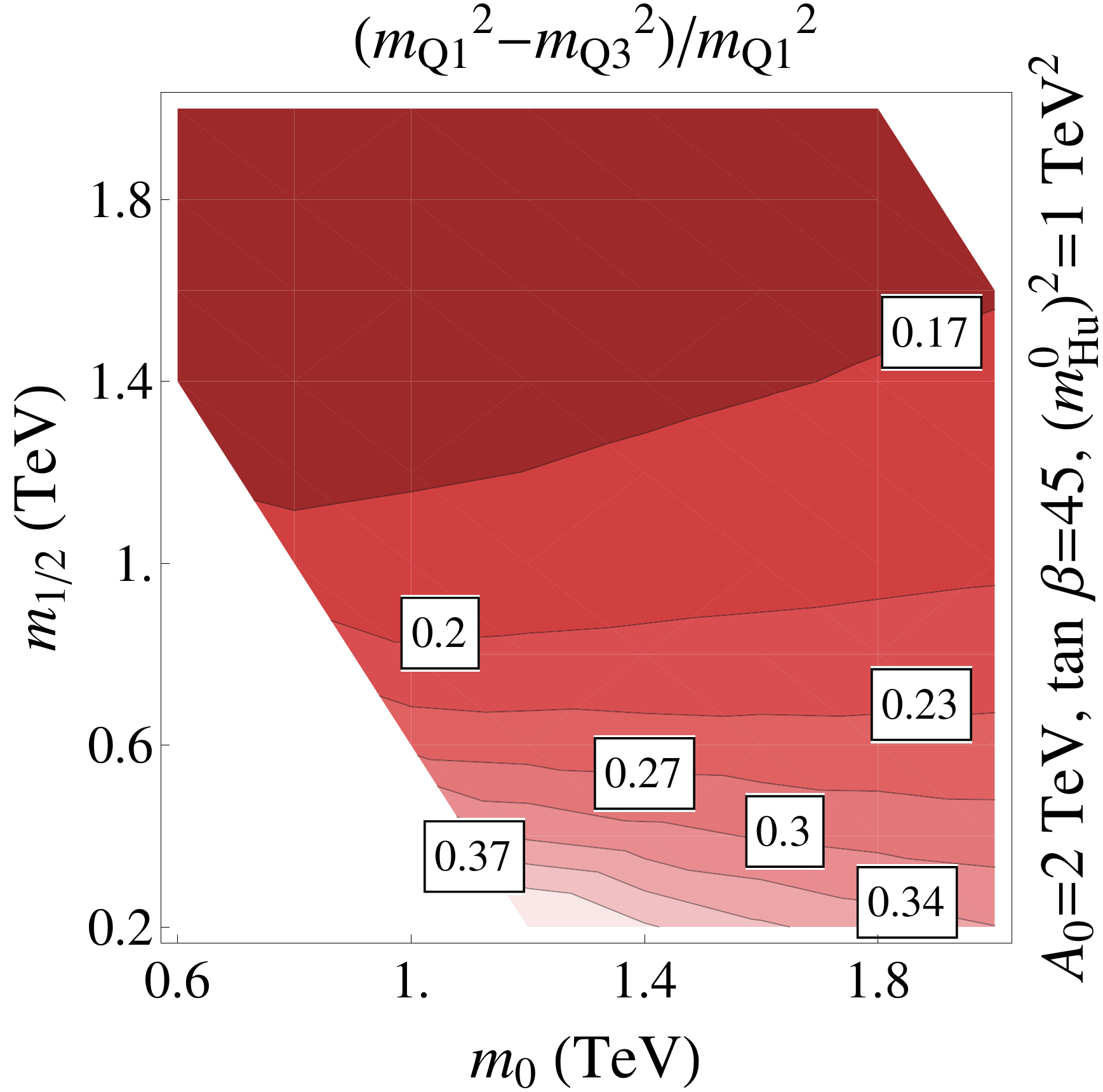} ~~~~
\includegraphics[width=0.30\textwidth]{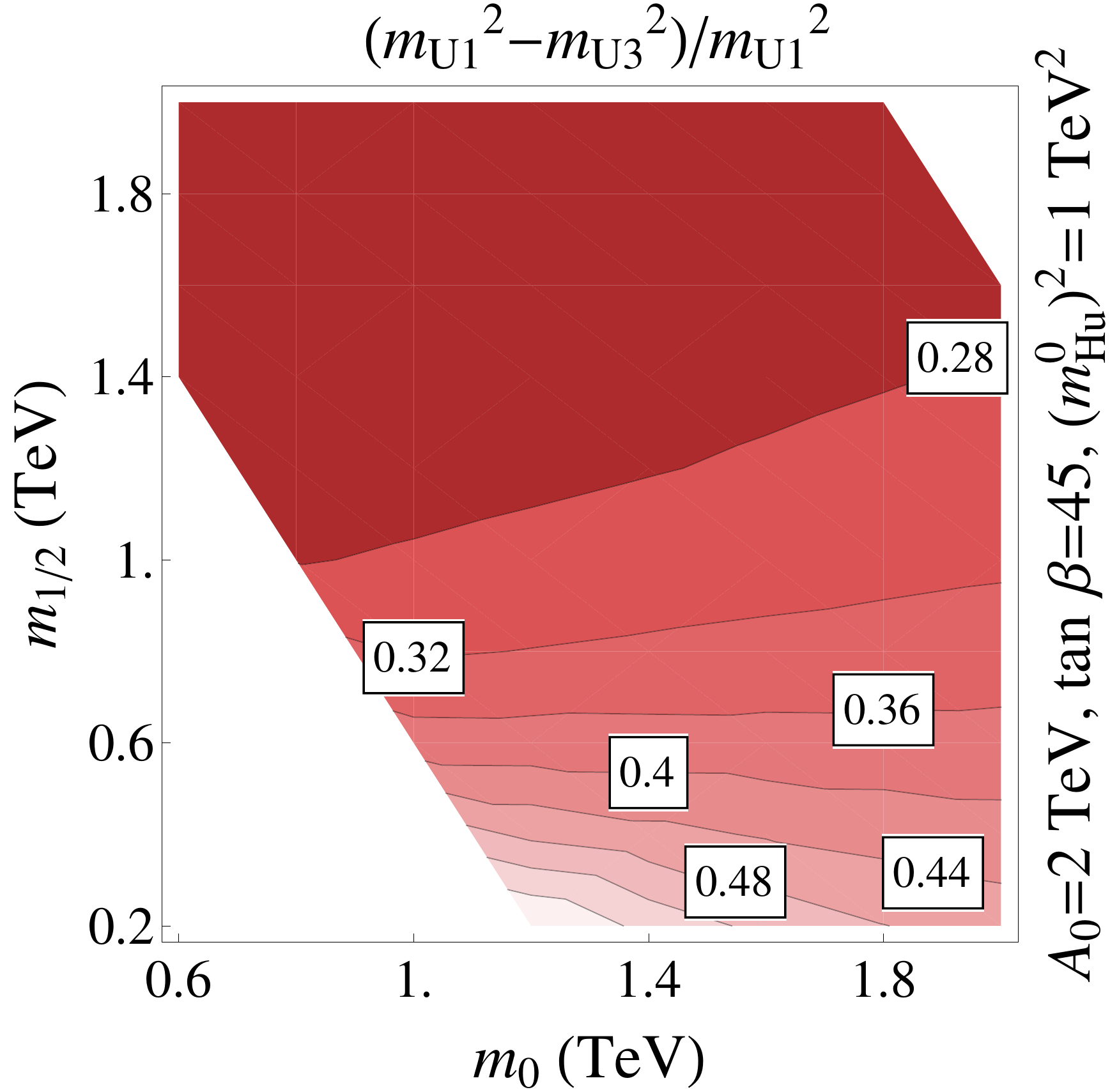} ~~~~
\includegraphics[width=0.30\textwidth]{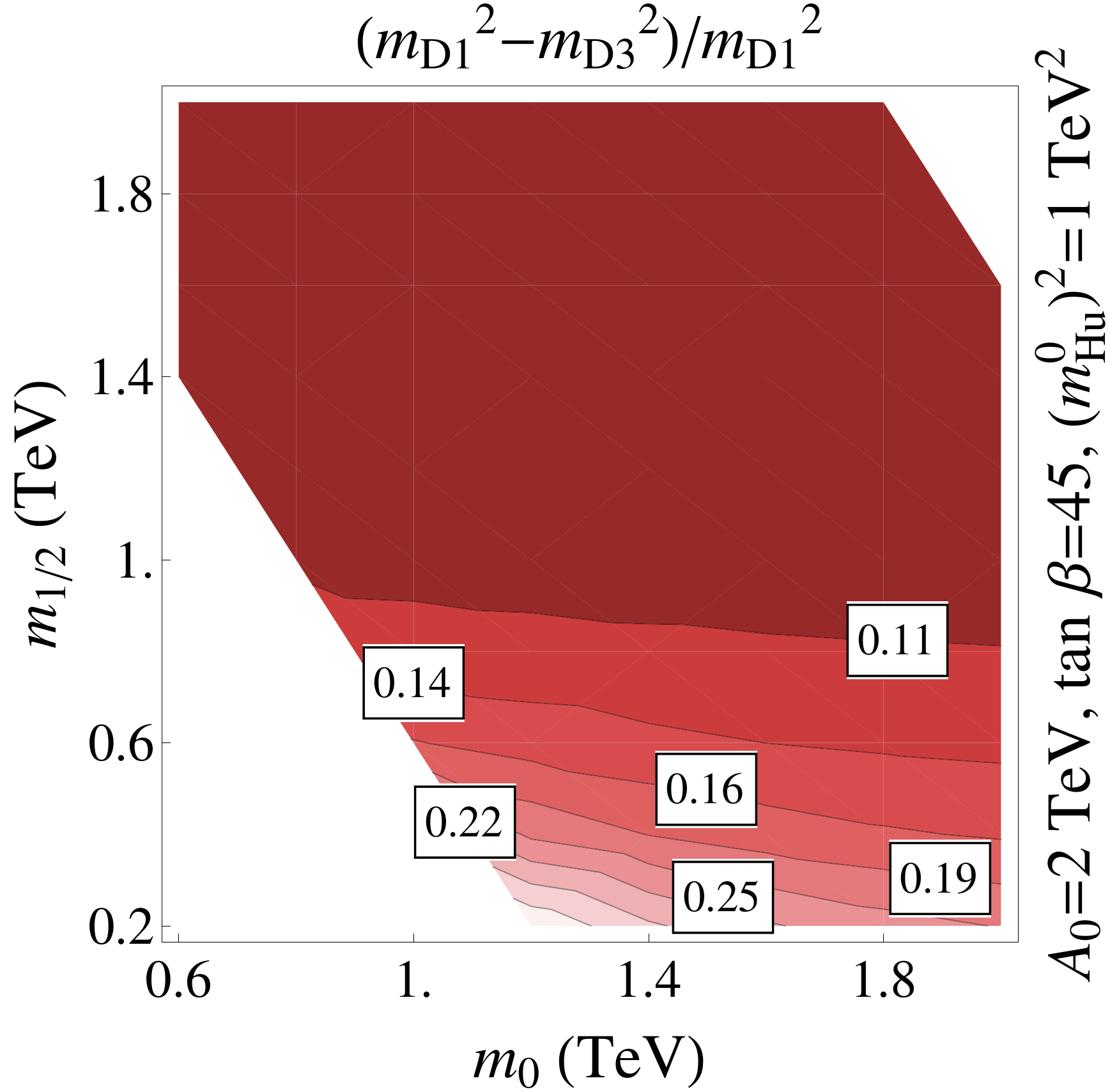} ~~~~ \newline
\includegraphics[width=0.30\textwidth]{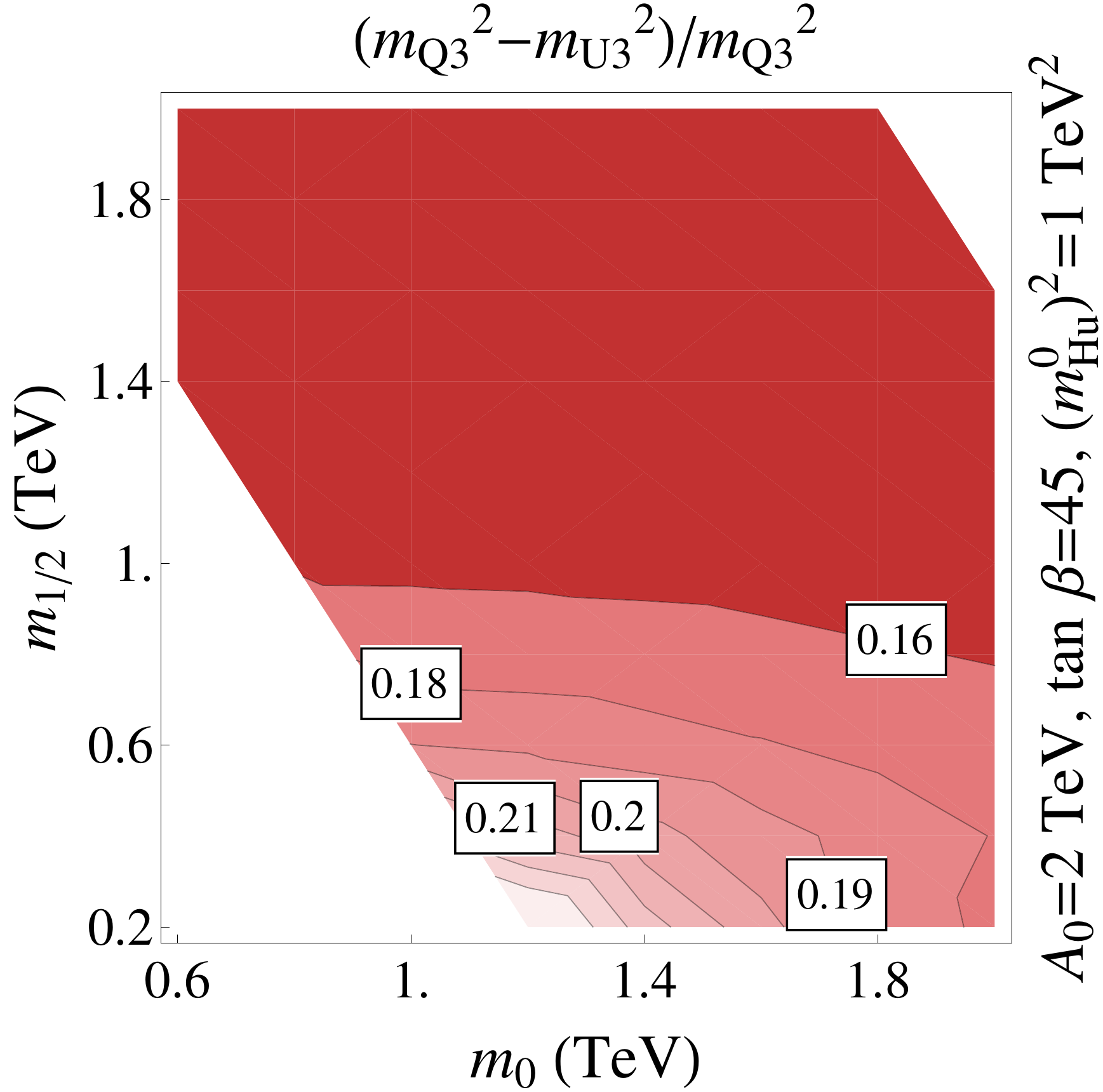} ~~~~
\includegraphics[width=0.30\textwidth]{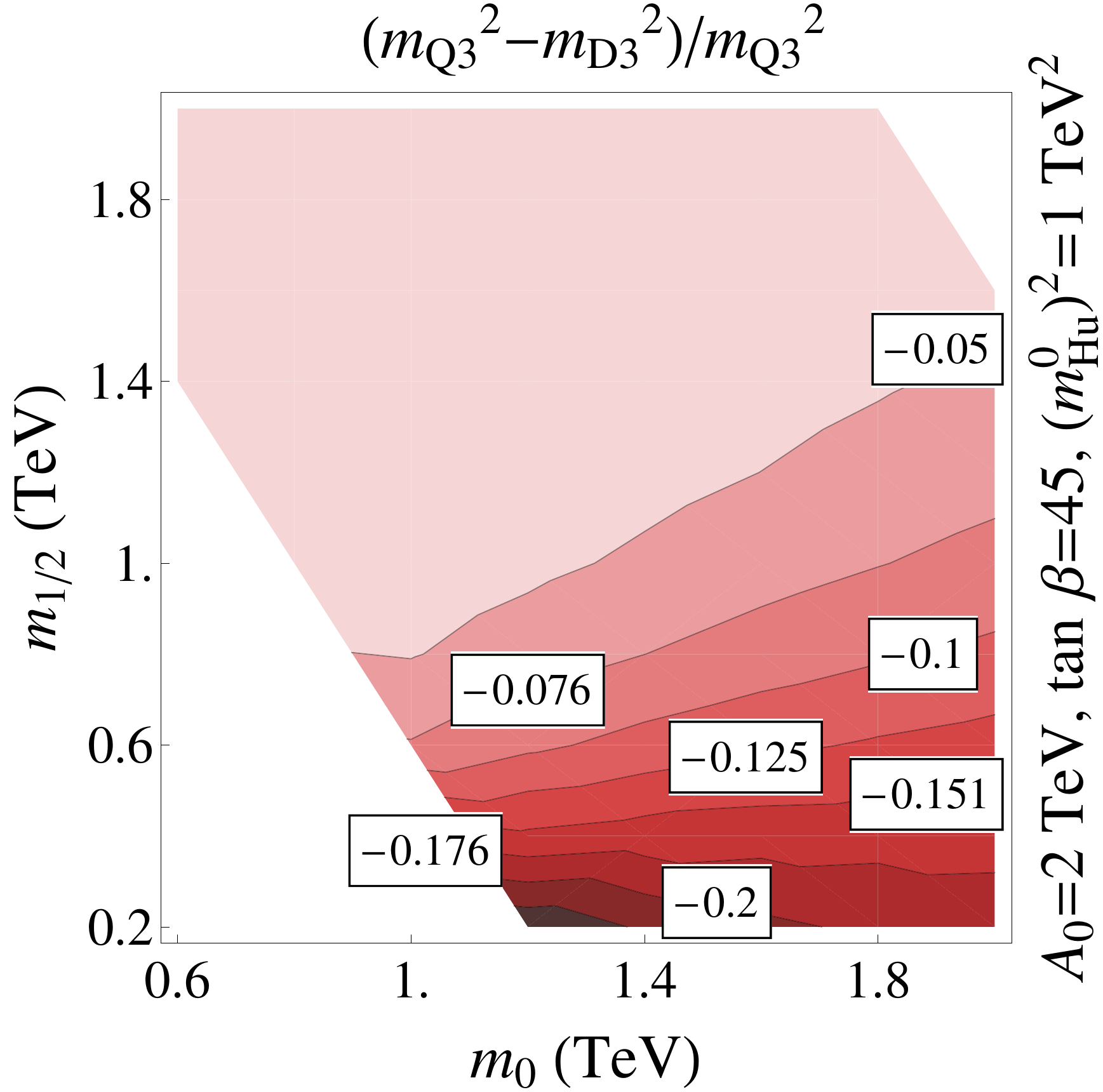} ~~~~
\caption{ Contours of $\Delta Q_{13}$ (top left), $\Delta U_{13}$ (top
  middle), $\Delta D_{13}$ (top right), $\Delta QU_3$ (bottom left),
  and $\Delta QD_3$ (bottom right) in the $(m_0, m_{1/2})$ plane,
  fixing $A_0 = 2$ TeV, $(m_{H_u}^0)^2 = 1$ TeV$^2$, $\tan \beta =
  45$, and requiring $M_A = 800$ GeV within 10\%. }
\label{fig:am3splittings}
\end{figure*}

\begin{figure*}[tb]
\centering
\includegraphics[width=0.30\textwidth]{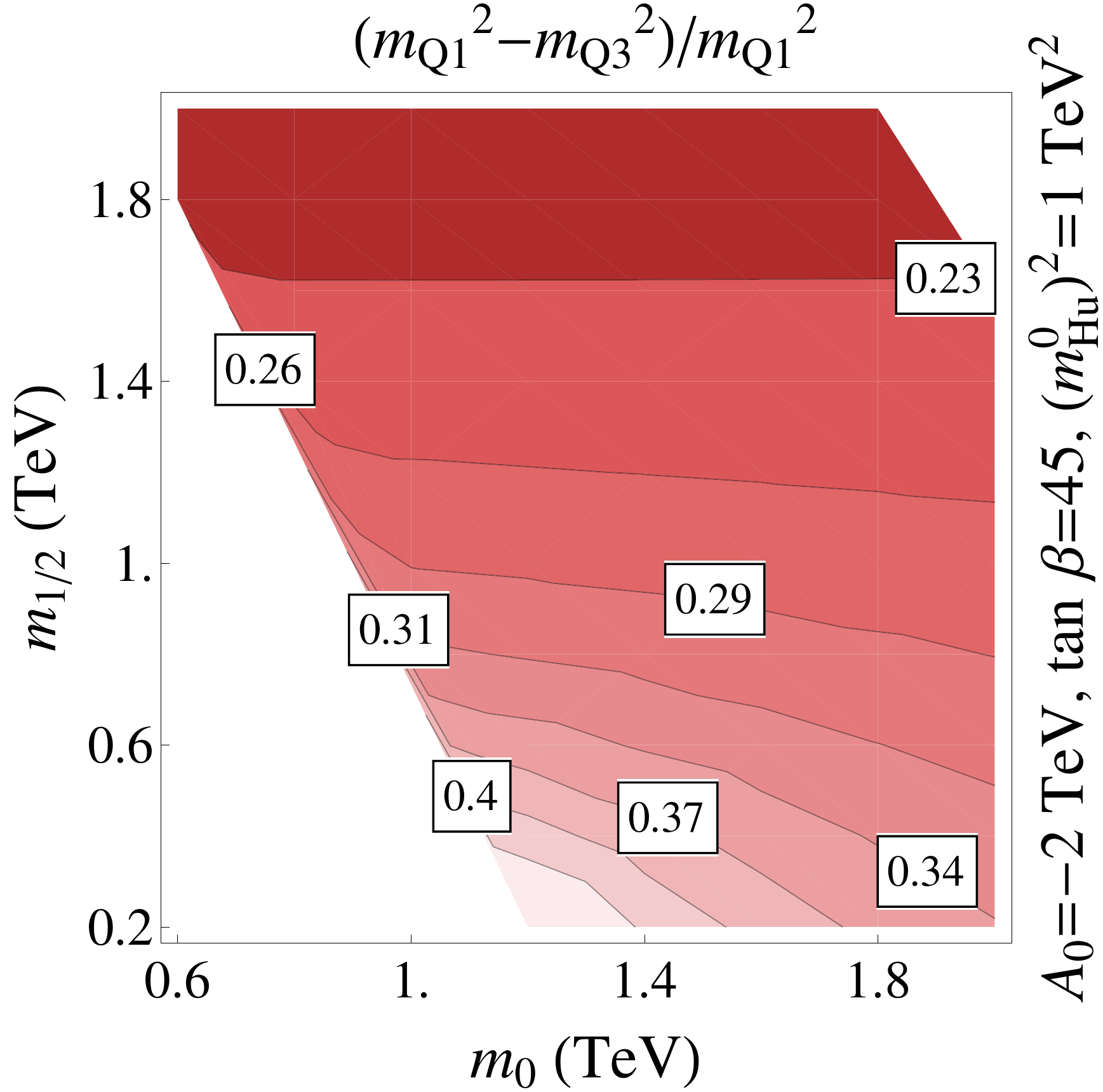} ~~~~
\includegraphics[width=0.30\textwidth]{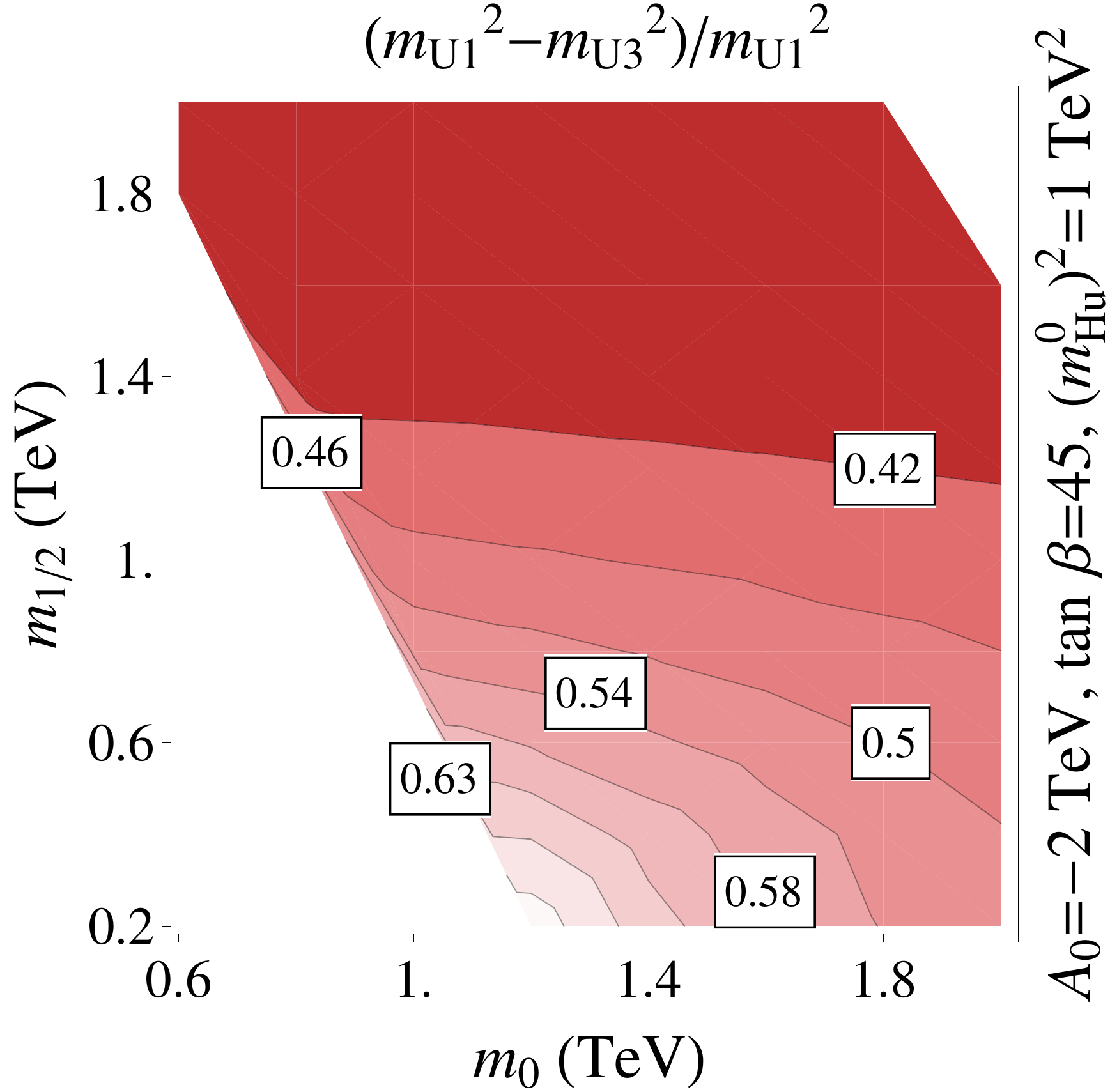} ~~~~
\includegraphics[width=0.30\textwidth]{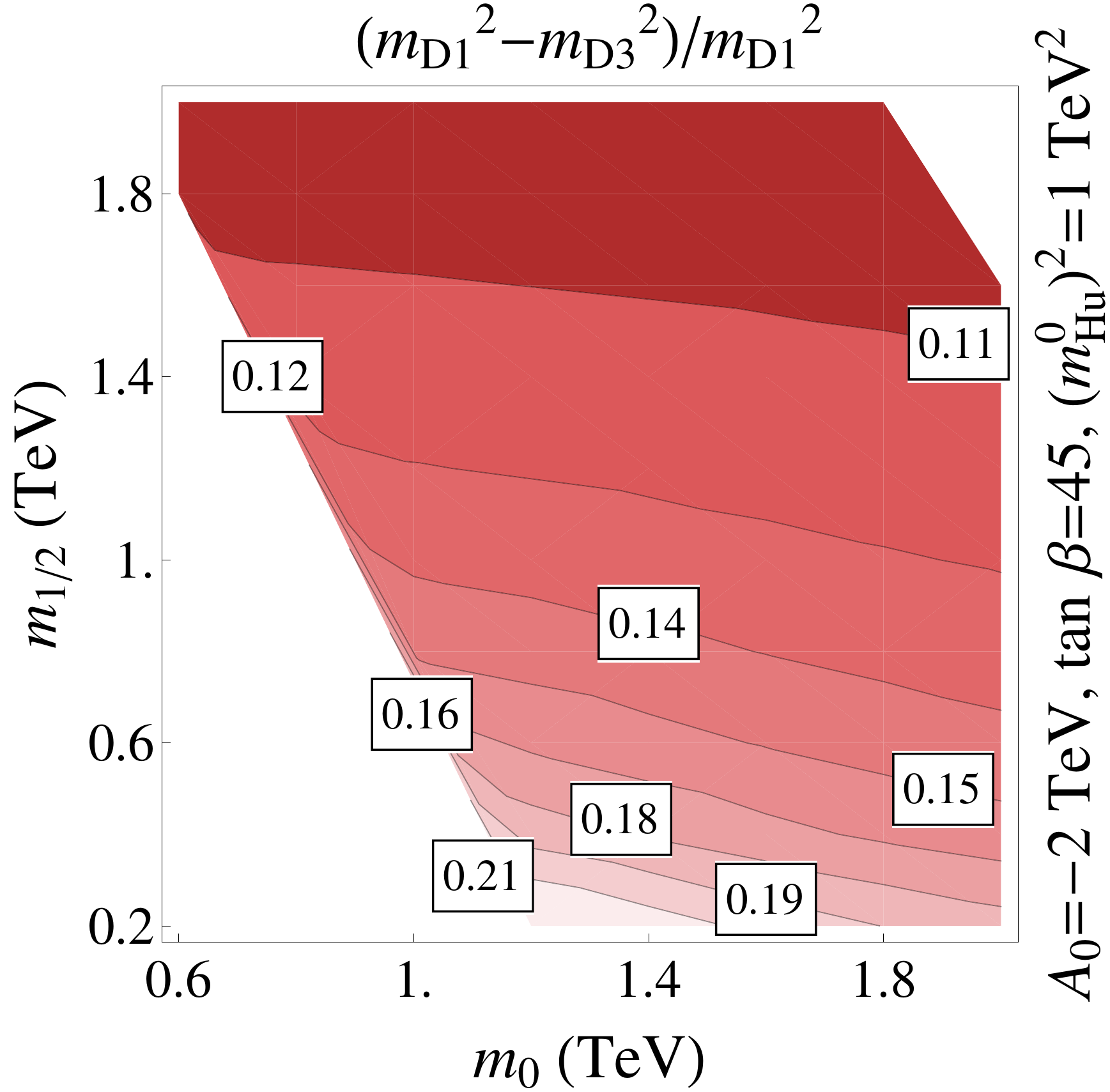} ~~~~ \newline
\includegraphics[width=0.30\textwidth]{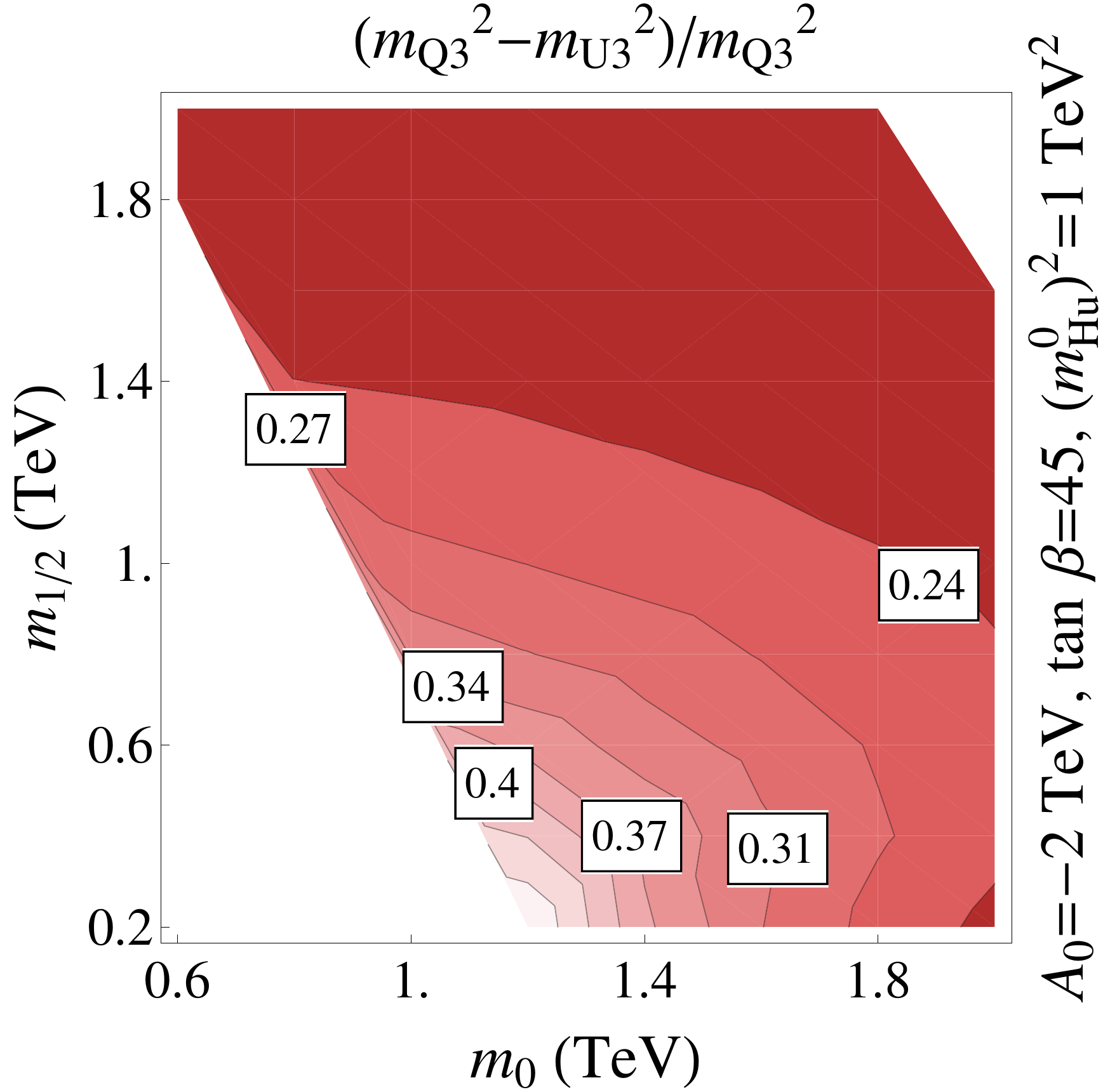} ~~~~
\includegraphics[width=0.30\textwidth]{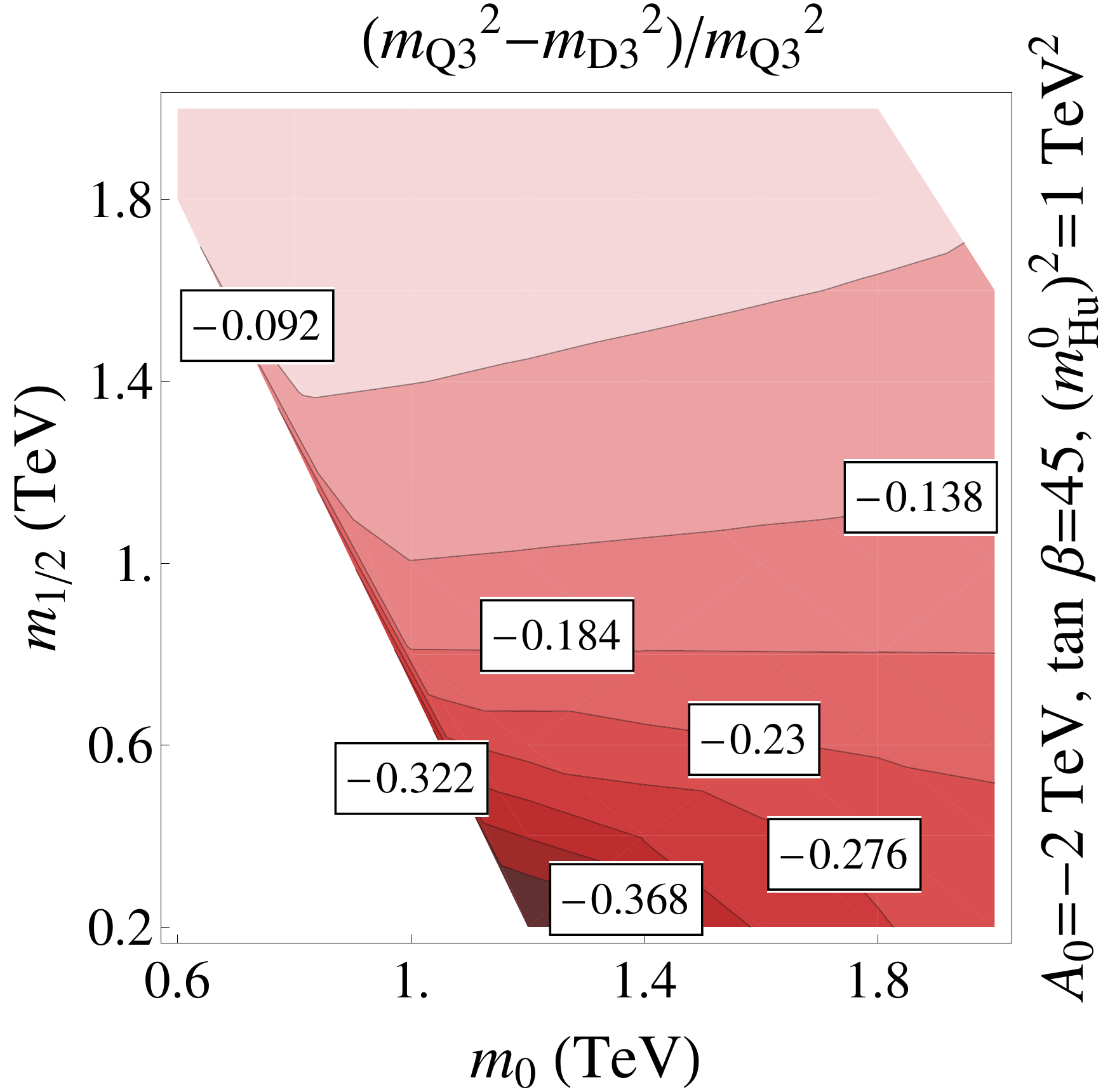} ~~~~
\caption{ Same as Fig.~\ref{fig:am3splittings} except $A_0 = -2$ TeV.}
\label{fig:ap3splittings}
\end{figure*}

\begin{figure*}[tb]
\centering
\includegraphics[width=0.45\textwidth]{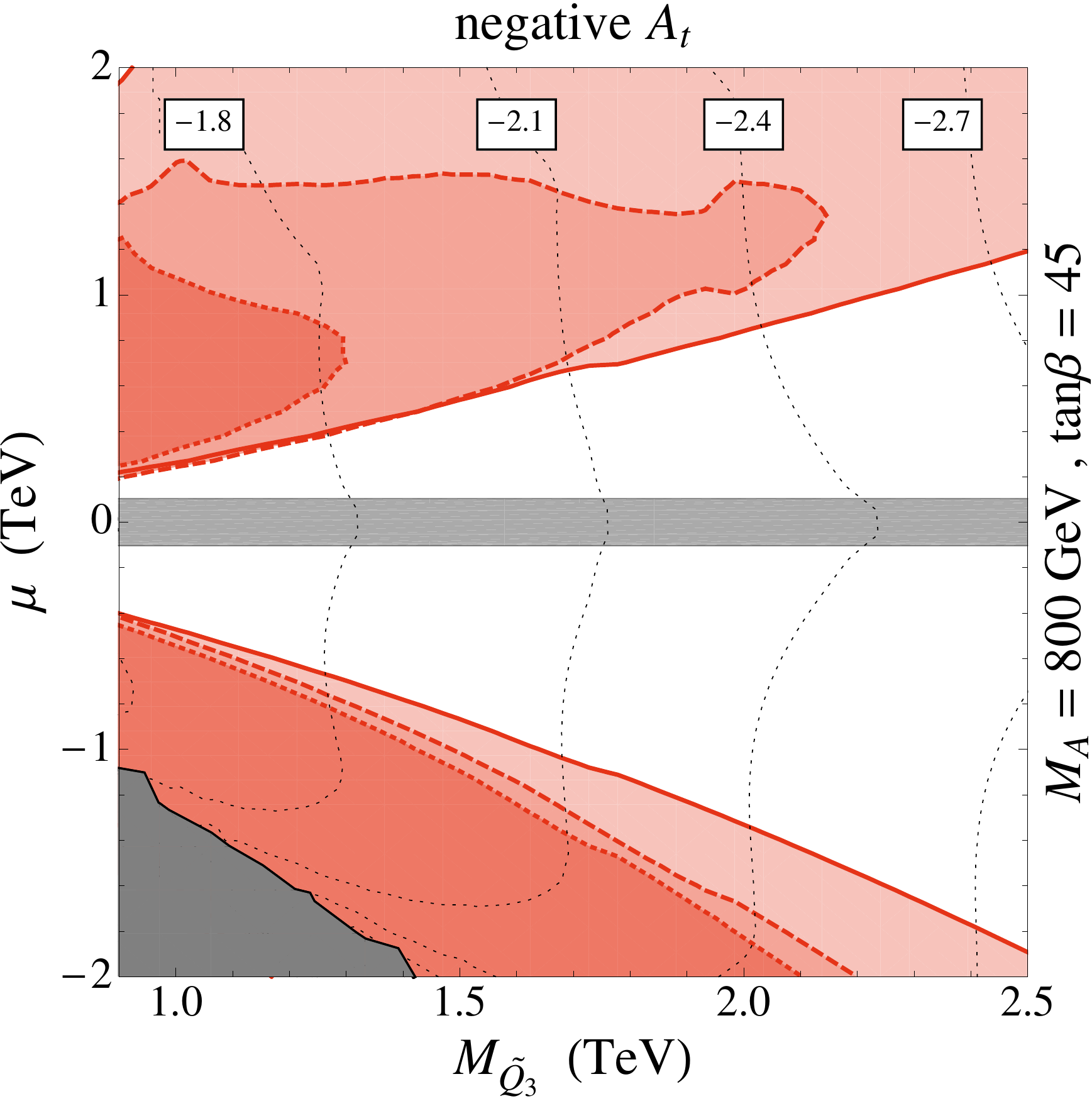} ~~~~
\includegraphics[width=0.45\textwidth]{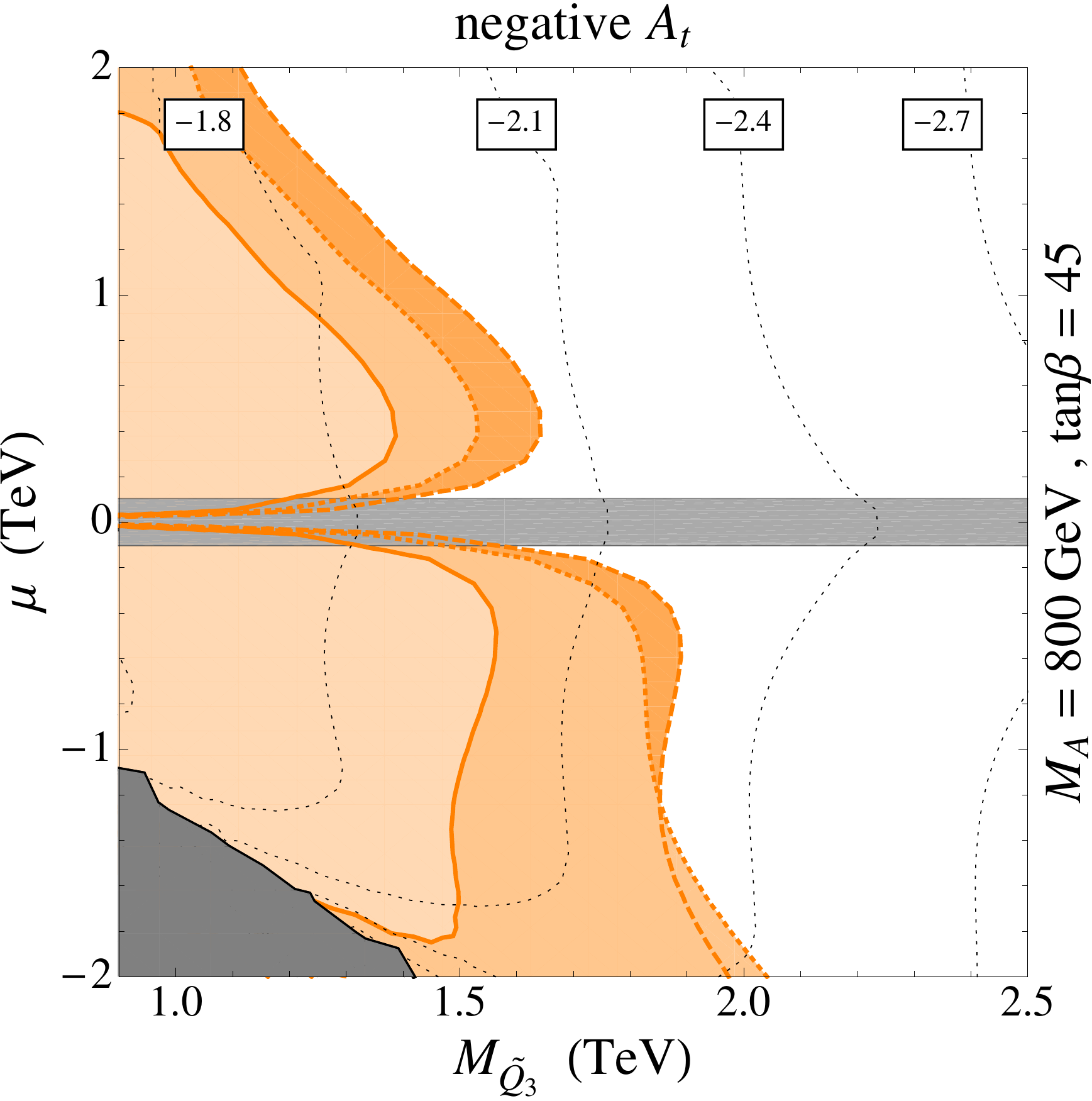}
\caption{Constraints in the $m_{Q_3}$--$\mu$ plane from $B_s \to \mu^+
  \mu^-$ (left) and $B \to X_s \gamma$ (right). The solid bounded
  regions correspond to a degenerate squark spectrum.  The dashed and
  dotted bounded regions correspond to mass splittings in the squark
  spectrum implied by RGE running. In particular $\Delta Q_{13} =
  0.35$, $\Delta U_{13} = 0.6$, $\Delta D_{13} = 0.15$, $\Delta QU_{3}
  = 0.35$, and $\Delta QD_{3} = -0.25$ for the dashed contours and
  $\Delta Q_{13} = 0.35$, $\Delta U_{13} = 0.6$, $\Delta D_{13} =
  0.15$, and $\Delta QU_{3} = \Delta QD_{3} = 0$ for the dotted
  contours. The gray horizontal band corresponds to the constraint
  from direct chargino searches. The vertical dotted lines show
  contours of constant $A_t$ such that $M_h = 125$~GeV. In the gray
  regions in the lower left corners the lightest Higgs mass is always
  below $M_h < 125$~GeV, taking into account a 3 GeV theory
  uncertainty.}
\label{fig:Bdecays_RGEs}
\end{figure*}

Our phenomenological analysis of MSSM mass parameters serves our
purpose of understanding the flavor constraints on the low energy MSSM
spectrum. However, we also want to connect these constraints to
parameters of a high scale SUSY parameter space. To this end, we
consider a typical example in the large $\tan \beta$ and $M_A$ region
compatible with direct $H / A \to \tau^+ \tau^-$ searches at the LHC.
We show typical mass differences between soft parameters for squarks
in the plane of the mSUGRA boundary conditions, $m_0$ and $m_{1/2}$,
fixing the remaining mSUGRA parameters to $A_0 = \pm 2$ TeV and $\tan
\beta = 45$. We also chose the SUSY breaking scale to be the GUT
scale, $10^{16}$~GeV.  We deviate slightly from the strict mSUGRA
prescription and work in a non-universal Higgs mass (NUHM) scenario by
fixing the Higgs soft mass $m_{H_u}^2 = 1$~TeV$^2$ and adjusting
$m_{H_d}^2$ at the high scale to obtain $M_A$ within 10\% of $800$ GeV
at the low scale of $Q = 1$~TeV.  Using these boundary conditions and
the usual low energy Yukawa constraints derived from fermion masses
run to $Q = 1$ TeV, we numerically solve the RGE system dictated by
2-loop running from~\cite{Martin:1993zk} and 1-loop radiative
corrections from~\cite{Pierce:1996zz}.  Our choice of $A_0$ typically
gives the lightest SM-like Higgs a mass of $122 \pm 2$~GeV. For the
bulk of the region in the $(m_0, m_{1/2})$ plane, adjusting $A_0$ (in
particular, $A_t$) to obtain a Higgs mass of 125 GeV changes the
quantitative picture by less than a few percent.  For very small $m_0$
and $m_{1/2}$, however, where some squarks or sleptons become close to
tachyonic, the mass splittings can vary significantly as result of
changing $A_0$.

We highlight that the $B$ observable constraints can vary
significantly as a result of Yukawa-induced squark mass splittings
inherent in RG running, as seen in Fig.~\ref{fig:Bsmumu2} and
Fig.~\ref{fig:bsgamma2}, respectively.  In particular, the most
significant mass splittings among the squarks occur as a result of the
top and bottom Yukawas, where a significant enhancement of the bottom
Yukawa occurs for large $\tan \beta$.

We can obtain a semi-analytic understanding of the resulting mass
splittings, following the simplified 1-loop RG analysis
of~\cite{Ibanez:1983di, Ibanez:1984vq, Carena:1993bs}.  Neglecting the
first- and second-generation Yukawa couplings and $\alpha_1^2$
contributions, we have
\begin{eqnarray}
m_{Q_3}^2 (t) &\approx& m_{Q_3}^2 (0) + I_{\alpha_3} + I_{\alpha_2} - I_t - I_b \ , \\
m_{U_3}^2 (t) &\approx& m_{U_3}^2 (0) + I_{\alpha_3} - 2 I_t \ , \label{eq:mU3} \\
m_{D_3}^2 (t) &\approx& m_{D_3}^2 (0) + I_{\alpha_3} - 2 I_b \ ,
\end{eqnarray}
and
\begin{eqnarray}
I_{\alpha_3} &\equiv& \int dt \left( 
\dfrac{16}{3} \dfrac{\alpha_3}{4 \pi} M_3^2 \right) \ , \\
I_{\alpha_2} &\equiv& \int dt \left( 
3 \dfrac{\alpha_2}{4 \pi} M_2^2 \right) \ , \\
I_t &\equiv& \dfrac{1}{16\pi^2} \int dt ~y_t^2 \left( 
m_Q^2 + m_U^2 + m_{H_u}^2 + A_t^2 \right) \ , \\
I_b &\equiv& \dfrac{1}{16\pi^2} \int dt ~y_b^2 \left(
m_Q^2 + m_D^2 + m_{H_d}^2 + A_b^2 \right) \ ,
\end{eqnarray}
where $t = 0$ corresponds to the GUT scale.  The analogous
$m_{Q_1}^2$, $m_{U_1}^2$, and $m_{D_1}^2$ approximations can be
obtained from the above by neglecting the $I_t$ and $I_b$
contributions.

For the trilinear couplings, neglecting $\alpha_1$ and
$A_\tau$,
\begin{eqnarray}
A_t &\approx& A_0 + \int dt ~\left[ \left( 
\dfrac{16}{3} \dfrac{\alpha_3}{4\pi} M_3 +
3 \dfrac{\alpha_2}{4\pi} M_2 \right) \right. \nonumber \\
&-& \left. 6 \dfrac{y_t^2}{16 \pi^2} A_t - \dfrac{y_b^2}{16 \pi^2} A_b 
\right] \\
A_b &\approx& A_0 + \int dt ~\left[ \left(
\dfrac{16}{3} \dfrac{\alpha_3}{4\pi} M_3 +
3 \dfrac{\alpha_2}{4\pi} M_2 \right) \right. \nonumber \\
&-& \left. \dfrac{y_t^2}{16 \pi^2} A_t - 6 \dfrac{y_b^2}{16 \pi^2} A_b 
\right] \ .
\end{eqnarray}
The most relevant mass splittings for our analysis are
\begin{eqnarray}
\Delta Q_{13} &\equiv& \dfrac{m_{Q_1}^2 - m_{Q_3}^2}{m_{Q_1}^2} 
\approx \dfrac{I_t + I_b}{m_{Q_1}^2} \\
\Delta U_{13} &\equiv& \dfrac{m_{U_1}^2 - m_{U_3}^2}{m_{U_1}^2} 
\approx \dfrac{2 I_t}{m_{U_1}^2} \\ 
\Delta D_{13} &\equiv& \dfrac{m_{D_1}^2 - m_{D_3}^2}{m_{D_1}^2} 
\approx \dfrac{2 I_b}{m_{D_1}^2} \\
\Delta QU_{3} &\equiv& \dfrac{m_{Q_3}^2 - m_{U_3}^2}{m_{Q_3}^2} 
\approx \dfrac{ I_{\alpha_2} + I_t - I_b}{
m_{Q_1}^2 - I_t - I_b} \\
\Delta QD_{3} &\equiv& \dfrac{m_{Q_3}^2 - m_{D_3}^2}{m_{Q_3}^2}
\approx \dfrac{ I_{\alpha_2} - I_t + I_b}{
m_{Q_1}^2 - I_t - I_b} \ .
\end{eqnarray}
From these relations we see that $\Delta U_{13} + \Delta D_{13}
\approx 2 \Delta Q_{13}$, where $\Delta D_{13}$ is small compared to
$\Delta U_{13}$ for small $\tan \beta$. We also expect $\Delta QU_{3}
= - \Delta QD_{3}$ for small $m_{1/2}$.  These relations for the
various mass splittings, based on 1-loop semi-analytic results, are
borne out in our numerical results, which are calculated from 2-loop
RG running, and are shown in Fig.~\ref{fig:am3splittings} and
Fig.~\ref{fig:ap3splittings}.

For a gluino with mass $M_3 = 1.5$~TeV as we considered in the
previous sections, we have $m_{1/2} \simeq 500$~GeV and therefore a
significant splitting is induced among the squark masses in running
down from the GUT scale. The most important splitting in the general discussion of the previous sections is $\Delta
Q_{13}$, as it leads to gaugino loop contributions to FCNCs. In our mSUGRA setup, it is
typically around $25\%$ for positive $A_0$ and $35\%$ for negative
$A_0$.  The splitting between the masses of the squarks decreases for
larger $m_{1/2}$. This is due to the universal $SU(3)$ contribution,
$I_{\alpha_3}$, to $m_{Q_3}$, $m_Q$, $m_{U_3}$, $m_U$, $m_{D_3}$, and
$m_D$, which dominates for large $m_{1/2}$.

From the approximate expressions above, we can also estimate the size
of $\zeta$ resulting from running. We have
\begin{equation}
\zeta \simeq \frac{I_t}{I_t + I_b} ~.
\end{equation}
Even though we chose a large value of $\tan\beta = 45$ for the
examples shown, the bottom Yukawa effects are limited.  Note that for
the parameter region explored here, $\zeta \sim 80\%$, which means that
the squark mass splitting is dominantly driven by the top Yukawa and
therefore aligned in the up-sector.  For smaller $\tan\beta$, the
alignment parameter $\zeta$ is even closer to 1.

Note that the gaugino loop contributions to FCNCs depend approximately on the product $\zeta \times \Delta
Q_{13}$. In the mSUGRA scenario discussed here, we find to a good approximation $\zeta \times \Delta
Q_{13} \simeq \Delta U_{13}/2$. In more generic setups however, this relation does not hold and we will continue to discuss the 
gaugino loop contributions to FCNCs in terms of $\zeta$ and $\Delta Q_{13}$ separately.

In the plots of Fig.~\ref{fig:Bdecays_RGEs}, we show again the
constraints from $B_s \to \mu^+ \mu^-$ and $B \to X_s \gamma$ in the
$m_{Q_3}$--$\mu$ plane, this time setting the various mass splittings
according to our results of the mSUGRA RGE running.  In particular, we
use $\Delta Q_{13} = 0.35$, $\Delta U_{13} = 0.6$, $\Delta D_{13} =
0.15$, $\Delta QU_{3} = 0.35$, and $\Delta QD_{3} = -0.25$, which are
typical values for $m_{1/2} \simeq 500$~GeV and negative $A_t$. As we saw in the previous
sections for positive $A_t$, the $B_s \to \mu^+ \mu^-$ constraint
depends very mildly on the squark mass splitting and the $B \to X_s
\gamma$ constraint is barely relevant. Therefore, we restrict
ourselves to negative $A_t$. For comparison, the solid contours
indicate again the constraints obtained for a degenerate squark
spectrum. The dotted contours corresponds to keeping all third
generation squarks degenerate and only implementing the splitting
between the first two and the third generation as given by the RGE
running. The dashed contours correspond to the situation where all
squark mass splittings are as dictated by the RGE running.  The former
case behaves as expected given the analysis of Secs.~\ref{sec:Bsmumu}
and~\ref{sec:bsgamma}.  For the latter case, however, once mass
splittings between the different types of third generation squarks are
also considered, an additional effect arises.  As can be seen
from~(\ref{eq:mU3}) and confirmed in the lower left plots of
Figs.~\ref{fig:am3splittings} and~\ref{fig:ap3splittings}, the right
handed stop is typically significantly lighter than the other third
generation squarks.  The light right-handed stop then increases the
chargino-stop loop contributions to $B_s \to \mu^+ \mu^-$ and to $B
\to X_s \gamma$ leading overall to stronger constraints compared to
the case of degenerate third generation squarks.

Two of the most important quantities dictated by RGEs for flavor 
observables are the values of $\Delta Q_{13}$ and $\zeta$. Within the 
assumption of flavor universality at the messanger scale, 
$\Delta Q_{13}$ and $\zeta$ depend mainly on the messenger scale, 
$\tan \beta$ and the ratio of gluino mass to squark masses. 
Lowering the messenger scale from the GUT scale as well as increasing 
the gluino mass decreases the splitting $\Delta Q_{13}$, but 
leaves $\zeta$ approximately invariant. Smaller (larger) values of 
$\tan \beta$ would decrease (increase) $\Delta Q_{13}$ and 
simultaneously increase (decrease) $\zeta$, leaving the product $\zeta \times \Delta Q_{13} \simeq \Delta U_{13}/2$ approximately invariant. As we saw, making the 
splitting smaller strengthens the BR$(B_s \to \mu^+ \mu^-)$ constraint 
for negative $A_t$, but increasing $\zeta$ will relax it. The effect 
of these two quantities is exactly opposite on the constraints coming 
from BR($B \to X_s \gamma$). This complimentary behavior implies that 
even varying the messenger scale and $\tan \beta$, these two flavor 
observables will be able to constrain the parameter space efficiently.

\section{Dark Matter Direct Detection} \label{sec:DM}

\begin{figure*}[tb]
\centering
\includegraphics[width=0.45\textwidth]{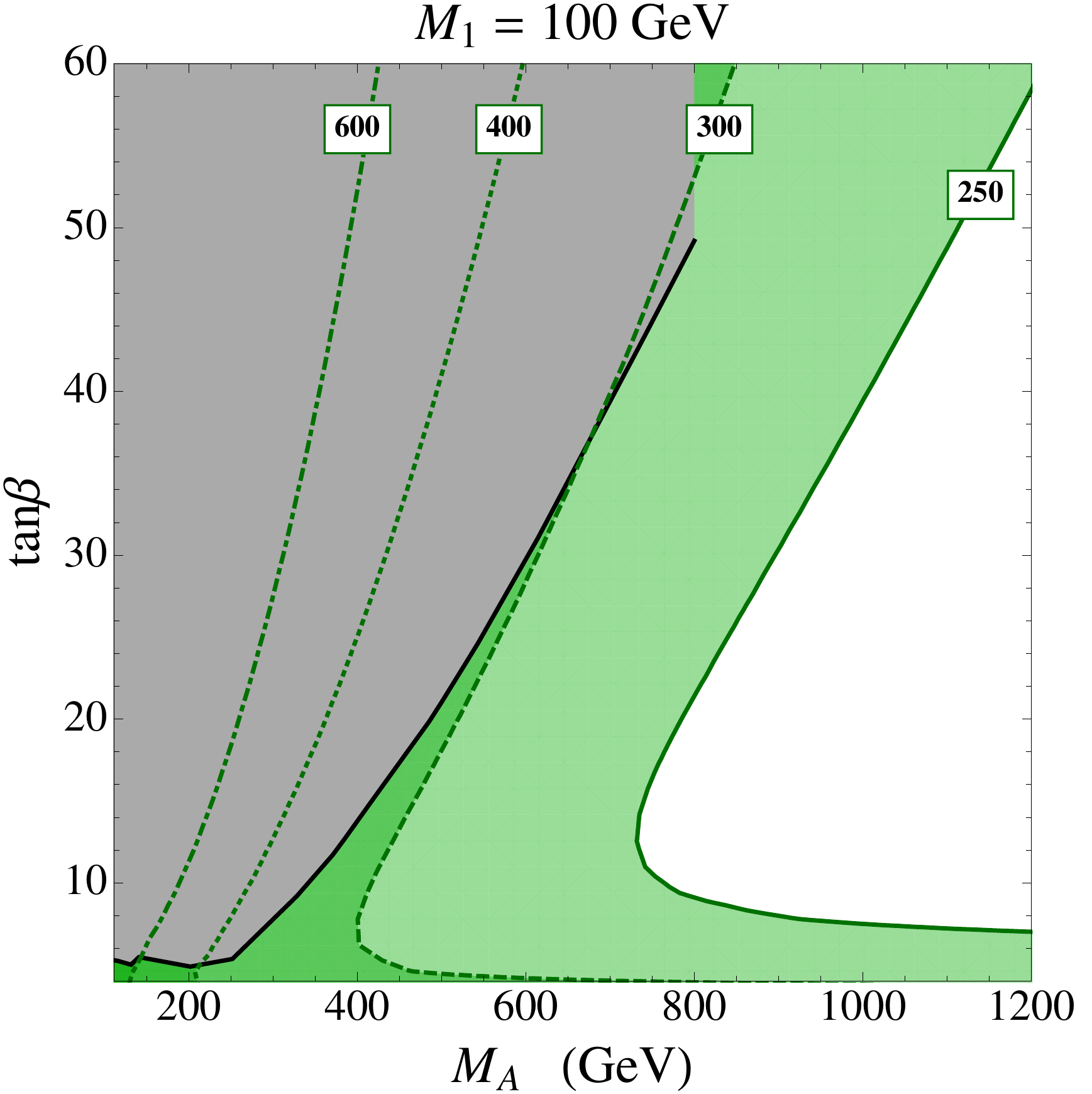} ~~~~
\includegraphics[width=0.45\textwidth]{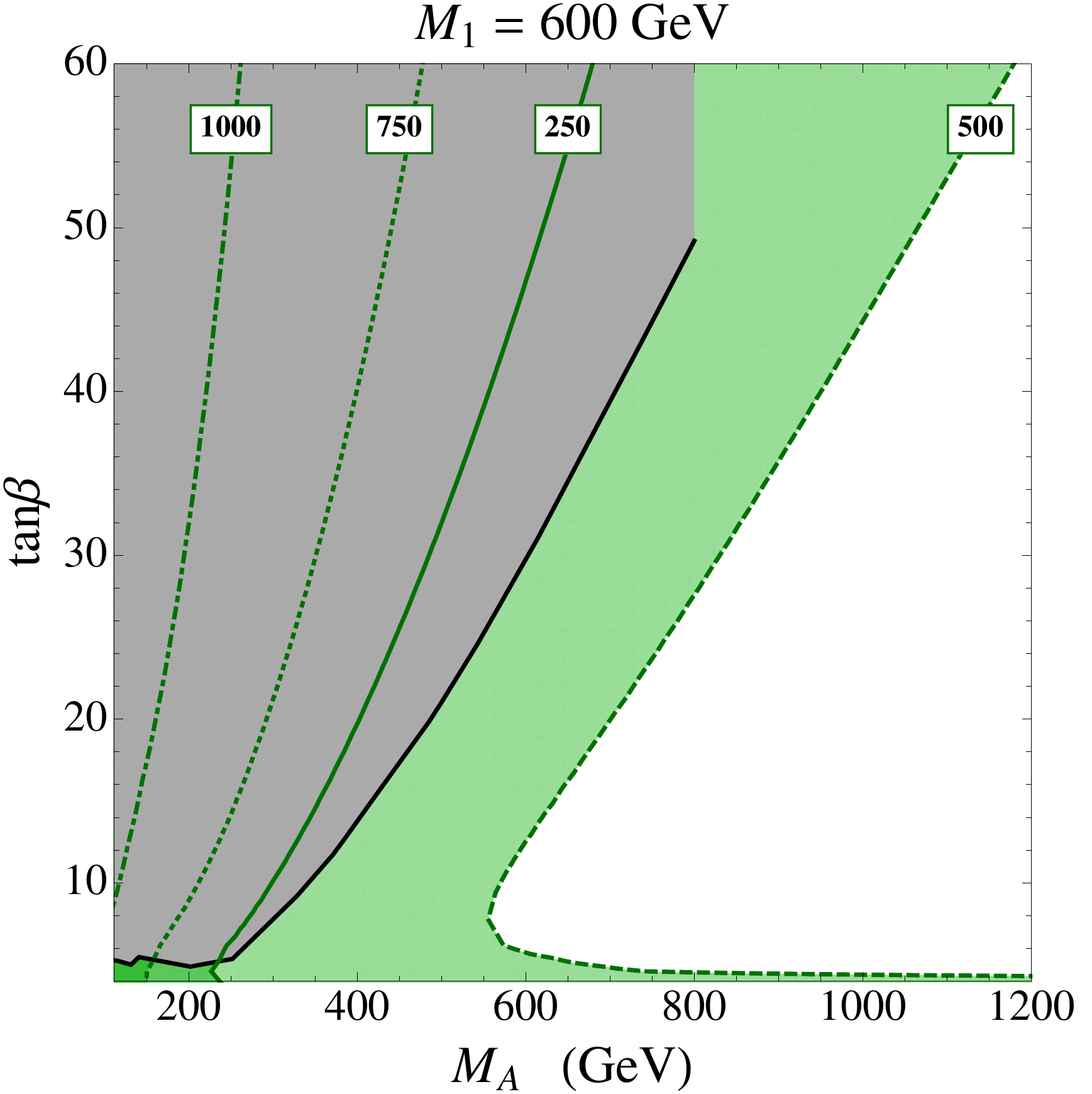}
\caption{ Constraints in the $M_A$--$\tan\beta$ plane from Dark Matter
  direct detection. The green solid, dashed, dotted and dash-dotted
  contours correspond to different values of $\mu$ as indicated.  The
  gray region is excluded by direct searches of MSSM Higgs bosons in
  the $H/A \to \tau^+ \tau^-$ channel.  }
\label{fig:darkmatter1}
\end{figure*}

The lightest neutralino in the MSSM is an excellent thermal dark
matter candidate. The lightest neutralino is a weakly interacting
massive particle (WIMP) and therefore generically leads to roughly the
correct order of magnitude for the observed dark matter relic density.
This is particularly true in the well-tempered neutralino
scenario~\cite{ArkaniHamed:2006mb}, where the lightest neutralino is a
mixture of the Bino and Wino or the Higgsino.  In the following we do not
assume any specific mechanism by which the correct dark matter relic
abundance is achieved, but simply assume that the lightest
neutralino in the MSSM accounts for the dark matter in the
universe~\cite{Gelmini:2006pw}.

Neutralinos interact with SM matter and therefore dark matter
direct detection limits can be used to put bounds on the MSSM
parameter space, complementary to the bounds from direct
searches and low energy flavor observables~\cite{Srednicki:1989kj,
  Gelmini:1990je, Drees:1993bu, Ellis:2000ds, Ellis:2005mb,
  Carena:2006dg, Carena:2006nv, Strege:2011pk, Fowlie:2012im,
  Buchmueller:2012hv, Perelstein:2012qg}.

The Xenon100 Collaboration recently set very stringent limits on the
spin-independent elastic dark matter nucleon scattering cross
section~\cite{Aprile:2011hi, :2012nq}. For dark matter masses of
$\mathcal{O}$(100 GeV), the bounds are as strong as $\sigma < 2 \times
10^{-45}$cm$^2$, assuming canonical values for the local dark matter
density, the local circular velocity and the Galactic escape velocity.
Interpreted in the context of the MSSM with neutralino dark matter,
these bounds are starting to probe significant parts of the parameter
space.

The spin-independent elastic neutralino-proton cross-section can be
written as
\begin{equation}
\sigma = \frac{4 M_\chi^2 m_p^2}{\pi (M_\chi + m_p)^2} f_p^2 ~,
\end{equation}
where $M_\chi$ is the mass of the lightest neutralino, $m_p$ is the
proton mass, and
\begin{equation}
\frac{f_p}{m_p} = \left( \sum_{q=u,d,s} f_{{\rm T}_q}^p c_q + 
\frac{2}{27} f_{\rm TG}^p \sum_{q=c,b,t} c_q \right) ~.
\end{equation}
The non-perturbative parameters $f_{{\rm T}_q}^p$ and $f_{\rm TG}^p =
1-f_{{\rm T}_u}^p-f_{{\rm T}_d}^p-f_{{\rm T}_s}^p$ come from the
evaluation of nuclear matrix elements. We use the latest
lattice determinations in our numerical analysis~\cite{Giedt:2009mr}
\begin{equation}
f_{{\rm T}_u}^p = f_{{\rm T}_d}^p = 0.028 ~,~ f_{{\rm T}_s}^p = 0.0689~.
\end{equation}
These values are expected to be affected by considerable
uncertainties.  We assume isospin symmetry when applying the Xenon100
bounds.

For large $\tan\beta$, the dominant contributions to the coefficients,
$c_q$, parametrizing the neutralino--quark couplings, typically
come from the $t$-channel exchange of the heavy scalar $H$ and read
\begin{eqnarray} \label{eq:DM1}
c_d^H = c_s^H &\simeq& 
\frac{g_1^2}{4 M_H^2} \frac{t_\beta}{1 + \epsilon_s t_\beta} 
\frac{\mu}{M_1^2 - \mu^2} ~,\\  \label{eq:DM2}
c_b^H &\simeq& 
\frac{g_1^2}{4 M_H^2} \frac{t_\beta}{1 + \epsilon_b t_\beta} 
\frac{\mu}{M_1^2 - \mu^2} ~.
\end{eqnarray}
The $t$-channel exchange of the SM-like Higgs affects all $c_q$
approximately equally:
\begin{equation} \label{eq:DMh}
c_q^h \simeq \frac{g_1^2}{4 M_h^2} ~ \frac{M_1}{M_1^2 - \mu^2} ~.
\end{equation}
While the $c_q^h$ are not enhanced by $\tan\beta$, bounds on the
direct detection cross section have become so strong that the
$t$-channel exchange of the SM-like Higgs is also probed.

The above expressions hold in the large $\tan\beta$ limit and assume
the lightest supersymmetric particle to be mainly a bino--higgsino
mixture with $M_1 \neq \mu$. In our numerical analysis, we go beyond
the large $\tan\beta$ limit: we work with neutralino mass eigenstates
and include the effects from $s$-channel squark exchange,
though these are always very suppressed by the squark masses.

As is evident from~(\ref{eq:DM1}),~(\ref{eq:DM2}), and~(\ref{eq:DMh}),
the neutralino-proton cross section depends strongly on $M_1$ and
$\mu$. This can be also seen from the plots of
Fig.~\ref{fig:darkmatter1}, which show in green the regions in
the canonical $M_A$--$\tan\beta$ plane that are excluded by the
Xenon100 constraints. In the left plot, the bino mass is set to $M_1 =
100$~GeV with $M_2 = 2 M_1$ and the solid, dashed, dotted and
dash-dotted contours correspond to $\mu =$ 250~GeV, 300~GeV, 400~GeV,
and 600~GeV, as indicated in the plot.  In the right plot, we choose a
larger bino mass of $M_1 = 600$~GeV, with $M_2 = 2 M_1$ again, and the
solid, dashed, dotted and dash-dotted contours correspond to $\mu = $
250~GeV, 500~GeV, 750~GeV, and 1000~GeV.
Dependence on other SUSY parameters enters at the loop level through
the $\epsilon_i$ factors in~(\ref{eq:DM1}) and~(\ref{eq:DM2}) and is 
therefore very moderate.
In these plots we fix a common squarks mass, $\tilde m = 2$ TeV, a
gluino mass of $M_3 = 1.5$ TeV and $A_t = A_b = A_\tau$ such that
the lightest Higgs mass 125 GeV.

The strongest constraints arise if binos and higgsinos are maximally
mixed, {\it i.e.} for $M_1 \simeq \mu$.  Indeed, if $M_1 = \mu$, we
find, independent of the values of $M_A$ and $\tan\beta$, that the
exchange of the SM-like Higgs leads to direct detection cross sections
that are already ruled out by the current bounds in the full range of
neutralino masses up to 1~TeV.  Away from bino-higgsino degeneracy,
regions of parameter space open up.  Still, for small heavy Higgs
masses and large values of $\tan\beta$, the heavy Higgs exchange
contributions can be sizable and lead to important constraints in the
$M_A$--$\tan\beta$ plane, as long as $\mu$ and $M_1 \lesssim
1$~TeV.  In the excluded regions with small $\tan\beta$ and a large
heavy Higgs mass, the constraint arises from the exchange of the light
Higgs.

\begin{figure}[tb]
\centering
\includegraphics[width=0.45\textwidth]{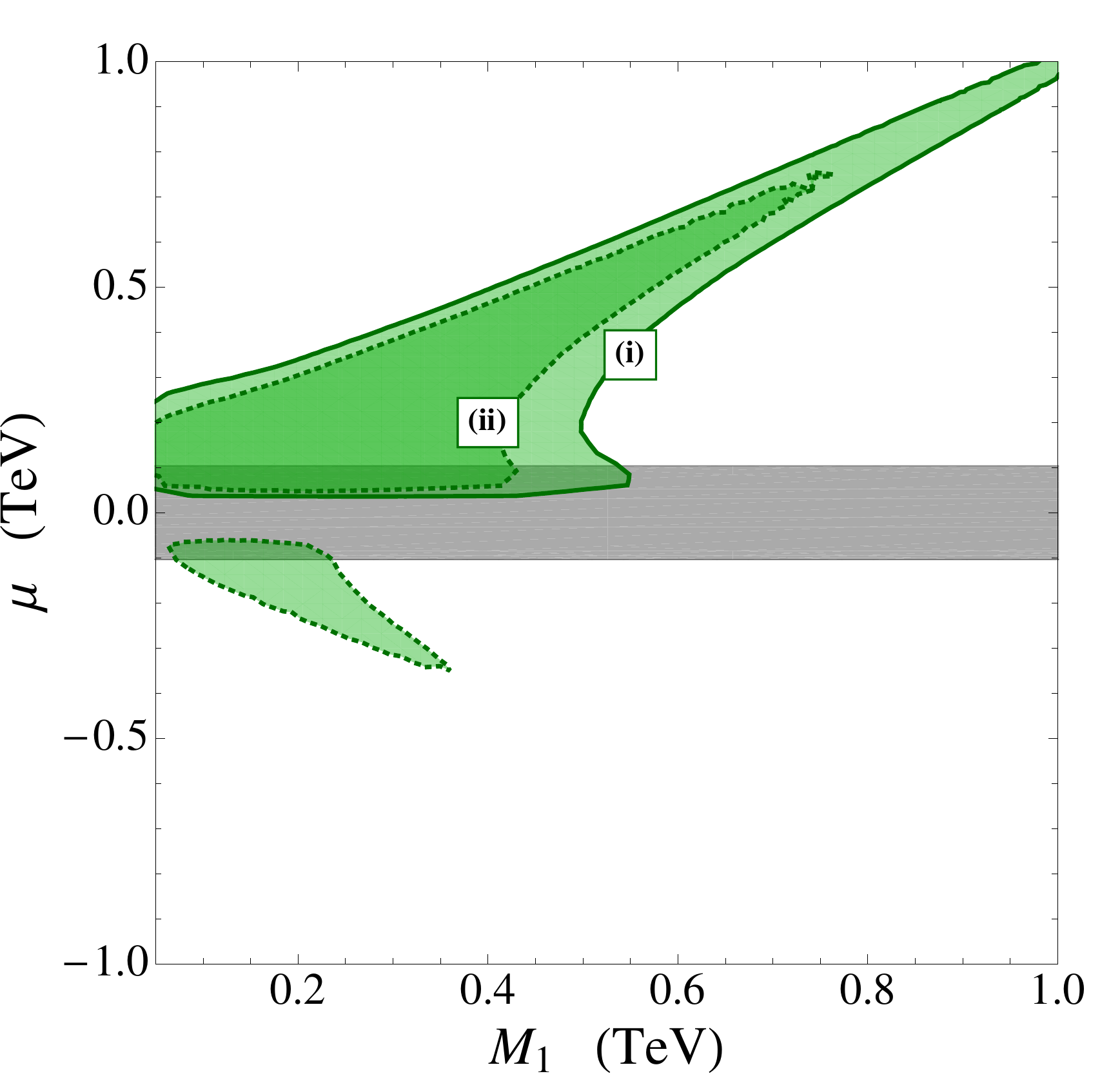}
\caption{Constraints in the $M_1$--$\mu$ plane from Dark Matter direct
  detection. The solid and dashed contours correspond to
  different choices for $M_A$ and $\tan\beta$ as defined in the
  text. The horizontal gray band is excluded by direct chargino
  searches.  }
\label{fig:darkmatter2}
\end{figure}

The plot of Fig.~\ref{fig:darkmatter2} shows the direct detection
constraints in the $M_1$--$\mu$ plane for 2 different points in the
$M_A$--$\tan\beta$ plane. The solid, dashed and dotted contours
correspond to $M_A = 800$~GeV and $\tan\beta = 45$ (scenario i) and
$M_A = 1$~TeV and $\tan\beta = 10$ (scenario ii), both compatible with
current direct searches.  As already mentioned, the strongest
constraints arise along the $M_1 \simeq \mu$ line.  Interestingly, the
constraints for negative values of $\mu$ are considerably weaker,
because for negative $\mu$, the heavy Higgs and SM like Higgs
contributions interfere destructively.  Observe that this behavior is
opposite to that of the constraints coming from $B_s \to \mu^+\mu^-$
and $B \to X_s \gamma$, which are currently weaker for positive $\mu$
(and positive $A_t$).

Note that the bounds from dark matter direct detection not only  depend 
very strongly on various MSSM parameters, but are also affected by
various uncertainties, {\it e.g.} from the nuclear matrix elements,
and astrophysical uncertainties, in particular the dark matter
velocity distribution.  Moreover, they also depend crucially on the
assumption that the dark matter of the universe indeed consist
entirely of MSSM neutralinos.  If neutralinos only make up a (small)
fraction of the dark matter, the bounds can be relaxed considerably
and even avoided completely.

\section{Conclusions} \label{sec:concl}

In this work, we evaluated the status of the minimal supersymmetric
standard model with minimal flavor violation in light of the recent
Higgs discovery as well as constraints from collider searches, flavor
measurements, and dark matter direct detection experiments.  In
concert, these complementary probes provide valuable constraints on
the MSSM parameter space.  In particular, we showed that flavor bounds
can be stronger than bounds from direct searches for heavy MSSM Higgs
particles or supersymmetric particles, even in the restrictive
framework of MFV.

Throughout our analysis, we consistently implemented the most general
structure of the soft SUSY breaking terms compatible with the MFV
ansatz, {\it i.e.} allowing splitting between the first two and the
third generations of squarks.  We demonstrated that, in addition to
the typical pMSSM parameters, an additional parameter, $\zeta$, reflective of
the alignment of the mass splitting of the left-handed squarks, is
required to discuss the flavor phenomenology of this framework.  In
the presence of such splitting, this parameter controls the size of
gaugino-squark loop contributions to FCNCs.  Possible cancellations 
between gaugino and higgsino loop contributions have a very strong 
dependence on $\zeta$. We showed its impact in the $B_s \to \mu^+ \mu^-$ 
and $B \to X_s \gamma$ decays and presented expectations for its 
magnitude as dictated by RGE running.

We discussed the constraints from direct searches of the heavy MSSM
Higgs bosons.  Bounds from $H/A \to \tau^+ \tau^- $ searches mainly
depend on $M_A$ and $\tan \beta$ and are robust against variations of
other SUSY parameters.  Separately, searches in the $H/A \to bb$
channel show a stronger dependence on the parameters under
consideration, in particular on the sign and magnitude of the Higgsino
mass parameter, $\mu$, and therefore provide complementary
information.  Currently, however, the $H/A \to \tau^+ \tau^-$ searches
are more strongly constraining for the considered scenarios.

On the flavor side, we considered the tree level decay $B \to \tau
\nu$ as well as the loop induced FCNC processes $B_s \to \mu^+ \mu^-$
and $B \to X_s \gamma$.  The recent experimental updates on the BR$(B
\to \tau \nu)$ show reasonable agreement with the SM prediction. At
tree level, charged Higgs contributions to $B \to \tau \nu$ interfere
destructively with the SM amplitude. At the loop level, a net
constructive interference is in principle possible for very large and
negative $\mu \tan\beta$. However, we find that the corresponding
regions of parameter space are excluded by vacuum meta-stability
considerations.  The $B \to \tau \nu$ decay can lead to constraints in
the $M_A$--$\tan\beta$ plane also for $M_A > 800$ GeV where current
direct searches for MSSM Higgs bosons end. For such heavy Higgs bosons
however, $B \to \tau\nu$ only probes very large values of $\tan\beta
\gtrsim 60$. The $B \to \tau\nu$ constraints depend only moderately on
SUSY parameters other than $M_A$ and $\tan \beta$.  In particular,
they depend only weakly on possible new sources of flavor violation
beyond the MFV ansatz.

The constraints from the FCNC decays on the $\tan \beta$--$M_A$ plane
depend crucially on several parameters, in particular the Higgsino
mass, $\mu$, the stop trilinear coupling, $A_t$, the gluino mass, $M_3$, the mass splitting of
the left-handed squarks, $\Delta Q_{13}$, and its alignment in flavor
space, $\zeta$.  The current experimental bounds on the BR$(B_s \to
\mu^+ \mu^-)$ lead to strong constraints in the large $\tan \beta$
regime of the MSSM with MFV.  Constraints are particularly strong if
the MSSM contributions interfere constructively with the SM, which
happens for sign$(\mu A_t)$ = -1.  In that case, even for moderately
large $\tan \beta \sim 30$, heavy Higgs masses of up to 1~TeV can be
probed.  Note that these bounds can have a strong dependance on
$\Delta Q_{13}$ and $\zeta$. For negative $A_t$, they become less
constraining for larger values of $\zeta$ and larger $\Delta Q_{13}$.
The main dependence is to a good approximation on the product $\zeta \times \Delta Q_{13}$.
In a mSUGRA setup this product is correlated with the mass splitting of the right-handed up squarks $\zeta \times \Delta Q_{13} \simeq \Delta U_{13}/2$.
From our RGE analysis of a simple mSUGRA model, we expect $\zeta=0.8$
for $\tan \beta =45$ and $\zeta$ even closer to 1 for smaller
$\tan\beta$.  We also find $\Delta Q_{13} \sim 20\%$ to $35\%$, which
should be approximately generic for SUSY breaking models with flavor
universal soft masses at the GUT scale and light gluinos $M_3 \lesssim
2$ TeV.  Such values have visible impact on the bounds derived from
BR$(B_s \to \mu^+ \mu^-)$.  If a lower bound on BR$(B_s \to \mu^+
\mu^-)$ above one half of the SM prediction is established in the
future, destructively interfering SUSY contributions will also be
highly constrained.

It is important to stress that the MSSM contributions to $B_s \to
\mu^+ \mu^-$ do not necessarily decouple with the SUSY scale, but can
probe masses of SUSY particles far above the scales that are currently
reached by direct searches.  On the other hand, the MSSM contributions
to the $B \to X_s \gamma$ decay do decouple with the SUSY scale, but
even so, the $B \to X_s \gamma$ decay can give nontrivial constraints
on the MFV MSSM parameter space.  If SUSY particles are heavier than
$\sim 2$ TeV, charged Higgs contributions to BR$(B\to X_s \gamma)$
still lead to a constraint for small $M_A$ which is almost independent
of all other parameters if $\tan\beta$ is not large. The corresponding
bound in the $M_A$--$\tan\beta$ plane can be stronger than the bounds
from direct searches for $\tan \beta \lesssim 5$ and rules out
$M_A\lesssim 300$ GeV if squarks are decoupled.  For a TeV scale SUSY
spectrum, SUSY loops can also contribute sizably to $B \to X_s
\gamma$.  This is particularly true for a sizable mass splitting
$\Delta Q_{13}$ and negative values of $A_t$, where Higgsino and
gluino loop contributions add up constructively.  Again, $\zeta$ can
impact the implied constraints significantly.  In contrast to $B_s \to
\mu^+ \mu^-$, however, the bounds become stronger for larger values of
$\zeta$, if $A_t$ is negative.  A main conclusion of our work is that
the current bounds from $B \to X_s \gamma$ and $B_s \to \mu^+ \mu^-$
are minimized if both $\mu$ and $A_t$ are positive.  In this region of
parameter space, $(g-2)_\mu$ generically prefers a positive sign of
$M_2$.

We remark that the discussed FCNC $B$ decays are also sensitive to
sources of flavor violation beyond MFV.  For the MSSM with generic
flavor violating structures, however, bounds from FCNC processes
become significantly more model dependent.

Finally, we analyzed the impact of the updated bounds from dark matter
direct detection searches.  We found that the parameter space region
where $M_1 \simeq \mu$ is ruled out throughout the whole
$M_A$--$\tan\beta$ plane.  Away from bino--higgsino degeneracy, the
current Xenon100 bounds still give strong constraints in the
$M_A$--$\tan\beta$ plane as long as $M_1$ and $\mu$ are below 1~TeV
and $\mu$ is positive.  The direct detection bounds are minimized for
negative $\mu$, where light and heavy scalar contributions to the
neutralino-proton cross section partially cancel.  These direct
detection constraints are the least robust among the considered
bounds, since they are subject to important nuclear and astrophysical
uncertainties and depend crucially on the assumption that the lightest
MSSM neutralino constitutes the entire dark matter in the universe.

In summary, we presented the viable MSSM parameter space using the MFV
assumption, incorporating the discovery of a Higgs state at 125 GeV,
the null direct search results for supersymmetric particles and for
$H/A \to \tau^+ \tau^-$ and $bb$, and constraints from $B$ and $K$
observables as well as dark matter direct detection searches.  We also
discussed and imposed electroweak vacuum meta-stability requirements, and
we illustrated expectations for $B$ flavor bounds arising from a
renormalization group running analysis of generic minimal supergravity
models.  Throughout, we have emphasized the connection between flavor
observables and direct collider searches in exploring the MSSM
parameter space.  This complementarity is not only important for
understanding the present status of the MSSM with MFV, but it is also
central to interpreting future experimental discoveries.

\section*{Acknowledgments}

We would like to acknowledge helpful discussions with Stefania Gori,
Arjun Menon and Carlos Wagner.  We thank the Aspen Center for Physics
for warm hospitality where part of this work was completed. The Aspen
Center for Physics is supported by the National Science Foundation
Grant No. PHY-1066293.  W.A. thanks the Galileo Galilei Institute for
Theoretical Physics for warm hospitality and the INFN for partial
support during the completion of this work.  Fermilab is operated by
Fermi Research Alliance, LLC under Contract No. De-AC02-07CH11359 with
the United States Department of Energy. N.R.S is supported by the DoE
grant No. DE-SC0007859.

\section*{Appendix: Loop Functions}

The loop induced ``wrong'' Higgs couplings involve a single loop
function
\begin{eqnarray}
 g(x,y,z) &=& \frac{x\log x}{(x-y)(x-z)} + 
\frac{y\log y}{(y-x)(y-z)} \nonumber \\ 
&& + \frac{z\log z}{(z-x)(z-y)}~. \nonumber
\end{eqnarray}

The loop functions $h_{7,8}$ enter the charged Higgs contributions to
the $b \to s \gamma$ transition
\begin{eqnarray}
h_7(x) &=& \frac{3-5x}{12(1-x)^2} 
+ \frac{2-3x}{6(1-x)^3} \log x~, \nonumber \\
h_8(x) &=& \frac{3-x}{4(1-x)^2} 
+ \frac{1}{2(1-x)^3} \log x~. \nonumber
\end{eqnarray}

The loop functions that enter the Higgsino, gluino, and Wino
contributions to the $b \to s \gamma$ transition can be written as
\begin{eqnarray}
f^{\tilde H}_7 &=& f_1 + \frac{2}{3} f_2 ~,~~ 
f^{\tilde g}_7 = -\frac{8}{9} f_2 ~,~~ 
f^{\tilde W}_7 = -f_3 - \frac{1}{2} f_2 ~, \nonumber \\
f^{\tilde H}_8 &=& f_2~,~~  
f^{\tilde g}_8 = - \frac{1}{3} f_2 - 3 f_1 ~,~~ 
f^{\tilde W}_8 = -\frac{3}{2} f_2 ~, \nonumber
\end{eqnarray}
with
\begin{eqnarray}
f_1(x,y,z) &=& -\frac{x^2\log x}{(x-y)(x-z)^3} 
-\frac{y^2\log y}{(y-x)(y-z)^3} \nonumber \\
&& - \frac{(x^2y^2-3xyz^2+(x+y)z^3)\log z}{(x-z)^3(y-z)^3} \nonumber \\
&& + \frac{x(z-3y)+z(y+z)}{2(x-z)^2(y-z)^2}~, \nonumber \\[8pt]
f_2(x,y,z) &=& \frac{xz\log x}{(x-y)(x-z)^3} 
+\frac{yz\log y}{(y-x)(y-z)^3} \nonumber \\
&& + \frac{z(xy(x+y)-3xyz+z^3)\log z}{(x-z)^3(y-z)^3} \nonumber \\
&& + \frac{z(y-3z)+x(y+z)}{2(x-z)^2(y-z)^2}~, \nonumber \\[8pt]
f_3(x,y,z) &=& -\frac{z^2\log x}{(x-y)(x-z)^3} 
-\frac{z^2\log y}{(y-x)(y-z)^3} \nonumber \\
&& - \frac{z^2(x^2+xy+y^2-3(x+y)z+3z^2)\log z}{(x-z)^3(y-z)^3} \nonumber \\
&& + \frac{x(y-3z)+z(5z-3y)}{2(x-z)^2(y-z)^2}~. \nonumber
\end{eqnarray}



\end{document}